\documentclass[12pt,a4paper]{report}
\usepackage{mathpazo}
\usepackage{amsfonts}
\usepackage{amssymb}
\usepackage[intlimits]{amsmath}
\usepackage{mathrsfs}
\usepackage{fancyhdr}
\usepackage{epsfig}
\usepackage{graphicx} 
\usepackage{bbm}
\usepackage{color}

\newcommand{\der}[2]{\ensuremath{\frac{\partial #1}{\partial #2}}}
\newcommand{\Dsl}{{D\hspace{-7pt}{/}}}

\newenvironment{narrowmargin}{
\begin{list}{}
    {
    \setlength{\topsep}{0pt}
    \setlength{\leftmargin}{1cm}
    \setlength{\rightmargin}{1cm}
    \setlength{\listparindent}{\parindent}
    \setlength{\itemindent}{\parindent}
    \setlength{\parsep}{\parskip}
    }\item[]}
{\end{list}}

\def\bea{\begin{eqnarray}}
\def\eea{\end{eqnarray}}

\def\hmcu{\mathcal{U}}
\def\hmcp{\mathcal{P}}

\begin{document}
\pagenumbering{gobble}

\begin{center}{\Large\bf Dynamical fermions in lattice quantum chromodynamics}\\
\vspace{2.0cm}
K\'alm\'an Szab\'o\\
\vspace{0.5cm}
{\em Theoretical Physics Department, University of Wuppertal\\
Wuppertal 42119 Gaussstrasse 20, Germany}\\
\vspace{4cm}
PhD thesis, WUB-DIS 2007-10\\
advisor: Zolt\'an Fodor
\vspace{2.0cm}
\end{center}
\begin{narrowmargin}
The thesis will present results in Quantum Chromo Dynamics (QCD) with
dynamical lattice fermions. The topological susceptibilty in QCD is
determined, the calculations are carried out with dynamical overlap
fermions. The most important properties of the quark-gluon plasma phase
of QCD are studied, for which dynamical staggered fermions are used.
\end{narrowmargin}

\newpage
\tableofcontents
\newpage
\pagenumbering{arabic}

\chapter{Introduction}

The theory of the strong interaction is known to be Quantum Chromo Dynamics
(QCD). It has all the features which are necessary for a successful
description of the strong interaction.  
There are very important reasons why it is necessary to invest a lot of
effort to solve QCD:
\begin{itemize}

\item
Validate or invalidate QCD by comparing its predictions with experiments.
Results in the high energy regime show very good agreement with the
experiments, however there are still many white areas with no results at
all, among them is the missing connection between nuclear physics and QCD.

\item 
Validate or invalidate the Standard Model of particle physics. Even if QCD
is the proper theory of strong interactions, it can happen that the weak
and electromagnetic interactions are not correctly described by Standard
Model. Examining weak decays can only be done by taking into account
low-energy strong interaction effects. The success of the
(in)validation is now mostly depends on the precision of QCD calculations.

\item 
Unfold the phase diagram and properties of QCD at finite temperature and
baryon densities. In parallel with the theoretical
developments intensive experimental work is done (and will be done) 
to produce and investigate the high temperature phase of QCD:
the quark-gluon plasma. Among these investigations the major goal is to
find signals of a first-order or second-order transition.

\end{itemize}

Solving the above problems is known to be extremely difficult. Currently
available methods (e.g.. weak coupling perturbation, $1/N_c$
expansion\footnote{$N_c$ is the number of colors, in QCD $N_c=3$.}, string
theory methods, lattice) are not able to provide us with rigorous solutions.
However some of these methods are believed to give us very good approximations
of these solutions. 

Today the lattice technique is the only which is (or will soon be) able to
calculate masses of hadrons, properties of low energy scattering processes, bulk and
spectral properties of the quark-gluon plasma and many more based only on the
Lagrangian of QCD. It contains systematical errors, however these can be
quantified, therefore can be kept under control. In the following short
introduction to the lattice technique we will highlight the role of "dynamical
fermions" in lattice QCD.

\section{Dynamical fermions in lattice QCD}
Lattice QCD discretizes\footnote{
Introductory materials
covering the extended literature
are \cite{Montvay:1994cy,Gupta:1997nd,Rothe:2005nw}. 
Annual review of the field can be found 
in the lattice conference proceedings. 
To avoid the proliferation
of citations in the introduction we refer to these, and cite articles 
only in special cases.
}
the path integral ($Z$)
\bea
\label{eq:z}
Z=\int [dU][d\psi_f][d\bar{\psi}_f] \exp (-S_{\rm gauge} - \sum_f \bar{\psi_f}D[U]\psi_f)
\eea
on a four dimensional Euclidean lattice. 
The Euclidean space formalism is useful for obtaining the spectrum of the
theory or for doing finite temperature calculations. The Minkowski space approach, which
is necessary to investigate real time processes, is not available
(however see \cite{Berges:2006xc}). 
In Eq. \ref{eq:z} we have an integral over the gauge ($U$) and
flavored fermion fields
\footnote{For simplicity we have same quark masses
for the different flavors in this introductory section.} 
($\bar{\psi_f},\psi_f$). The $S_{\rm gauge}$ is the gauge action, the fermion action is 
bilinear in the fermion fields, so the fermion integral can be easily carried out. We end
up with the determinant of the Dirac-operator ($D[U]$) under the path integral:
\bea
Z=\int [dU] \det D[U]^{n_f} \exp (-S_{\rm gauge}),
\eea
with $n_f$ being the number of fermion flavors.
The minimal distance on the lattice is called lattice spacing ($a$).  The final
results are obtained by sending the lattice spacing to zero together with doing some
necessary renormalization.  

There are two important observations: firstly the Euclidean path integral is
equivalent with a Boltzmann sum of a statistical mechanical system and secondly
the discretized path integral in a finite volume can be put on a computer.
These properties made lattice QCD a multidisciplinary science: it is a mixture
of quantum field theory, statistical physics, numerical analysis and computer
science.  The development in computer algorithms and  the exponential rise of
the available computational capacity made lattice QCD from a toy model to a
powerful predictive tool, giving us high precision pre- and postdictions in a
huge number of areas.
 
Nowadays the calculations are reaching the \% level precision thanks to the
gradual elimination of the so called quenching effects. Quenching means approximating
the fermion determinant with a constant, $U$ independent value in Eq. \ref{eq:z} and keeping fermions only in
the correlation functions. It is used to decrease the computational
requirements, since taking into account the fermion determinant in the path
integral (in other words dealing with the fermions dynamically) is a hard task.
One can look on the system of \ref{eq:z} as a gauge system but with a highly
nonlocal\footnote{Nonlocality is driven by the smallness of the quark mass.}
effective action: \[ S_{\rm eff}=S_{\rm gauge} - n_f \log \det D.\] Developing
efficient algorithms for such systems is nontrivial, however there is a
considerable progress in the last years. 

There is a huge arbitrariness in choosing the type of the discretization, only
a few requirements are to be fulfilled: eg. it should have appropriate
symmetries or there should exist an equivalent local formulation. Universality,
a well-known concept from statistical physics ensures that in the zero lattice
spacing limit the results will not depend on the choice of the discretization.
Since the cost of algorithms usually goes with an enormous power of the inverse
lattice spacing, in practice it is desirable to improve the lattice actions,
that is to reduce their lattice artefacts to make the continuum extrapolations
easier from the available lattice spacings. However one should be careful with
the improvement: overimproving can lead to several practical problems (loss of
locality, unitarity, irregular continuum limit, slowing down of algorithms
etc.).

Even there is a possibility that for the fermion determinant and for the
fermion correlation functions one uses different discretizations (the
first are called sea, the latter are the valence fermions). This is the so
called mixed approach. Then the expensive, improved fermion is used
in the valence sector, whereas for the sea fermions a faster, less
improved is chosen. The correct continuum limit is again ensured by universality.

The design of lattice fermion actions is hindered by the fermion doubling
problem.  Naive discretization of the continuum Dirac action yields 16
fermions on the lattice. There are three different ways to cure this:
staggered, overlap and Wilson fermions. Let us take a brief look on all of
them.

\subsection*{Fast, but ugly\footnote{For a recent
review on the staggered controversy see \cite{Sharpe:2006re}.}: staggered fermions}

The naive fermion determinant containing 16 fermions has an $U(4)\times U(4)$
symmetry, which can be eliminated by the staggering transformation. This
reduces the degeneracy from 16 to four.  To get one out of the remaining
four fermions one applies the fourth root trick: 
\bea
\det D[U] \Longrightarrow \det D_{\rm st}[U] \Longrightarrow (\det D_{\rm
st}[U])^{1/4}.
\eea
The quark correlation functions are 
usually\footnote{Other than staggered quark
discretizations are often used in the correlation functions,
these are the so called mixed approaches with staggered sea quarks.}
calculated
with the four flavor operator
$D_{\rm st}$.

The main advantage of staggered fermions is that the $D_{\rm st}$ operator
has a $U(1)_{\epsilon}$ symmetry in the massless case at any finite
lattice spacing. This symmetry corresponds to the flavor non-singlet axial
symmetries in the continuum, an important organizing principle in
low-energy QCD. Thanks to this symmetry the spectrum of $D_{\rm st}$ is
bounded from below, making staggered algorithms well-conditioned.  The bare
quark mass is only multiplicatively renormalized. These
make the simulations fast and convenient.

The major problem is that no local theory is known to correspond to the
fourth-root trick, which can be a danger for the universality of the
theory. Which means that it can happen that results of staggered lattice
QCD differ from that of other discretizations. However any attack against
staggered QCD is in a hard position, it should give account for the
remarkable agreement between staggered lattice results and real world.

There is an explicit example \cite{Durr:2003xs}, where fourth root trick gives
an incorrect result: case of one massless fermion.  According to the anomaly
the chiral condensate should acquire some nonvanishing value in the continuum
limit with a proper fermion discretization.  However due to the
$U(1)_{\epsilon}$ symmetry of staggered fermions, the staggered chiral
condensate is always exactly zero at any finite lattice spacing (in a finite
volume), making staggered fermions fail at this setup.
 
Usual calculations are done with unphysical, large pion masses, and then
an extrapolation to the physical pion mass is carried out (staggered chiral
perturbation theory). These extrapolations are controlled by several
parameters ($O(40)$ in NLO for the kaon bag constant), which can make them
very ill-conditioned. 
 
\subsection*{Glorious, but slow: chiral fermions}

According to the Nielsen-Ninomya theorem eliminating the doubling problem
is equivalent with violating continuum chiral symmetry
($\{D,\gamma_5\}=0$) on the lattice. 
The idea of
Ginsparg and Wilson was to find a Dirac-operator satisfying 
\bea
\label{eq:gw}
\{D,\gamma_5\}=2D\gamma_5D.
\eea
Only much later was it realized, that 
this relation makes possible to maintain the chiral symmetry at finite lattice
spacing, for which an $O(a)$ redefinition of the chiral transformation is
needed. All continuum relations related to chiral symmetry
(Ward identities, index theorem, low energy theorems,
continuum chiral perturbation theory) are one to one applicable at finite
$a$ using a fermion satisfying this relation.
The overlap fermion and the fixed-point fermion are the known solutions of Eq.
\ref{eq:gw}, whereas the domain-wall fermion provides an approximation to such
operators.

The price of these nice properties is quite high:
one ends up with a non-ultralocal Dirac-operator (there
are interactions between points at any distance), which results in O(100)
times or more slower algorithms compared to other discretizations.

The determinant of a Ginsparg-Wilson Dirac-operator will have
discontinuities in the space of gauge fields at the topological sector
boundaries, just as the continuum Dirac-operator. It is a feature from one
hand, on the other hand these jumps makes the conventional dynamical fermion
algorithms with chiral Dirac-operators to slow down considerably. 

\subsection*{Robust: Wilson fermions}

Wilson fermions are curing the doubling problem by violating the chiral
symmetry drastically. One gets several inconvenient features at a first
sight: additive quark mass renormalization, $O(a)$ lattice artefacts (the
previous two discretizations have $O(a^2)$), loss of strict spectral bound
on the Dirac-operator. Latter yields ill-conditioned, slow algorithms.
The absence of the chiral symmetry makes necessary to evaluate
renormalization constants at those places where it is trivial in case of
staggered or overlap fermions. 

Much theoretical and numerical work was done to improve these properties:
using smeared gauge links in the fermion action the additive
renormalization can be decreased by two orders of magnitude, the $O(a)$
lattice artefacts can be reduced to $O(a^2)$ via the Symanzik-improvement
program, the spectral bound is reported to be recovered in the infinite
volume limit \cite{DelDebbio:2005qa}.

These made possible to be competitive with the staggered discretization in
speed and since it is in much better theoretical shape it might be the
choice for the near future.

\section{Overview of the thesis}

The use of dynamical fermions is nowadays obligatory in lattice QCD.
Developing new algorithms for dynamical fermions is still an active area of
research. In this thesis I will present two dynamical fermion projects in which
I have participated. 
 
The first topic, discussed in Chapter 2, deals with the dynamical overlap fermion project in
detail. 
This chapter is partially based on the articles:
\begin{itemize}
\item \cite{Fodor:2003bh} Z. Fodor, S.D. Katz, K.K. Szabo JHEP 0408:003,2004
\item \cite{Egri:2005cx} G.I. Egri, Z. Fodor, S.D. Katz, K.K. Szabo JHEP 0601:049,2006
\end{itemize}
First I will present our dynamical overlap algorithm, then I will show how the
naive algorithm fails to change topological sectors. A new algorithm is proposed
and tested, which solves the problem. Finally the topological sector changing
of the new algorithm is examined, and attempts to improve it are proposed. I am
trying to give a comprehensive review of the field, which also means that only
a part of the results belong to me. My contributions are the following:
\begin{itemize}
\item Writing and developing a 5000 line C program for generating
overlap fermion configurations.
\item Modifying the conventional HMC algorithm to circumvent its
failure at topological sector boundaries.
\item Improving the stepsize dependence of this algorithm.
\item Examining the tunneling behavior of this algorithm.
\item Performing simulations and measuring the topological susceptibility
in two flavor QCD.
\end{itemize}

The second topic is the dynamical staggered project (Chapter 3). This is a large
scale computation of thermodynamical properties of the quark gluon plasma.
I will describe our choice of action and the algorithmic improvements first.
Then I will show our determination on the order of the finite temperature
QCD transition in continuum limit and with physical quark masses. Next the
transition temperature in physical units is calculated, again in the
continuum limit and with physical quark masses. These results can be
considered as final ones modulo the uncertainty in the staggered
discretization.  Finally I will present the equation of state, but there
the calculations were only done with two lattice spacings, the continuum
limit is missing. The chapter is partially contained in these articles:
\begin{itemize}
\item \cite{Aoki:2006we} Y. Aoki, G. Endrodi, Z. Fodor, S.D. Katz, K.K. Szabo Nature 443:675-678,2006
\item \cite{Aoki:2006br} Y. Aoki, Z. Fodor, S.D. Katz, K.K. Szabo Phys. Lett. B643:46-54,2006
\item \cite{Aoki:2005vt} Y. Aoki, Z. Fodor, S.D. Katz, K.K. Szabo JHEP 0601:089,2006
\end{itemize}
Again not all results belong to me, my contributions are:
\begin{itemize}
\item Writing and developing a 5000 line C program for generating
staggered fermion configurations.
\item Performing large scale zero and finite temperature simulations. 
\item Analyzing and renormalizing the data.
\end{itemize}

\chapter{Dynamical overlap fermions}

Chiral symmetry is one of the most important feature of the strong
interaction. Lattice regularization and chiral symmetry were
contradictious concepts for many years. Fermionic operators satisfying the
Ginsparg-Wilson relation \cite{Ginsparg:1981bj} 
\bea
\label{eq:gw2}
\{D,\gamma_5\}=\frac{1}{m_0}D\gamma_5D.
\eea
made possible to solve the chirality problem of
four-dimensional QCD at finite lattice spacing \cite{Hasenfratz:1998jp,
Neuberger:1997bg,Niedermayer:1998bi,
Luscher:1998pq}. 

Several numerical studies with exact chirality operators were done in the
quenched approximation \cite{DeGrand:2000tf,Gattringer:2003qx,Babich:2005ay}.
The results were really compelling, but people were forced to work with
nonlocal Dirac-operators. This made the algorithms more complicated and slowed
them down by large factors. At the same time one could reach rather small quark
masses, which was unimaginable with Wilson-type discretizations before.  

The life becomes even more complicated when introducing dynamical fermions with
exact chirality. This chapter is devoted to this problem. We will going to work
with overlap fermions \cite{Neuberger:1997fp,Neuberger:1998wv}, it is an
explicit solution of the Ginsparg-Wilson relation. The other type of solutions,
the fixed-point Dirac operators \cite{Hasenfratz:1998ri} are defined via a
recursive equation, they are considerably harder to implement\footnote{For
dynamical simulation of an approximate fixed-point Dirac operator see
\cite{Hasenfratz:2005tt}.}. 

\section{Overlap Dirac operator}

First we fix our notations. The massless Neuberger-Dirac operator (or overlap 
operator) $D$ can be written as 
\begin{equation}\label{eq:overlap}
D=m_0 [1+\gamma_5 {\rm sgn}(H_W)],
\end{equation}
This $D$ operator satisfies Eq. (\ref{eq:gw2}).
$H_W$ is the hermitian Dirac operator, $H_W=\gamma_5 D_W$,
which is built from the massive Wilson-Dirac operator, $D_W$, 
defined by
\begin{eqnarray}\label{eq:wilson}
[D_W]_{xy}=(4-m_0)\delta_{xy}-\frac{1}{2} \sum_{\mu}
\left\{
U_{\mu}(x)(1+\gamma_\mu)\delta_{x,y-\mu}+
U_{\mu}^{\dagger}(y)(1-\gamma_\mu)\delta_{y,x+\mu}
\right\}.
\nonumber
\end{eqnarray}
One fermion is obtained in the continuum limit, if $m_0$ takes any value between $0$ and
$2$. 
In a finite volume one should be careful, that
the physical branch of the
spectrum of the $D_W$ operator is to be projected to the physical part 
of the overlap circle. 

The mass is introduced in the overlap operator by
\begin{equation}
D(m)=(1-\frac{m}{2m_0})D+m.
\nonumber
\end{equation}

Sometimes it is useful to consider the hermitian version of the overlap 
operator. Let us review its properties. The massless hermitian overlap operator is
$H=\gamma_5 D =m_0(\gamma_5 + {\rm sgn}(H_W))$. The eigenvalues are real, the
eigenvectors ($|\lambda \rangle$) are orthogonal and
span the whole space. Due to the Ginsparg-Wilson
relation
\bea
\{H,\gamma_5\}=\frac{1}{m_0}H^2,
\eea
the matrix elements of $\gamma_5$ satisfy:
\bea
\langle \lambda | \gamma_5 | \mu \rangle (\lambda+ \mu)=\frac{\lambda^2}{m_0}
\delta_{\lambda \mu}.
\eea
From this equation it follows, that the zeromodes can be 
chosen to be chiral, furthermore eigenvectors
with eigenvalues $\pm2m_0$ have positive/negative chirality.
The difference in the number of left and right handed zeromodes is proportional
to the trace of $H$:
\bea
\frac{1}{2m_0}{\rm Tr} H = \sum_\lambda \frac{\lambda}{2m_0} =
\sum_\lambda \langle \lambda | \gamma_5 | \lambda \rangle 
- (n(+0)-n(-0)) = n(-0)-n(+0).
\eea
This difference is called the index of $H$, it
can be considered as the definition of the topological charge on the lattice.
This is supported by the fact, that in the continuum limit the density 
${\rm tr} H$ converges to $\sim F\tilde{F}$. 
Since the right hand side can take only integer values,
we can immediately conclude that the overlap operator cannot be continuous function of
the gauge fields. It should be nonanalytic on the boundaries of topological sectors.

Further eigenvalues
are always coming in pairs,
the vector 
\bea
|-\lambda \rangle = \left( 1-\frac{\lambda^2}{(2m_0)^2}\right)^{-1/2} \left(\gamma_5 -\frac{\lambda}{2m_0} \right)|
\lambda \rangle
\eea
is an eigenvector with eigenvalue 
$-\lambda$. The $\gamma_5$ matrix leaves the subspace $\{|\lambda \rangle, |-\lambda \rangle\}$
invariant, it can be written as
\bea
\gamma_5=
\begin{pmatrix}
\frac{\lambda}{2m_0} &  \left( 1-\frac{\lambda^2}{(2m_0)^2}\right)^{1/2}\\
\left( 1-\frac{\lambda^2}{(2m_0)^2}\right)^{1/2} & -\frac{\lambda}{2m_0}
\end{pmatrix}.
\eea
Since $\gamma_5$ is traceless on the subspace $\{|\lambda \rangle, |-\lambda \rangle\}$ with 
$\lambda\neq \{0,\pm2m_0\}$, it should be also traceless for the $\lambda= \{0,\pm2m_0\}$ space. This 
requires that  
the numbers of $\pm2m_0$ eigenmodes satisfy 
\bea
\label{eq:nzero}
n(+2m_0)-n(-2m_0)=n(-0)-n(+0).
\eea

The overlap operator squared $H^2(m)$ commutes with $\gamma_5$. This is a trivial fact in
the $\lambda=\{0,2m_0\}$ subspaces, where even $[H(m),\gamma_5]=0$ is true. In $\{|\lambda \rangle, |-\lambda \rangle\}$ subspace it is proportional to the identity matrix 
\bea
H^2(m)=
\begin{pmatrix}
\left(1-\frac{m^2}{(2m_0)^2}\right)\lambda^2 + m^2& 0\\
0 & \left(1-\frac{m^2}{(2m_0)^2}\right)\lambda^2 +m^2
\end{pmatrix}.
\eea

\subsection{Numerical implementation}
In the sign function of Eq. (\ref{eq:overlap}) one uses ${\rm sgn}
(H_W)=H_W/\sqrt{H_W^2}$.  We usually need the action of the ${\rm sgn}
(H_W)$ operator on a given vector, which was studied many times in the
literature \cite{Edwards:1998yw,vandenEshof:2002ms}. The common in all algorithms is that their speed is
proportional to the inverse condition number of the matrix $H_W$. To make
the algorithms better conditioned, one can project out the few low-lying
eigenmodes of the matrix $H_W$ and calculate the ${\rm sgn}$ operator in
this space exactly:
\bea
\label{eq:sgn}
{\rm sgn}(H_W)=\sum {\rm sgn}(s) P_s + {\rm sgn} (QH_W),
\eea
where $P_s$ is the projector to the $s$ eigenspace of $H_W$ and
$Q=1-\sum_s P_s$. 
The projections were done by the 
ARPACK code. To speed up the projections we preconditioned the problem
with a Chebyshev-polynomial transformation \cite{Struckmann:2000bt}. 
We have taken the $n$-th order
approximation of the $\tanh (80(x+1))-1$ function in $[-1,1]$ interval:
\bea
T_n(x) \approx \tanh (80(x+1))-1.
\eea
This blows up very fast around $-1$. If we concentrate the interesting
part of the spectrum of $H_W$ there, then we make the job of the
eigenvector projecting algorithm considerably easier (its speed usually
depends on the distance between consecutive eigenmodes). That is we were
calculating the eigenvectors of $T_n(H_W^2/s_{\rm max}^2-1)$
instead of those of $H_W$. Since this is only a polynomial transformation,
only the eigenvalues are different, 
the eigenvectors should be the same.
The result is that the problem is much better conditioned using Chebyshev-polynomials.
The speed gain was almost an order of magnitude.

For the rest of the sign function (${\rm sgn}(QH_W)$) one can take her/his
favorite approximation: $\sigma(QH_W)$. We have considered two of them: 
Zolotarev rational function and Chebyshev polynomial.

The $n^{\rm th}$ order Zolotarev optimal rational approximation 
for $1/\sqrt{x}$ in some interval $[x_{\rm min},x_{\rm max}]$ can be expressed 
by elliptic functions (see e.g. \cite{Chiu:2002eh}), the coefficients can
be also determined by a Remes-algorithm.
A particularly useful form of the
approximation for the sign function is given by the sum of 
partial fractions
\begin{equation}\label{eq:parcfrac}
 \sigma_{\rm Zol}(x)
=x\left(a_0 +\sum_{i=1}^n {a_i \over x^2+b_i}\right),
\end{equation}
in usual cases $n\sim O(10)$. To get the approximation
of the sgn of a matrix $A$ one has to plug the $A$ into Eq.
\ref{eq:parcfrac}: ${\rm sgn} (A) \approx \sigma_{\rm Zol}(A)$.
The inversions appearing in this approximation all contain the
same matrix but with different shifts ($b_i$). There are so-called
multishift Krylov-space methods \cite{Frommer:1995ik,Jegerlehner:1996pm}, 
which can solve the system of $n$
inversions with the same number of matrix multiplications as needed to
solve only one system. Since matrix multiplications dominate such
algorithms, this means that practically for the cost of one inversion one
obtains the solutions of $n$ systems. When inverting the multishift system
it might be desirable to project out even more eigenvectors than before in
Eq. \ref{eq:sgn} and calculate
the inverse in this subspace exactly: $(s^2+b_i)^{-1}P_s$.
 
The Chebyshev-polynomial approximation ($\sigma_{\rm Cseb}$)
in principle performs similarly as
the Zolotarev rational function. Here one has to make multiplication
with the Wilson matrix many times ($\sim O(100)$), there is no
need for global summations as in Zolotarev case. There are architectures
(eg. Graphical Processing Units \cite{Egri:2006zm}), where global summation is a  bottleneck,
it should be avoided everywhere if possible. In these cases the
Chebyshev-approximation is a good choice.

\section{Hybrid Monte-Carlo}
Hybrid Monte-Carlo (HMC, \cite{Duane:1987de}) is the most popular method for simulating dynamical
fermions. There are many other choices possible, but HMC outperforms from
all of these. We will base our work on the HMC algorithm. One would like to
generate gauge configurations via a Monte-Carlo update with the following weight
(see \ref{eq:z}):
\bea
\label{eq:w}
w[U]=\exp(-S_{\rm gauge})\det (D^\dagger D),
\eea
where we have $D^\dagger D$ instead of $D^2$ for the weight of two fermion
flavors. This substitution is legal, since the one flavor fermion determinant
is real.
The standard procedure to implement the fermion
determinant is to rewrite it using bosonic fields, so-called pseudofermions
($\phi(x)$):
\bea
w[U]=\exp(-S_{\rm gauge}) \int [d\phi^\dagger d\phi] \exp (-\phi^\dagger
(D^\dagger D)^{-1}
\phi)
\eea

Now let us consider the following steps:
\begin{enumerate}
\item Choose Gaussian distributed momenta $P_\mu(x)$.
\item Choose a $\phi$ field according to the distribution $\exp
(-\phi^\dagger(D^\dagger D)^{-1}\phi)$.
\item At a fixed $\phi$ background evolve the $U_\mu(x)$ gauge fields 
and $P_\mu(x)$ momenta using the equations of motion derived from the
Hamiltonian 
\bea
H=\frac{1}{2}P^2+S_{\rm gauge}+\phi^\dagger(D^\dagger D)^{-1}\phi,
\eea
and from the structure of the manifold. The evolution from $(U,P)$ fields
to some $(U',P')$ via the equations of motion is usually called a 
trajectory.
\end{enumerate}
Iterating steps from 1. to 3. one obtains
a chain of gauge configurations with the distribution
Eq. \ref{eq:w}. Along a trajectory the energy and the area are conserved, moreover 
these trajectories are exactly reversible.
To determine the trajectories is an infinitely hard 
problem, it requires an exact solution of the equations of motion.

However an approximate solution of these equations can be also
used to build a chain of gauge configurations with the exact $w[U]$
distribution. One just has to find an area preserving and reversible
integrator and integrate the equations approximately with it. With such
an integrator at hand one
trajectory will be generated via the following procedure:
\begin{enumerate}
\item As before.
\item As before.
\item Integrate the equations of motion with an approximate integrator.
\item Calculate the energy difference: $\Delta H=H(U',P')-H(U,P)$ and
accept/reject the configuration with the probability ${\rm
min}(1,\exp(-\Delta H))$.
\end{enumerate}
This is one iteration of the HMC algorithm. This iteration also produces
configurations with the same equilibrium distribution (Eq. \ref{eq:w}) as the previous one.

The easiest area conserving and reversible integrator is the leapfrog. The
leapfrog integration consist of making $1/\epsilon$ of the following
steps\footnote{
In case of unit length trajectories.}:
\bea
\label{eq:lf}
\mathcal{U}(\epsilon/2) \mathcal{P}(\epsilon) \mathcal{U}(\epsilon/2),
\eea 
where $\hmcu$ operator evolves the gauge fields with the actual (fixed)
momenta, wheres $\hmcp$ operator evolves the momenta using the force calculated at
the actual (fixed) gauge field. The $\epsilon$ is called the stepsize,
obviously in the $\epsilon \to 0$ limit one arrives to the exact solution of the
equations of motion. The energy is violated
by $O(\epsilon ^3)$ by making one leapfrog step, during $1/\epsilon$ steps of a unit
length trajectory this error grows up to $O(\epsilon ^2)$. The average
energy conservation violation can be shown to be $\langle \Delta H
\rangle= C\epsilon^4 + \dots$. Since $C$ is most generally proportional to the
volume, the stepsize should be decreased as $\epsilon \sim V^{-1/4}$ to
keep constant acceptance.

Using area conservation and reversibility one can show that the condition
\bea
\langle \exp (-\Delta H) \rangle =1
\eea
should be satisfied, this relation provides an easy 
check of the consistency of the algorithm. This condition also shows us that if one has large 
energy conservation violations ($|\Delta H|\gtrsim 1$), then the acceptance should be small.  In this
case one should decrease the stepsize to obtain a good acceptance. The optimal 
acceptance rate depends on the type of the integrator,
in case of the leapfrog and its variants the optimum 
is around $80\%$.  

Even one can drop away the area preservation property for
the evolution of the gauge fields and momentum, in this case one has to
include the Jacobian of the mapping into the accept/reject step. 

\subsection{HMC for two flavors of overlap fermions}

In case of the overlap fermion the gauge field and momentum evolution is
the following:
\bea
\label{eq:update}
\hmcu(\epsilon): U \to \exp(\epsilon P) U, &&
\hmcp(\epsilon): P \to P - \epsilon \mathcal{A} \left[U\frac{\partial}{\partial U}
\left(S_{\rm gauge} +
S_{\rm pf}\right)\right],
\eea
where in the force term the $\mathcal{A}$ operator projects onto traceless,
antihermitian matrices (in color indices). The complication arises in
the derivative of $S_{\rm pf}$, which schematically can be written as 
\bea
\delta S_{\rm pf} =-\psi^\dagger \delta (D^\dagger D) \psi 
\qquad\qquad {\rm with}\qquad\qquad \psi=(D^\dagger D)^{-1} \phi.
\eea
The inversion of the fermion operator 
$\psi=(D^\dagger D)^{-1}\phi$ 
is done by $n_o$ conjugate gradient\footnote{
For studies with inverters other than conjugate gradient see \cite{Arnold:2003sx}.}
steps (''outer inversion''). Note, 
however, that each step in this procedure needs the calculation of 
$(D^\dagger D)\phi$. The operator $D$ contains 
$\sigma(H_W)$, which is given for example  by the partial fraction expansion 
(see Eq. \ref{eq:parcfrac}). Thus, at each ''outer'' conjugate
gradient step one needs the inversion of the $H_W^2$ matrix (''inner
inversion''). This nested type of the inversions is the price one has to pay for an
exactly chiral Dirac-operator, in other formulations one only has one
matrix inversion per force calculation.
No method is known to avoid the nested inversions. 

The $\delta (D^\dagger D(m))=\delta (H(m)^2)$ derivative has a complicated form.
Let us assume, that we treat the sgn function in the overlap operator as in Eq. \ref{eq:sgn}.
Then the derivative of $H(m)$ will have two parts
\bea
\delta H(m)= \delta_1 H(m) + \delta_2 H(m).
\eea
The first part will contain the derivative of the projectors ($P_s=|s\rangle \langle
s |$), this term comes not only as a derivative of $\sum {\rm sgn}(s) P_s$, but the
projectors are also there in the $Q{\rm sgn}(H_W)$ term. All together one obtains
\bea
\delta_1H(m)=\left( m_0 -m/2 \right)\sum_{s} \delta P_s ( {\rm sgn}(s) - \sigma(H_W)
).
\eea
The derivative of the projector can be derived with the tools of
quantum mechanical perturbation
theory \cite{Narayanan:2000qx}:
\bea
\delta P_s =-\left( \frac{1-P_s}{H_W-s}\delta H_W P_s 
+ P_s \delta H_W \frac{1-P_s}{H_W-s} \right).
\eea
As we can see
each projected mode brings an extra inversion of the (shifted) Wilson matrix. It might be 
safe to treat the few lowest lying modes of $H_W$ this way, 
but it is meaningless to calculate exactly the force of modes 
from the bulk of the spectrum.

The second term in $\delta H(m)$ will be the derivative of the sgn function approximation 
($\delta_2 H(m)$).
In case of the Zolotarev approximation one has
\bea
\delta \sigma_{\rm Zol}(A)= \delta A \left( a_0+\sum_{i=1}^n \frac{a_i}{A^2+b_i} \right)-
\sum_{i=1}^n \frac{a_i}{A^2+b_i}(A^2\delta A + A\delta A A)\frac{a_i}{A^2+b_i},
\eea
therefore the contribution to the derivative of $H(m)$ is
\bea
\delta_2H(m)=(m_0-m/2)Q \delta \sigma_{\rm Zol}(H_W).
\eea
In case of the polynomial approximation the formula is more complicated.

The {sgn} function is needed as three different places during the HMC
trajectory: in the force calculation one needs $H(m)^{-2}$ and $\delta H(m)$
and in the action calculation one needs $H(m)^{-2}$ again.  The presence of the
accept/reject step allows one to use different approximations at different
places. One can speed up the algorithm with large factors by carefully choosing
and tuning the approximations.  We were using the Chebyshev polynomial
approximation in the inversions with O(20) projected modes whereas for the
$\delta H(m)$ we chose the Zolotarev rational approximation with 2 projected
modes. The relative precision was always set to $10^{-6}$ everywhere. 

For somewhat different implementations of the standard HMC and for various
improvement techniques we refer here to the work of two other groups
\cite{Cundy:2004pza,DeGrand:2004nq,Cundy:2005pi,Schaefer:2005qg}.

\subsection{Criticism of the HMC for one fermion flavor}

The HMC described above works only for a positive definite fermion matrix. This
is suitable for two flavors. For one flavor one can take the square root of
the squared operator (RHMC algorithm).  A different way to get the square root
of the fermion determinant is to exploit the exact chiral symmetry of the
Dirac-operator \cite{Bode:1999dd,DeGrand:2006ws}.  As we have seen before
$H(m)^2$ has $n(+0)+n(-0)$ chiral modes with eigenvalue $m^2$, $n(2m_0)+n(-2m_0)$
chiral modes with eigenvalue $(2m_0)^2$ and all the other modes are doublets
with eigenvalue $\left(1-\frac{m^2}{(2m_0)^2}\right)\lambda^2 +m^2$. It is a
conventional wisdom that only those configurations contribute to the path
integral where either $n(+0)=0$ or $n(-0)=0$. If this is true then we can write
the two flavor fermion determinant as
\begin{eqnarray*}
n(\pm0)=0: \quad \quad
\det H(m)^2={\rm det} _\pm H(m)^2 {\rm det} _\mp H(m)^2 =  
[{\rm det} _\pm H(m)^2 ]^2 \left(\frac{m^2}{(2m_0)^2}\right)^{n(\mp0)}, 
\end{eqnarray*}
where the ${\rm det} _\pm$ determinants have to be restricted to positive/negative chirality
subspaces.
The numerical factor at the end of the formula takes into account that 
the zeromodes and $\pm 2m_0$
modes are not coming in chirality pairs
(see Eq. \ref{eq:nzero}).
At this point it is easy to perform the square root
\bea
\det H(m) = \left(\frac{m}{(2m_0)}\right)^{n(\mp0)} {\rm det}_\pm H(m)^2,
\eea
where the sign depends on the chirality of the zero modes of $H$.  Since
$H(m)^2$ is positive definite even on the definite chirality subspaces,
there is no obstacle to introduce
pseudofermions for its determinant. The contribution of the zeromodes and
$\pm2m_0$ modes can be taken into account by reweighting the observables with
them. 
 
One has to face with the following problem. Let us consider a trajectory which
starts with a gauge configuration where $n(+0)=0$ and ends where $n(-0)=0$.
The pseudofermion is generated at the beginning of the trajectory according to the distribution:
\bea
\exp (- \phi^{\dagger} H(m)^{-2} P_{+} \phi),
\eea
where $P_{+}$ projects on positive chirality.
If one consider the reversed trajectory (starting from the sector, where $n(-0)=0$)
the pseudofermion distribution is
\bea
\exp (- \phi^{\dagger} H(m)^{-2} P_{-}\phi),
\eea
with $P_{-}$ negative chirality projector.
This means that the reversed trajectory uses a different pseudofermion distribution 
than the original one. For the proof of the detailed balance one needs to have the
same pseudofermion distribution on both ends.
This algorithm
yields the violation of the detailed balance for trajectories where the ends are in topological
sectors with different signs. The usual
way out is that the trajectories are constrained to that part of the phase
space where e.g. $n(+0)=0$ is always satisfied. However this choice opens the
way of a possible ergodicity breaking, which is also hard to keep under control.

\section{Reflection/refraction}
\label{sec:rr}
In the previous section we have shown how to set up the traditional HMC for
overlap fermions.  Performing simulations on rather small ($6^4$) lattices the
acceptance rate was almost zero. The strange thing was that decreasing the
stepsize of the integrator did not help at all. Tracing down the problem, one
finds that there are sudden jumps in the microcanonical energy during the
trajectories. These jumps are usually in the order of O(10) or larger, the
trajectories where the energy violations are of this size are practically never
accepted in the final accept/reject step. 

These jumps occur at the discontinuity of the overlap operator, that is at the
topological sector boundaries.  The phenomena can be nicely observed in the
spectrum of the hermitian Wilson-Dirac operator. The topological charge can be written as
\bea
Q=\frac{1}{2m_0}{\rm Tr} H = \frac{1}{2}\sum_s \langle s| \gamma_5 + {\rm sgn} (H_W) | s \rangle =
\frac{1}{2} \sum_s {\rm sgn} (s),
\eea
where $s$'s are the real eigenvalues of $H_W$. The charge changes when an eigenvalue of $H_W$ crosses
zero. At this point most presumably the overlap operator itself
is discontinuous, since its trace is discontinuous. This means that the pseudofermion action also
has a discontinuity, which means
a Dirac-delta in the fermion
force. Obviously a finite-stepsize integrator will never notice the presence of
a Dirac-delta in the force. Without any correction one will end up with an
energy violation which is roughly the discontinuity in the fermion action. 

One can improve
on this situation.
This feature is already present in a classical
one-dimensional motion of a point-particle in a step
function potential. 
During the integration one should check whether the particle moved from
one side to the other one of the step function. If it is necessary,
one corrects its momentum and position.
This correction has to be done also in the case of the overlap fermion.
The microcanonical energy,
\begin{equation}
H=
\frac{1}{2}P^2+S_{\rm gauge}[U]+S_{\rm pf}[U,\phi]=\frac{1}{2}P^2
+S[U,\phi]
\label{eq:ener}
\end{equation}
has a step function type non-analyticity on the
the zero eigenvalue surfaces of the $H_W$
operator in the space of link variables coming from the pseudofermion action.
When the microcanonical trajectory reaches one of these surfaces,
we expect either reflection or refraction.
If the momentum component,
orthogonal to the zero eigenvalue surface,
is large enough to compensate the change of the action
between the two sides of the singularity ($\Delta S$)
then refraction should happen,
otherwise the trajectory should reflect off the singularity surface.
Other components of the momenta are unaffected.
The anti-hermitian normal vector ($N$)
of the zero eigenvalue surface
can be expressed with the help of the gauge derivative
as
\bea
N\sim \langle s | \mathcal{A}\left( U \der{H_W}{U^T}\right)| s \rangle, 
\eea
where $\mathcal{A}$ projects to the antihermitian, traceless matrices in color space. 
Table \ref{tab:frref} summarizes the
conditions of refraction and reflection and the new momenta.
\begin{table}[t!]
\begin{center}
\begin{tabular}{|c|c|c|}
\hline
& When & New momenta \\
\hline
Refraction & $ ( N,P) ^2 > 2\Delta S$ &
$P-N (N, P)  + N (N,P)\sqrt{1-2\Delta S/ ( N,P )^2}$ \\
\hline
Reflection & $(N,P) ^2 < 2\Delta S$ &
$P-2N (N,P) $ \\
\hline
\end{tabular}
\end{center}
\caption{\label{tab:frref} Refraction and reflection can happen to the system when approaches a zero eigenvalue
surface of $H_W$. The conditions and the new momenta are indicated. $P$ is
the momentum before the refraction/reflection.}
\end{table}

\subsection{Modified leapfrog}
We have to modify the standard leap-frog integration of the equations of motion
in order to take into account reflection and refraction. This can be done in
the following way. The standard leap-frog consists of three steps: an update of
the links with stepsize $\epsilon/2$, an update of the momenta with $\epsilon$
and finally another update of the links, using the new momenta, again with
$\epsilon/2$, where $\epsilon$ is the stepsize of the integration.  The system
can only reach the zero eigenvalue surface during the update of the links. We
have to identify the step in which this happens.  After identifying the step in
which the zero eigenvalue surface is reached, we have to replace it with the
following three steps:

\begin{enumerate}
\item Update the links with $\epsilon_c$, so that we reach exactly the zero eigenvalue
surface. $\epsilon_c$ can be determined with the help of $N$.
\item Modify the momenta according to Table \ref{tab:frref}.
\item Update the links using the new momenta, with stepsize $\epsilon/2-\epsilon_c$.
\end{enumerate}
This means that in leapfrog step of Eq. \ref{eq:lf} we have to substitute the appropriate
$\mathcal{U}(\epsilon/2)$ operator
with
\bea
\label{eq:lfmod}
\mathcal{U}_{\rm mod}(\epsilon/2)=
\mathcal{U}(\epsilon_c) \mathcal{R} \mathcal{U}(\epsilon/2-\epsilon_c), 
\eea
where $\mathcal{R}$ is the refraction/reflection operator, which changes the
momenta according to Tab. \ref{tab:frref}.  This procedure is trivially reversible and it
also preserves the integration measure as shown in the appendix of this
chapter. 

where $P'=\mathcal{R} P$ is the reflected/refracted momentum.
$F=\mathcal{A}\left( U \der{S}{U^T}\right)$ is the force evaluated at the
topological sector boundary on the starting side, which means that the
undefined ${\rm sgn}(0)$ in $F$ is interpreted as ${\rm sgn}(s-)$. $F'$ is also
evaluated on the boundary, but with ${\rm sgn}(0)={\rm sgn}(s+)$. Obviously
$F'=F$ in case of a reflection. The energy violation of Eq. \ref{eq:deltah} can be a serious problem,
since a $(P,F)$ type quantity is in general proportional to the volume. Which
means that one has to decrease the stepsize with $\epsilon \sim V^{-1}$ to keep
the acceptance constant. This is much worse than the original $\epsilon \sim
V^{-1/4}$ scaling of the leapfrog.

\begin{figure}[t!]
\begin{center}
\includegraphics*[width=8cm,bb=18 202 575 717]{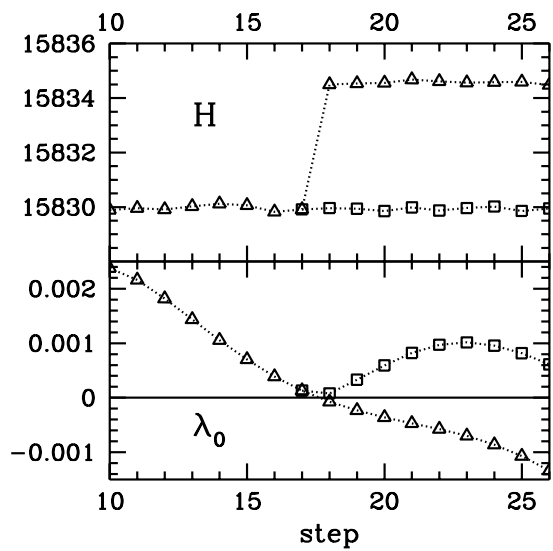}
\caption{Comparison of the unmodified (triangles) and modified (boxes)
leapfrogs.
Upper part is the energy, lower part is the lowest
mode of $H_W$.}
\label{fig:ref}
\end{center}
\end{figure}

Fig. \ref{fig:ref} compares the evolution of the energy and lowest $\lambda$
for the usual and for the modified leapfrog. In the unmodified case there is
a huge energy jump at the crossing, the trajectory is most probably
rejected.
Whereas in the modified case a reflection
happens, and $\mathcal{H}$ is much better conserved.
There is one bottleneck however.  If we correct only for the discontinuity in
the potential as above, then the microcanonical energy violation will be
proportional to the stepsize. That is for a $\mathcal{U}_{\rm
mod}(\epsilon/2)\mathcal{P}(\epsilon/2)$ step the energy violation is
\bea
\label{eq:deltah}
\Delta H = \epsilon_c ( (P',F')- (P,F)) + O(\epsilon^2),
\eea

\subsection{Improving the modified leapfrog}
There are several ways out of the problem. 
We will consider two
variants in this section:
\begin{itemize}
\item Reflection/refraction step with $C\epsilon$ energy violation, the $C$
coefficient being only an $O(1)$ size number instead of $O(V)$, which means that
the bad scaling is eliminated. 
\item Reflection procedure with $O(\epsilon^2)$ energy violation.
\end{itemize}
In the literature there exists a reflection/refraction procedure with $O(\epsilon^2)$ energy
violation \cite{Cundy:2005pi}, however it is somewhat
more complicated than these two improvements.

The large energy conservation violation of $\mathcal{U}_{\rm mod}$ arises,
since not the appropriate momentum is reflected/refracted. If one inserts two extra
momentum updates into 
Eq. \ref{eq:lfmod} 
\bea
\mathcal{U}(\epsilon_c) \mathcal{P}(\epsilon_c)\cdot 
\mathcal{R} \cdot
\mathcal{P}(\epsilon/2-\epsilon_c) 
\mathcal{U}(\epsilon/2-\epsilon_c), 
\eea
then the energy is conserved upto $O(\epsilon^2)$. This is due to the fact that
both $\mathcal{U}(\epsilon)\mathcal{P}(\epsilon)$ and
$\mathcal{P}(\epsilon)\mathcal{U}(\epsilon)$ conserves the
energy upto $O(\epsilon^2)$ and $\mathcal{R}$ conserves the energy exactly.
This step however violates the area conservation upto
$O(\epsilon)$, there is no free lunch. 

The first idea is based on the fact, that if one makes the extra momentum updates 
only in the space, which is orthogonal to the normalvector $N$
\bea
\mathcal{U}(\epsilon_c) \mathcal{P}_{\bot}(\epsilon_c)\cdot 
\mathcal{R} \cdot
\mathcal{P}_{\bot}(\epsilon/2-\epsilon_c) 
\mathcal{U}(\epsilon/2-\epsilon_c), 
\eea
then the area is exactly conserved again (as shown in the appendix). Now the energy
is still violated by $O(\epsilon)$ terms, but since the update in the orthogonal 
space is done $O(\epsilon^2)$ correctly, their coefficients is only an $O(1)$
number. The stepsize should be decreased only as $\epsilon \sim V^{-1/2}$. 

The second idea applies only for the reflection. It uses the following observation.
In a one dimensional case if the time required to reach the boundary ($\epsilon_c$)
and the time which is required to step away from the boundary ($\epsilon/2
-\epsilon_c$) are the same, then as a result of the reflection step, the
trajectory has been exactly
reversed. The area conservation, reversibility are obviously preserved, the energy
is conserved exactly. We can try whether these remain true in arbitrary
dimensions (the area conservation is proven in the appendix,
the exact reversibility and energy conservation upto $O(\epsilon ^2)$ are obvious).  Then we should insert a
\begin{equation}
\label{eq:corr}
\mathcal{U}(\epsilon_c)\mathcal{P}(\epsilon_c) \mathcal{R}
\mathcal{P}(\epsilon_c)\mathcal{U}(\epsilon_c)
\end{equation} 
step into the chain of leapfrogs, when the boundary is hit. 
In Eq. \ref{eq:lf} we have written the elementary leapfrog step in the $UPU$ form,
now let us write it in the $PUP$ order:
\begin{equation}
\mathcal{P}(\epsilon/2)\mathcal{U}(\epsilon) \mathcal{P}(\epsilon/2),
\end{equation} 
Now we split the evolution of the
links into two parts:
\begin{equation}
\label{eq:lf2}
\mathcal{P}(\epsilon/2)\mathcal{U}(\epsilon/2) \cdot \mathcal{U}(\epsilon
/2)\mathcal{P}(\epsilon/2).
\end{equation} 
Consider that the boundary would be crossed during one of the evolutions of
the links in Eq. \ref{eq:lf2}. Then replace the original leapfrog with the
following:
\begin{equation*}
\mathcal{P}(\epsilon/2)\mathcal{U}(\epsilon/2) \cdot \mathcal{U}(\epsilon_c)
\mathcal{P}(\epsilon_c) \mathcal{R} \mathcal{P}(\epsilon_c) \mathcal{U}(\epsilon_c) \cdot
\mathcal{U}(\epsilon/2) \mathcal{P}(\epsilon/2).
\end{equation*}
Now $\epsilon_c$ is the time to reach the boundary surface measured from the
midpoint of the leapfrog. Thus if the crossing would happen in the first
evolution then $\epsilon_c<0$, if in the second, then $\epsilon_c>0$.  

\subsection{Tracing the evolution of low lying eigenmodes}
We have to
trace the evolution of the low lying eigenmodes, since we are looking for the
moment when one eigenvalue crosses zero.  The eigenvectors and eigenvalues are
available at discrete times only (once or twice per time step),
therefore one has to pair the eigenvectors at time $t$ and time $t+\epsilon$.
We calculated the scalar products $\langle s'(t+\epsilon) | s(t) \rangle$ after
each link update, and our recipe was the following: the $s(t)$ has evolved to that
$s'(t+\epsilon)$ with which the scalar product was maximal. Of course this can
break down if the time step is too large. It is easy to show, that to make this
naive method work one has to decrease the stepsize as the volume is increased
with $\sim V^{-1}$. Expanding the $\langle s'(t+\epsilon) | s(t) \rangle$
one obtains:
\bea
\langle s'(t+\epsilon) | s(t) \rangle=
\delta_{s's}-\epsilon \langle s'(t) | \frac{dH_W}{dt}\frac{1-P_{s'}}{H_W-s'}| s(t) \rangle+ O(\epsilon^2),
\eea
where the derivative of $H_W$ is simply 
\bea
\frac{dH_W}{dt}=\left(P,\mathcal{A}(U\der{H_W}{U^T})\right).
\eea 
Clearly the $O(\epsilon)$ term is proportional to the volume, therefore the $\epsilon \sim
V^{-1}$ relation should hold to be able to keep track of the evolution of the eigenvectors.
If we use the derivatives of the eigenvectors, then we can get a considerably better scaling.
That is, instead of $\langle s'(t+\epsilon) | s(t) \rangle$ we calculate 
the scalar products at time $t+\epsilon/2$ using the eigenvectors and their derivatives
at $t$ and $t+\epsilon$:
\begin{eqnarray*}
\langle s'(t+\epsilon/2) | s(t+\epsilon/2) \rangle = \\
\left(\langle s'(t+\epsilon)| -\epsilon/2 \frac{d}{dt} \langle s'(t+\epsilon)| \right)
\left(|s(t)\rangle +\epsilon/2 \frac{d}{dt} |s'(t)\rangle \right) + O(\epsilon^2).
\end{eqnarray*}
We have not used this formula in practice, it is expensive (to monitor $N$ low lying eigenmodes
one has to make $N(N-1)$ Wilson-matrix inversions to apply the above formula).
Fortunately in all cases our stepsizes were always small enough that the eigenvector
identification with the naive procedure was no problem.

The stepsize should be also small to avoid the crossing of two or more
eigenvalues in a microcanonical time step. This happened very rarely, and
since the energy violation were usually very large in these cases, these
configurations were simply rejected.

\section{Numerical simulations}

In this section we will detail the particular implementation of our overlap HMC
variant. Afterwards we will show results on the topological susceptibility as
obtained from these simulations.

\subsubsection*{Gauge action}

For testing purposes the standard Wilson action was chosen as gauge action,
later on we moved to the tree-level Symanzik-improved action.  Apart from
decreasing the scaling violations in the gauge sector, the improvement is
beneficial from the overlap operator point of view, too.  Experience in the
quenched case shows that improved gauge actions can drastically reduce the
eigenvalue density of the negative mass hermitian Wilson-Dirac operator
\cite{DeGrand:2002vu}. Since this operator is the kernel of the overlap
operator, gauge action improvement speeds up the overlap inversion algorithms.
Since the topological sector change happens when an eigenvalue of the
Wilson-Dirac operator changes zero, improvement reduces the tunneling events at
the same time. Therefore one has to be careful not to overimprove the gauge
action (as it is done for the DBW2 action). The Symanzik tree level improved
action is the simplest improved gauge action.

\subsubsection*{Fermion action}

The fermion action is of two overlap fermions with standard
Wilson-operator as a kernel.  After test runs with thin-links, we started smearing
the links in the kernel operator via stout smearing procedure. The smearing
reduces the fluctuations of the gauge configuration, which is again helps
reducing the density of zeromodes of the 
Wilson-operator \cite{Kovacs:2002nz}.  It is an $O(a^2)$
redefinition of the gauge fields, so keeping the smearing recipe constant as
the lattice spacing goes zero will not change the continuum limit of the
theory. The stout smearing has the particular advantage compared to other
smearing techniques, that it is an analytic function of the thin 
gauge field \cite{Morningstar:2003gk}.
Therefore its derivative (which is needed to obtain the HMC force) can be
calculated exactly. In our simulations we were using two levels of stout
smearing with smearing parameter $\rho=0.15$. The speedup was almost an order
of magnitude compared to the unsmeared case.\\ The negative mass of the
Wilson-kernel was chosen to be $-m_0=-1.3$ in the smeared link case. For
smaller $m_0$ values would have been no small eigenvalues of the overlap
operator, the topology of gauge fields would have been always trivial. The
$-m_0=-1.3$ was chosen to be from a small $8^4$ run, where the topological
susceptibility started increasing from its zero value at $m_0=0$ and reached
its plateau value around $-m_0 \sim -1.3$.

\subsubsection*{Algorithm}

We have tried several variants of the HMC algorithm, which were discussed
in the previous section. For all of them we were using the
reflection/refraction modification in some way.  In addition to the standard
consistency tests (reversibility of the trajectories, $\epsilon^2$ scaling of
the action and $\langle \exp (-\Delta H)\rangle=1$) we performed a brute force
approach on $2^2$ and $4^4$ lattices. We generated quenched configurations,
then we explicitely calculated the determinants of $H(m)$.  These determinants
were used in an additional Metropolis accept/reject step. The hybrid
Monte-Carlo results agree completely with those of the brute force approach.

\subsubsection*{Results}

\begin{figure}
\begin{center}
\includegraphics*[width=10cm,bb=20 170 570 700]{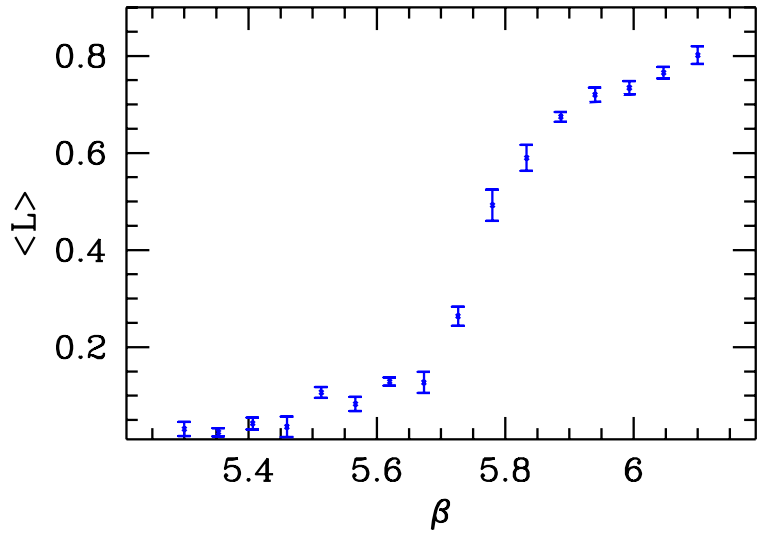}
\caption{
The $\beta$ dependence (right panel)
of the Polyakov-loop on $4\cdot 6^3$ lattices at $m=0.1$.}
\label{fig:ovl}
\end{center}
\end{figure}

\begin{figure}
\begin{center}
\includegraphics*[width=16cm,bb=46 522 573 707]{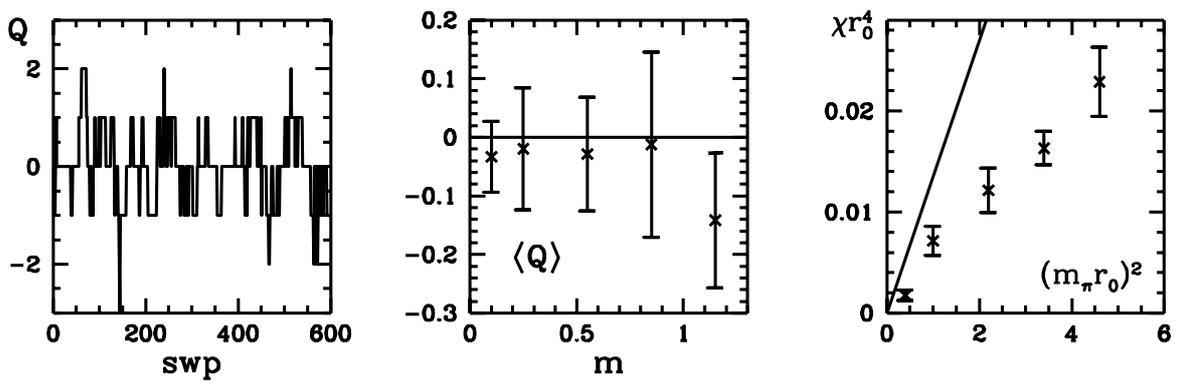}
\caption{Topology on $N_s=6$ lattices. See text.}
\label{fig:ts}
\end{center}
\end{figure}

Now let us take a closer look on the results obtained with standard Wilson
gauge action with standard Wilson fermion kernel in the overlap operator
(results with improved action will be discussed later).  On $4\cdot 6^3$
lattices there is a sharp increase in the Polyakov loop at $\beta=5.7$ (see Fig. \ref{fig:ovl}), which
can give a hint on the lattice spacing, since the finite temperature transition
is usually around $200$~MeV temperature ($T=1/(4a)$). 
This value of the coupling was used for measuring the topology
on $6^4$ lattices, which were considered as zero temperature lattices.  The
negative quark mass was set to $m_0=1.6$, the bare fermion mass was in the
range $m=0.1..1.15$, the stepsize was $\epsilon=0.025$ in average.  Using the
conventional HMC one would have no acceptance at all on these lattices, but
modifying the leapfrog step according to the previous section the acceptance
becomes $>70\%$ for these stepsizes.  At each bare mass roughly 800
trajectories were generated.  The results are plotted on Fig.  \ref{fig:ts}.
The left panel shows the charge history. The average topological charge is
consistent with zero for the total mass range (middle panel).  $12\cdot 6^3$
lattices were used to fix the scale using $r_0$ from Wilson-loops. The result
is $a \sim 0.25$ fm for small masses.  Pion masses were also measured and
$m_\pi^2=Am$ with $A\approx 1$ is found in lattice units.  Using the scale and
the pion mass, it is possible to get the topological susceptibility in physical
units (right panel of Fig.  \ref{fig:ts}).  $\chi(m)$ tends to zero for small
quark masses. One can compare these results with the continuum expectation in
the chiral limit (solid line of the figure):

\bea
\lim_{m \to 0} \chi(m) = \lim_{m\to 0}\frac{\langle Q^2 \rangle
}{V}=\frac{f_\pi^2m_\pi^2}{2n_f}.
\eea


\newpage 
\section{Topological sector changing}

In the previous two sections we have described a HMC algorithm for overlap
fermions.  The nonanalytic behavior of the overlap-operator at topological
sector boundaries requires non-trivial modification of the original HMC. Our
modification is able to handle the discontinuity problem as shown in the
numerical results section.

As we have started simulating even larger volumes ($8^4$) with the modified
algorithm, we had to face a new problem. In majority of the
reflection/refraction steps reflection happened, which means that the
trajectories were confined to a given topological sector for long times. This
dramatic increase of the autocorrelation time of the topological charge makes
the measurement of the topological susceptibility very hard and effectively
also means the violation of the ergodicity. 

One can come up with the solution to let the trajectories continue their
evolution as if the discontinuity in the action would not be present. This
algorithm would obviously allow the system to tunnel between topological
sectors. The price is that at the end of each trajectory  one has to keep the
$\exp(-\Delta H)$ factors to finally reweight the configurations with them.
The energy conservation violation is dominantly coming from the sum of the
discontinuities in the action $\Delta H= \sum_a \Delta S_a + O(\epsilon)$ along
a trajectory. If the system moves in a fixed potential, then $\Delta H$ will
take positive and negative values equal times, since some times the system goes
up sometimes goes down the same discontinuity.  The reweighting would have the
following form:
\bea
r[U_n]=\exp(-\sum_{i}^n \Delta H_i), \quad \quad \quad
\langle A \rangle = \frac{\sum_n A[U_n] r[U_n]}{\sum_n r[U_n]}.
\label{eq:rew}
\eea
If $\Delta H_i$'s are roughly equal times positive and negative, then the
reweighting works well: the configurations have nearly the same weight.
Unfortunately this turned out to be not true for the overlap HMC case, almost all
$\Delta H$'s were positive, the configurations were becoming unimportant very
fast in the sum of Eq. \ref{eq:rew}. 

How could this happen? The answer is that the evolution of the trajectories is
not done in a fixed fermion potential $S_{\rm exact}=-\log \det H(m)^2$, but in
a pseudofermionic one $S_{\rm pf}=\phi^{\dagger} H(m)^{-2} \phi$. The
pseudofermion is not fixed, it is regenerated at the beginning of each
trajectory.  Let us take a closer look on how the pseudofermions approximate
the fermion determinant. This will help us to understand the slowing down of
the tunneling between topological sectors.  In particular, we show that the
jump in the pseudofermionic action overestimates \(\Delta S_{\textrm{exact}}
\). 

Let us
assume that the trajectory crosses the boundary. Let \(H_{-}\) and \(H_{+}\)
be the overlap operator evaluated on the two sides of the boundary right
before and after the crossing, respectively. Clearly \(H_{-}\) and \(H_{+}\)
contain the same gauge configuration, but they differ, since one eigenvalue of
\(H_W\) changes sign on the boundary. In the HMC algorithm one chooses the
pseudofermion field as
\begin{equation*}
\phi=H_{-} \eta,\ \ \ \ \ \phi^\dagger=\eta ^\dagger H_{-}, 
\end{equation*}
where \(\eta, \eta ^\dagger\) are random vectors with Gaussian distribution,
in order to generate \(\phi, \phi^\dagger\) with the correct distribution. (In
a real simulation one chooses new pseudofermion configurations only at the
beginning of each trajectory, but for simplicity let's consider, that \(\phi\)
and \(\phi ^ \dagger\) are refreshed when hitting the boundary.)  The jump of
the pseudofermionic action now reads:
\begin{equation*}
\Delta S_{\textrm{pf}} =S_{\textrm{pf} +} - S_{\textrm{pf} -}= \eta^ \dagger (H_{-}
H^{-2}_{+} H_{-} -1) \eta
\end{equation*}
The relation between \(\Delta S_{\textrm{exact}}\) and \( \Delta
S_{\textrm{pf}}\) can be obtained by the following straightforward calculation:

\begin{equation*}
e^{- \Delta S_{\textrm{exact}}} = \frac{\det H^2_{+}}{\det H^2_{-}} =
\frac{\int [d \eta ^ \dagger] [d \eta] e^{- \eta^ \dagger \eta} e^{- \eta^ \dagger (H_{-} H^{-2}_{+} H_{-}
  -1) \eta}}{\int [d \eta ^ \dagger] [d \eta] e^{- \eta^ \dagger \eta}} =
\end{equation*}
\begin{equation*}
= \langle e^{-\eta^ \dagger (H_{-} H^{-2}_{+} H_{-} -1) \eta} \rangle 
_{\eta^{\dagger} \eta} \geq e^{- \langle \eta^ \dagger (H_{-} H^{-2}_{+} H_{-}
-1) \eta \rangle _{ \eta^{\dagger} \eta}} = e^{- \langle \Delta
S_{\textrm{pf}} \rangle } 
\end{equation*}
The inequality in the second line is a consequence of the concavity of the
\(e^{-x}\) function. So we conclude to:
\begin{equation*}
\langle \Delta S_{\textrm{pf}} \rangle \geq \Delta S_{\textrm{exact}}.
\end{equation*}

We can examine this relation in realistic simulations, if we take into
account, that there is a simple relation between \(H_{+}\) and
\(H_{-}\). Let's denote by \(\lambda_0\) the eigenvalue of \(H_W\) which
crosses zero at the boundary, and by \(|0 \rangle\) the eigenvector belonging
to \(\lambda_0\). With this notation:
\begin{equation*}
\label{eq:Hpm}
H_{+}=H_{-} + c |0 \rangle \langle 0 |,
\end{equation*}
where
\begin{equation*}
c=\Delta \textrm{sgn} \lambda_0 \ m_0 (1-\frac{m}{2 m_0}), 
\end{equation*}
with $\Delta \textrm{sgn} \lambda_0 = \pm 2$ being the jump of ${\rm sgn}
\lambda_0$ on the boundary.  The expectation value of the discontinuity in the
pseudofermionic action is:
\begin{equation*}
\langle \Delta S_{\textrm{pf}} \rangle = \langle \eta^ \dagger (H_{-}
  H^{-2}_{+} H_{-} -1) \eta \rangle _{ \eta^{\dagger} \eta} =
  \textrm{Tr}(H_{-} H^{-2}_{+} H_{-} -1) =
\end{equation*}
\begin{equation} 
\label{eq:Spf}
= \textrm{Tr} \big( (1-c |0 \rangle \langle 0 | H_{+}^{-1})(1-c\ H_{+}^{-1} |
0 \rangle \langle 0 | )-1 \big) = -2 c \langle 0 | H_{+}^{-1} | 0 \rangle + c^2
\langle 0 | H_{+}^{-2} | 0 \rangle. 
\end{equation}
In a similar way one can get a simple formula for the exact value of the jump
on the boundary:
\begin{equation}
\label{eq:Sex}
e^{-\Delta S_{\textrm{exact}}}=\frac{\det H_{+}^{2} }{\det H_{-}^{2}}=
\frac{1} {\det(H_{+}^{-1} H_{-})^2} = \frac{1}{\det(1- c H_{+}^{-1} |0 \rangle
\langle 0 |)^2} = \frac{1}{(1-c \langle 0 | H_{+}^{-1} | 0 \rangle)^2}.
\end{equation}
Eq. (\ref{eq:Spf}) and Eq. (\ref{eq:Sex}) offers a numerically fast way to
determine both action jumps, since one needs only one inversion of the overlap
operator to obtain both of them.

\begin{figure}[t!]
\includegraphics*[width=15.0cm,bb=18 433 591 718]{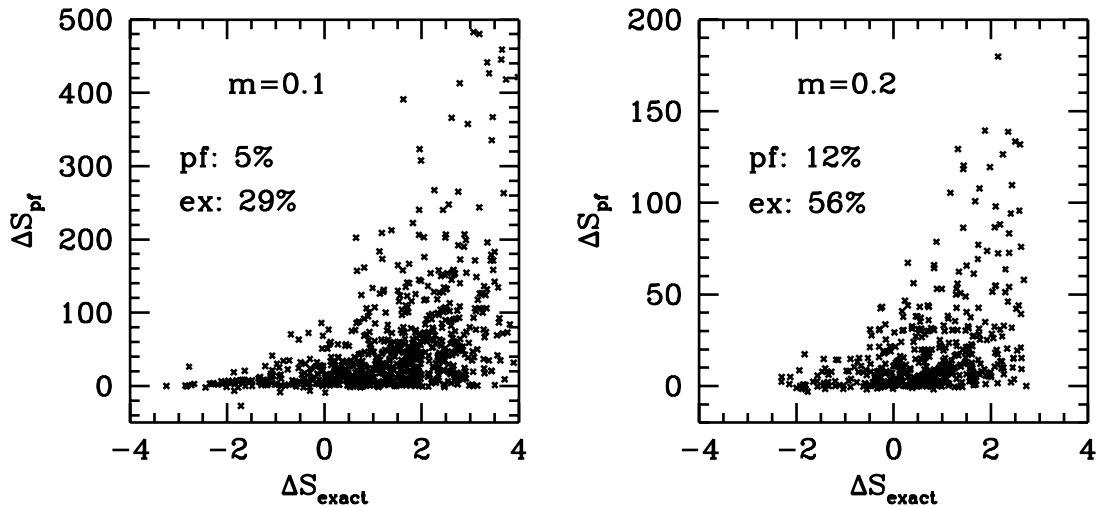}
\caption{\label{fi:sc}
The jump in the exact vs. pseudofermionic action at $\beta=4.05$ and
  $m=0.1, 0.2$. Since the average of $\langle n,p \rangle^2$ is around
  $\approx 1$, topological sector changing would happen considerably
  frequently using $S_{\rm exact}$, than with $S_{\rm pf}$. We also indicated
  the probability of topological sector changing with the pseudofermionic
  action, and an estimate on the probability using the exact action (assuming
  that the two algorithms would behave the same way except for the
  boundaries).  }  
\end{figure}
For illustration we made a scatter plot (Fig. \ref{fi:sc}) from a $6^4$
lattice at two different masses. 
From the joint distribution of ${\Delta S_{\rm exact},\Delta S_{\rm pf}}$
we can understand why are the tunneling events are so rare. Topological 
sector changing occurs when the HMC momentum of the system
in direction of the topological sector boundary surface is large
enough to "climb" the discontinuity (see Tab. \ref{tab:frref}). The 
momentum squared is usually an $O(1)$ number. As we can see
on Fig. \ref{fi:sc} the $\Delta S_{\rm pf}$ distribution
overestimates the real discontinuity $\Delta S_{\rm exact}$
with orders of magnitude. 
Therefore a crossing which would be possible with $\Delta S_{\rm exact}$
becomes impossible with $\Delta S_{\rm pf}$. The HMC which uses
$\Delta S_{\rm pf}$ stucks into a given topological sector.
The overestimation
becomes worse with lowering the quark mass.

One way to cure this is to use several pseudofermion estimators instead of one
\cite{DeGrand:2004nq}. More pseudofermions mean smaller spread of the
pseudofermionic action distribution, therefore the overestimation is smaller,
too.  However the computational time also increases with the number of extra
fields.  Obviously the best would be to use the exact action in the
simulations, but only its discontinuity on the boundary can be calculated
easily (the calculation of the exact fermion determinant is an $O(V^3)$
operation in general). In the following two subsections we show two different
ways to use the exact action jump instead of its pseudofermion estimator in the
simulations. Both of them are inexact, the errors present in the measured
quantities are of $O(\epsilon^2)$.

\subsection{Using $\Delta S_{\rm exact}$ to sew together simulations with fixed
topology }
\label{ssec:sew}

Let us write the partition function in the form (assuming a
vanishing \(\theta\) parameter):

\begin{equation*}
Z=\sum_{Q=-\infty}^{\infty} Z_Q,
\end{equation*}
where $Z_Q$ is the partition function of the topological sector $Q$.
The expectation value of an observable:
\begin{equation*}
\langle O \rangle = \frac{\sum_{Q} Z_Q \langle O   \rangle_Q}{\sum_{Q} Z_Q}
= \frac{\sum_{Q} \frac{Z_Q}{Z_0} \langle O \rangle_Q}{\sum_{Q}
\frac{Z_Q}{Z_0}}, 
\end{equation*}
where the restricted expectation value \(\langle O \rangle_{Q}\) is
\begin{equation*}
\langle O \rangle_{Q}=\frac{1}{Z_Q}\int [dU]_Q O[U] \det H^2_Q \exp (-S_g).
\end{equation*}
For reasons which will be clear later the integration goes not only over the
configurations with $Q$ charge, but also over the boundary of the topological
sector as well (though the boundary has only zero measure in this case).  When
calculating the partition function in a given topological sector the following
boundary prescription is used: we define the determinant on the boundary as
the limit of determinants approaching the wall from the $Q$ side ($\det
H^2_Q$).  If the measurement of the quantities \(Z_{Q+1}/Z_{Q}\) would be
possible, then we could recover \(Z_{Q}/Z_{0}\) for any \(Q\). With these in
hand, we would need only the restricted expectation values \(\langle O
\rangle_{Q}\), whose measurement doesn't require topological sector changings.

\subsubsection*{Measuring \(Z_{Q+1}/Z_{Q}\) using \(\Delta S_{\textrm{exact}} \)}

Now we will show a way to measure \(Z_{Q+1}/Z_{Q}\).  It will make use of the
fact, that we can calculate easily \(\Delta S_{\textrm{exact}} \) on the
boundary of topological sectors (see Eq. (\ref{eq:Sex})).  The pseudofermionic
action is only used to generate configurations in fixed topological sectors,
so its bad distribution for the jump of the action will not effect us.  (In
the following formulae $\Delta S$ will automatically mean $\Delta
S_{\textrm{exact}}$.)  The main idea is the following: an observable measured
in sector $Q$ is inversely proportional to $Z_Q$ and an observable in $Q+1$ is
to $Z_{Q+1}$. If the observables in the two sectors are concentrated only to
the common wall separating the two sectors, then from the ratio of the two
expectation values one can recover the ratio of the two sectors.

First let us measure in the $Q$ sector an operator, which is concentrated to
the boundary:
\begin{equation}
\label{eq:fd}
\langle \delta_{Q,Q+1} F\rangle_Q = \frac{1}{Z_Q}\int [dU]_{Q} \delta_{Q,Q+1}F[U]
\det H^2_Q
\exp (-S_g) 
, 
\end{equation}
where we introduced the distribution \(\delta_{Q,Q+1}\), a Dirac-\(\delta\),
which is equal to zero everywhere but on the \(Q, Q+1\) boundary.  Then let us
measure another operator $G$ on the same wall (thus on the boundary separating
sectors $Q$ and $Q+1$), but now from the $Q+1$ sector:
\begin{equation}
\label{eq:gd}
\langle \delta_{Q,Q+1} G\rangle_{Q+1} = \frac{1}{Z_{Q+1}}\int [dU]_{Q+1} \delta_{Q,Q+1}G[U]
\det H^2_{Q+1}
\exp (-S_g) 
. 
\end{equation}
The wall is the same (i.e. $[dU]_{Q}\delta_{Q,Q+1}=[dU]_{Q+1}\delta_{Q,Q+1}$)
in both cases, however due to our boundary prescription the determinants are
different on it. Therefore if $F$ and $G$ satisfies
\begin{equation}
\label{eq:db1}
F[U]\det H^2_{Q}[U]=G[U]\det H^2_{Q+1}[U]
\end{equation}
for configurations on the boundary,
then
the ratio of Eq. (\ref{eq:fd}) and Eq. (\ref{eq:gd})  gives us 
\begin{equation}
\label{eq:db2}
\frac{\langle \delta_{Q,Q+1} F\rangle_Q}{\langle \delta_{Q,Q+1}
G\rangle_{Q+1}}=\frac{Z_{Q+1}}{Z_Q}.
\end{equation}

\subsubsection*{Choosing $F[U]$ and $G[U]$ functions}

The easiest choice is $G(U)=1$ and $F(U)=\det H^2_{Q+1}/\det H^2_Q = \exp
(-\Delta S)$, the ratio of sectors becomes:
\begin{equation}
\label{eq:zq}
Z_{Q+1}/Z_Q=\frac{\langle \delta_{Q,Q+1} \exp (-\Delta S) \rangle_Q}{\langle \delta_{Q,Q+1}\rangle_{Q+1}}.
\end{equation}
This choice is still not optimal, since the measurement of the numerator is
problematic, if the distribution of $\Delta S$ extends to negative values.
The exponential function amplifies the small fluctuations in the negative
$\Delta S$ region, which can destroy the whole measurement: a very small
fraction of the configurations will dominate the result.  As a consequence one
ends up with relatively large statistical uncertainties.  With a slightly
different choice of $F$ and $G$ we can improve on the situation. With
$F(U)=\Theta (\Delta S - x) \exp (-\Delta S)$ and $G(U)=\Theta (\Delta S - x)$
we can omit the problematic part of the $\Delta S$ distribution (the values
smaller than $x$) from the measurement, and we get:
\begin{equation}
\label{eq:zq1}
Z_{Q+1}/Z_Q=\frac{\langle \delta_{Q,Q+1} \exp (-\Delta S) \rangle_Q^{\Delta S >x}}{\langle
\delta_{Q,Q+1}\rangle_{Q+1}^{\Delta S >x}}.
\end{equation}  
The price of this choice of $F,G$ is that we do not make use of the $\Delta
S<x$ part of our data set.  The value of $x$ can be tuned to minimize the
statistical error.

Let us note that Eq. (\ref{eq:db1}) can be viewed as a detailed balance
condition on a given $U$ configuration between $Q$ and $Q+1$ sector ($F$ and
$G$ are just the ``transition probabilities'').  This can give us a hint, that
the Metropolis-step is a good a solution for $F,G$: $F=\min(1,\exp(-\Delta
S))$ and $G=\min(1,\exp(\Delta S))$.  The ratio of sectors is simply:
\begin{equation}
\label{eq:zq2}
Z_{Q+1}/Z_Q=\frac{\langle \delta_{Q,Q+1} \min(1,\exp (-\Delta S)) \rangle_Q}{\langle \delta_{Q,Q+1} \min(1,\exp(\Delta S))\rangle_{Q+1}}.
\end{equation}
The inconvenient part of the distribution ($\Delta S<0$) is cut off, however
in contrast to Eq. \ref{eq:zq1} all configurations are used to get the
expectation values.

\subsubsection*{Expectation value of a Dirac-delta type operator}

Let us discuss briefly that in the framework of HMC, how to measure an
expectation value, which contains a Dirac-delta on the surface. The important
observation is that one can use the pseudofermionic action in the HMC to get
the fixed topology expectation values in Eq. \ref{eq:zq}, \ref{eq:zq1},
\ref{eq:zq2}. Inside a topological sector the behavior of the pseudofermionic
estimator is not an issue, we can use it instead of $S_{\rm exact}$ as usual.
In practice it is not possible to measure an operator containing a Dirac-delta
on the boundary surface on configurations generated by the pseudofermionic HMC,
because none of them will be exactly located on it.  If we would be able to
exactly integrate the equations of motion, then all inner points of the
trajectories could have been taken into the ensemble.  Those ones also, which
are located exactly on the surface. Here one would pick up a contribution from
the Dirac-delta to the above expectation values, at the inner points the
contribution would be zero. In the real case the trajectories differ by
$O(\epsilon^2)$ from the exact ones\footnote{In order to have only an
$O(\epsilon^2)$ difference one has to use an improved modified reflection step
as described in the previous section.}. Here using the above procedure
(measuring the $F[U]$ and $G[U]$ operators on the boundary and summing them up
along the trajectories) one makes $O(\epsilon^2)$ errors in expectation values.

\subsubsection*{Summarizing the new technique}

We have achieved our main goal: without making expensive topological sector
changes we can obtain the ratio of sectors (see Eq. \ref{eq:zq}, \ref{eq:zq1},
\ref{eq:zq2}).  The key point is to make simulations constrained to fixed
topological charge, and match the results on the common boundaries of the
sectors.  Since no sector changing is required, the inconvenient distribution
of the pseudofermionic action jump on the boundary will not effect the
measurement of the ratios of sectors.  The exact action is needed only on the
boundary: the formulas \ref{eq:zq}, \ref{eq:zq1}, \ref{eq:zq2} require
\(\Delta S\).

Obviously an important issue for this new method is whether topological sectors
defined by the overlap charge are path-connected or not.  In
\cite{Luscher:1998du} it has been proven that Abelian lattice gauge fields
satisfying the admissibility condition can be classified into connected
topological sectors.  No result is known for non-Abelian groups or
non-admissible gauge fields.  (Though there are some concerns on the structure
of the space of non-Abelian lattice gauge fields \cite{Adams:2002ms}.) If
configurations with the same $Q$ would not be continuously connectable in
sector $Q$, then our assumption that we make measurements on the common
boundary of sectors could be violated. It could happen, that the wall sampled
from sector $Q$ does not coincide with the wall sampled from $Q+1$.  Moreover
the fixed sector simulations would also violate ergodicity in this case. Let us
note here that the large autocorrelation time for the topological charge in the
conventional pseudofermionic HMC effectively also causes the breakdown of
ergodicity. In case of non-connected sectors one can cure these problems by
releasing the system from a sector after a certain amount of time and closing
it to another.

\subsection{Using $\Delta S_{\rm exact}$ in R-algorithm}
In the following we will describe another technique, which uses the $\Delta
S_{\rm exact}$ and can circumvent the critical slowing down of the topological
sector change.
If one does not insist on an exact algorithm, then an R-algorithm \cite{Gottlieb:1987mq}
where the $\Delta S_{\rm exact}$'s are taken into account can be a particularly
good choice.  Let us describe it shortly.  Instead of evolving the trajectory
in a pseudofermion potential (see Eq. \ref{eq:update}), one can try to estimate the exact
force by a random vector:
\bea
\der{}{U}\log \det H(m)^2 \sim R^{\dagger} H(m)^{-2}
\der{H(m)^2}{U}R.
\eea
Usually one estimator ($R$) per integrator step is used, so the
approximation might be poor.  If the stepsize goes to zero, then on a fixed
time interval the number of estimators will diverge making the approximation
exact.  Since there is no recipe, how to make the R-algorithm at finite
stepsize exact (like the accept/reject step in the HMC algorithm) the stepsize
extrapolation is a necessary ingredient. The stepsize error scales with
$O(\epsilon^2)$.  When a trajectory hits the topological boundary surface, then
one just has to modify the trajectory according to the reflection/refraction
rules, but now one can use the $\Delta S_{\rm exact}$ discontinuity instead of
a badly behaving estimator (eg. $\Delta S_{\rm pf}$). The modified leapfrog
step is not necessarily to be an exactly area conserving one (since stepsize
errors are already present).  But still it is required,  that the errors caused
either in the energy or in the area conservation are minimal (a good candidate
is the leapfrog-in leapfrog-out, which conserves the energy upto
$O(\epsilon^3)$ and the area upto $O(\epsilon^2)$).

\section{Numerical simulations 2.}
In the previous section we described two methods, to solve
the topological sector changing problem of pseudofermionic HMC simulation.
We were extensively using the first one (see subsection \ref{ssec:sew}). 
Here we describe the details of these simulations, and finally
give the topological susceptibility in physical units measured on $8^4$ and
$8^3\times 16$ lattices.

Simulations were done using unit length trajectories, separated by momentum and
pseudofermion refreshments. The system was confined to a fixed topological
sector in each run, we reflected the trajectories whenever they reached a
sector boundary.  The end points of the trajectories obviously follow the exact
distribution in a given sector, usual quantities can be measured on them.
We compared a few observables (plaquette, size of the potential wall) in a
given topological sector, but in different runs. We have not found any sign
indicating that the sectors were disconnected.
When calculating the ratio of sectors using Eq. \ref{eq:zq} or Eq.
\ref{eq:zq1} or Eq.  \ref{eq:zq2} we integrated along the trajectories,
this quantity will be burdened by a step size error.
We carried out simulations at one stepsize.

\begin{figure}
\includegraphics*[width=15.0cm,bb=18 433 591 718]{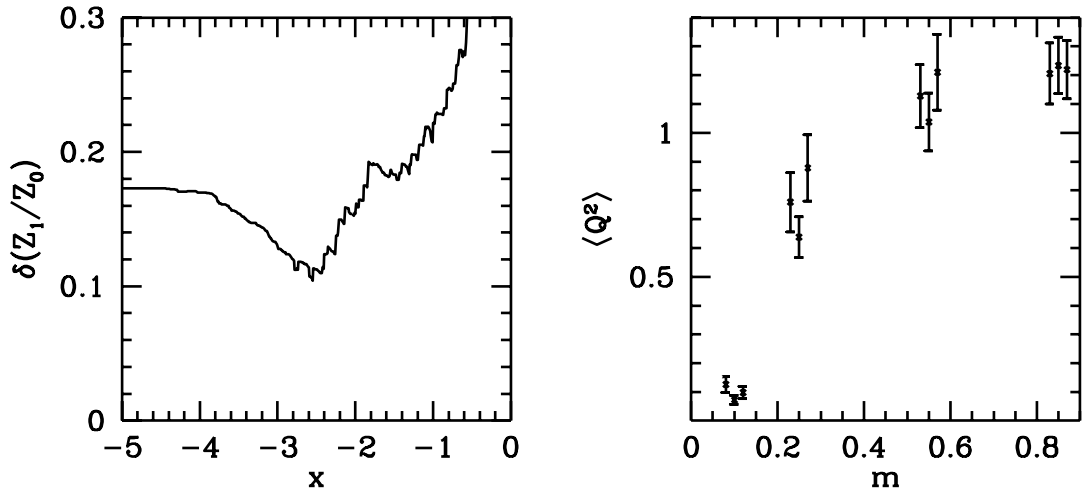}
\caption{
\label{fi:test}
{\it Left panel:} a typical optimization procedure of the lower limit ($x$) on
$\Delta S$ in the formula (\ref{eq:zq1}).  The statistical error of the ratio
$Z_1/Z_0$ shows a minimum as the function of $x$, which is considered as the
optimal value. {\it Right panel:} Bare mass dependence of topological
susceptibility using three different methods on $6^4$ lattices.  The points
corresponding to the same mass were slightly shifted vertically for
clarity. Result based on our new technique and Eq. \ref{eq:zq1} is on the
left, based on Eq. \ref{eq:zq2} is in the middle, the standard
pseudofermionic HMC is on the right.  The simulation parameters are from
\cite{Fodor:2004wx}.} 
\end{figure}

In case of large enough statistics the value of $Z_{Q+1}/Z_Q$ should be the
same, independently which of the three formula \ref{eq:zq}, \ref{eq:zq1} and
\ref{eq:zq2} was used to calculate it.  We omit Eq. \ref{eq:zq} in the
following, since it is hard to give a reliable error estimate on the
expectation value of $\exp (-\Delta S)$, if $\Delta S$ can be arbitrary
negative number. Eq. \ref{eq:zq1} still measures $\exp (-\Delta S)$, but with
a lower limit ($x$) on $\Delta S$. Smaller limit yields a smaller and more
reliable error, however the statistics is decreased at the same time.  One can
tune the value of $x$, so that the statistical error takes its minimum. A
result of a typical optimum search can be seen on the left panel of Fig.
\ref{fi:test}.  The optimal value can be compared to the one obtained from Eq.
\ref{eq:zq2}.  On the right panel of Fig.  \ref{fi:test} the two new
topological susceptibilities and the one calculated by using traditional
pseudofermionic HMC \cite{Fodor:2004wx} are shown. The agreement is perfect.
Comparing these results with those of the HMC, we conclude that the stepsize
effect is negligible (at least at our present statistics).  Let us compare the
amount of CPU time of the two different methods for roughly the same
statistical errors (see Fig. \ref{fi:test}): the conventional HMC consisted
500-1000 trajectories (500 for the smallest, 1000 for the largest mass),
whereas we generated less than 200 at each mass for the new method. Moreover it
is important to emphasize in this context that the new method can be
efficiently parallelized.

\begin{table}[h!]
\begin{center}
\begin{tabular}{|c||c|c||c|c|c|c|}
\hline
m      & $\langle Q^2 \rangle$     & $\langle Q^2\rangle r_0^4/V$ &
$r_0$ & $m_\pi$ &  $L m_\pi$  & \#traj  \\
\hline
$0.03$ & $0.13(2)$   &  $0.0047(9)$ &              $3.52(13)$  & $0.29(11)$  &
$2.4$   & $39$ \\
\hline
$0.1$  & $0.41(6)$   &  $0.010(1)$ &               $3.17(5)$  & $0.53(4)$ &
$4.3$ & $51$ \\
\hline
$0.2$  & $0.97(19)$  &  $0.017(3)$ &              $2.89(2)$ & $0.74(6)$    &
$5.9$ & $63$ \\
\hline
$0.3$  & $1.59(18)$  &  $0.027(3)$ &             $2.88(6)$ & $0.99(8)$    &
$7.9$ & $54$ \\
\hline
\end{tabular}
\caption{
\label{tb:spect} Topological susceptibility measured on $8^4$ lattices in the
second and third column.
The further columns contain the
Sommer-scale, pion mass, pion mass times box size
and number of trajectories
on $8^3\times 16$ lattices.
}
\end{center}
\end{table}

\begin{figure}
\includegraphics*[width=15.0cm,bb=18 429 591
717]{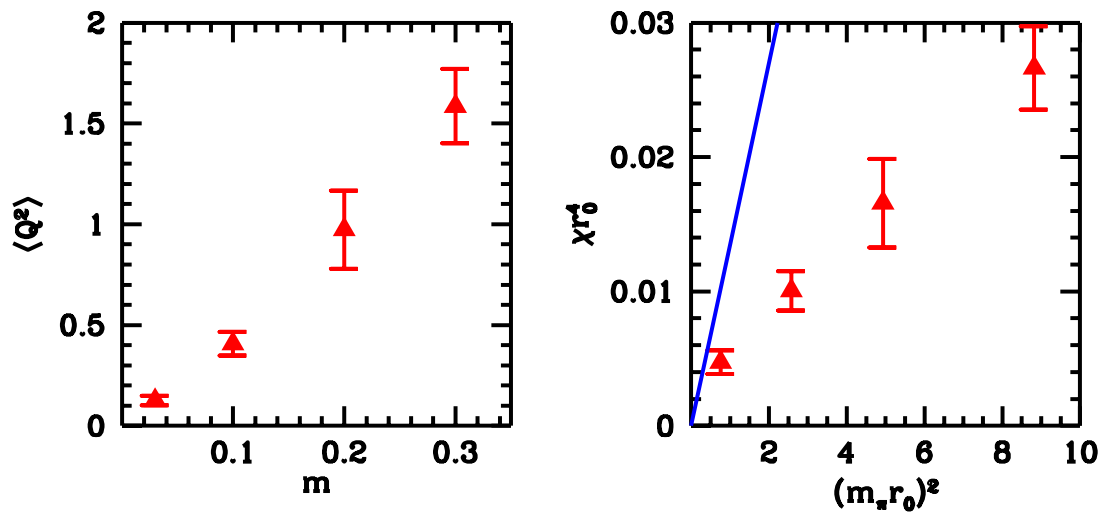}
\caption{
\label{fi:ts}
Topological susceptibility as the function of quark mass on $8^4$ lattices in
lattice units (left), and in physical units as the function of pion mass
(right).  Scale fixing and mass measurements were done on $8^3\times 16$.  The
error bars on the right plot do not contain the errors of scale fixing.  The
line is the leading order chiral behavior in the continuum.
}
\end{figure}

To measure the topological susceptibility on $8^4$ lattices we generated
configurations with tree-level Symanzik improved gauge action ($\beta=4.15$
gauge coupling) and 2 step stout smeared overlap kernel ($\rho=0.15$ smearing
parameter, the kernel was the standard Wilson matrix with $m_0=1.3$). We
performed runs in sectors $Q=0\dots 3$ (based on the measured $Z_3/Z_2$ we can
conclude, that the contribution of $Q \ge 4$ sectors are small compared to
statistical uncertainties).  For the negatively charged sectors we used the
$Q\to -Q$ symmetry of the partition function. The bare masses were $m=0.03,
0.1, 0.2$ and $0.3$, at each mass approximately 1000 trajectories were
collected. The average number of the topological sector boundary hits was
around $1.5$ per trajectory. We calculated the ratio of sectors using Eq.
\ref{eq:zq1} and Eq.  \ref{eq:zq2}. The result for the topological
susceptibility can be seen on Fig. \ref{fi:ts} (see also Table \ref{tb:spect}).
It is nicely suppressed for the smallest mass.  To convert it into physical
units, we made simulations on $8^3\times 16$ lattices. We measured the static
potential by fitting the large time behavior of on and off-axis Wilson-loops.
Then fitting it at intermediate distances we extracted the value of
Sommer-parameter. We also measured the pion mass (see Table \ref{tb:spect}).
Since our statistics was quite small on these asymmetric lattices, the errors
are large.  Note, that in order to get the mass-dimension 4 topological
susceptibility in physical units, one has to make very precise scale
measurements.

When
interpreting the results, one should keep in mind, that the volume is small,
and the lattice spacing is large. Note however, that smeared kernel overlap
actions show nice scaling behavior and good locality properties already at
moderate lattice spacings \cite{Kovacs:2002nz,Durr:2005an}.

\section{Discussion}

In this chapter we have given a summary of the work to implement a dynamical
overlap fermion algorithm. The lattice index theorem of the overlap
Dirac-operator is a very nice feature, however it has its bottleneck. The
operator is nonanalytic at the topological sector boundaries, which makes the
conventional dynamical fermion algorithm (HMC) break down.  We have proposed,
implemented and tested a modification which is able to handle this
nonanaliticity.  Examining the properties of the modified algorithm carefully,
we have made a few improvements on it. One of them was an improvement of the
acceptance ratio, the other is connected to the slow topological sector
changing of the algorithm.
 
Even with these improvements the simulation with dynamical overlap fermions is
in an exploratory phase. Other fermion formulations are considerably faster
than the overlap.  There are two major problems at the moment.
\begin{enumerate}
\item The first bottleneck is that the construction of the overlap operator is
a very expensive procedure, it scales with $V^2$ (as one can see in 
Ref. \cite{Chiu:2002ak},
but the extra $V$ factor can be expected, since the number of zeromodes
increases with the volume). Therefore it is very hard to imagine a dynamical
fermion algorithm with better scaling behavior. As it was mentioned
in the introduction the algorithms for conventional fermion formulations
scale usually with $V^{5/4}$.

\item The second bottleneck is handling the nonanaliticity of the overlap
operator.  The most simple modification of the conventional HMC (as described
in the chapter) can easily bring extra $V$ factors in the scaling. There exists
modifications improving the situation as we have seen, but they are really
cumbersome.  The problem is that the more sophisticated an improvement is,
there are more ways to go wrong. The nice feature of the HMC, the robustness
will be lost.
\end{enumerate}
Without a solution of the first issue at hand (which would mean to get rid of
the nested inversion), one can simply accept that the overlap dynamical fermion
algorithm will scale at least with $V^2$.  At this point new algorithms might
come into play where presumably different problems have to be solved. If the
only gain is that one can forget the discontinuities in the overlap operator
(second issue), it might worth to change. 

\section{Appendix: area conservation proof}
The leapfrog is trivially an area conserving mapping in the phase space, since
the increase of the momentum depends only on the actual coordinates, and the
change in the coordinates depends only on the momentum.  In case of the
modified leapfrog the difficulty arises since e.g. in the first step the updates of
the link variables  are depending on the actual links through $\epsilon_c$.
Similarly the momentum update also depends on the momentum through the normal
vector.

In order to keep the discussion brief, first let us start with a Hamiltonian
system in the $N$-dimensional Euclidean coordinate space. This shows the basic
idea of the proof in a transparent way.

We solve the equations of motion with a finite stepsize integration
of the following Hamiltonian:
\begin{equation*}
{\cal H} = \frac{1}{2}p_a p_a +S\left({\rm sgn}M(q) \right),
\end{equation*}
where $q_a, p_a$ $(a=1\dots N)$ are the coordinates and the momenta. 
$M$ depends only on the coordinates and the action $S$
is a smooth function (note that $q_a$, $M$ and $S$ are analogous
to the links, the fermion matrix and the fermionic action, respectively). 
The standard leap-frog algorithm can be effectively 
applied to this system, as long
as the trajectories do not cross the 
zero eigenvalue surface of $M$ ($\lambda(q)=0$, where $\lambda(q)$ is the 
eigenvalue with smallest magnitude\footnote{We do not deal with the 
possibility of degenerate zero eigenvalues which appears only
on a zero measure subset of the zero eigenvalue surface.}).
                          
We have to modify the leap-frog algorithm, when the coordinates  reach
the zero eigenvalue surface.  Instead of the original leap-frog update
of the coordinates, where the constant $p_a$ momenta are used for the
time $\epsilon/2$, we first update the coordinates with $p_a$ until the
surface, then we change the momentum to $p'_a$, which is used to evolve
$q_a$ for the remaining time. In case of {\bf\it refraction} one has the
following phase space transformation:
\begin{eqnarray}
\label{eq:transform}
q'=q+\epsilon_c p+(\epsilon/2-\epsilon_c)p' \\
p'=p-n(np)+n(np') + h, \nonumber
\end{eqnarray}
where $n$ is the normalvector of the surface, $\Delta S$ is the
potential jump along the surface, and  $(np')^2=(np)^2-2\Delta S$.
$\epsilon_c$ is the time required to reach the surface with the incoming
momenta $p$. $h$ is a vector orthogonal to $n$ and depending on
$q,p$ only through $\epsilon_c$ or quantities which measured on the
eigenvalue surface. The $h$ might be needed to improve the energy
conservation of the leapfrog (see Sec. \ref{sec:rr}), eg. one can use
\bea
\label{eq:lfh}
h=-\epsilon_c QF_{-} - (\epsilon/2-\epsilon_c)
QF_{+},
\eea
where the $F_{\pm}$ forces are measured on the eigenvalue surface with
setting ${\rm sgn}(\lambda(\epsilon_c))={\rm
sgn}(\lambda(\epsilon_c\pm0))$.  $Q_{ab}=\delta_{ab}-n_an_b$ is simply the orthogonal
projector to the surface.  

First let us concentrate on the $q,p$ dependence of $\epsilon_c$. 
$\epsilon_c(q,p)$ is determined from the  condition
\(
\lambda(q+\epsilon_c(q,p)p)=0
\).
One obtains the partial derivatives of $\epsilon_c$ with respect to $q,p$
by expanding this zero eigenvalue 
condition to first order in $\delta q$ or $\delta p$.
First take the $\delta q$ variation:
\bea
\lambda(q_a+\epsilon_cp_a+\delta q_a+\der{\epsilon_c}{q_b}\delta q_b p_a)=\lambda(q+\epsilon_cp)+\left.\der{\lambda}{q_a}\right|_{q+\epsilon_cp}
(\delta_{ab}+\der{\epsilon_c}{q_b}p_a)\delta
q_b=0
\eea
Since the normalvector is just 
\[
n_a=\left.\der{\lambda}{q_a}\right|_{q+\epsilon_cp}/||\der{\lambda}{q}||, 
\]
we have for the partial derivative of $\epsilon_c$ with respect to $q$:
\[
\der{\epsilon_c}{q_a}=-\frac{n_a}{(np)}.
\]
Similarly one gets for the partial derivative with respect to $p$:
\[
\der{\epsilon_c}{p_a}=-\epsilon_c\frac{n_a}{(np)}.
\]

There is an important identity between the $q$ and $p$ derivatives of a
function, which depends only on $q+\epsilon_c(q,p)p$. (Two examples are
$n$ and $\Delta S$.) Let us evaluate $p$ and $q$ derivatives of an
arbitrary $g(q+\epsilon_c(q,p)p)$ function:
\bea
\der{g}{q_a}=\left.\der{g}{q_b}\right|_{q+\epsilon_cp}(\delta_{ab}+\der{\epsilon_c}{q_a}p_b)=\left.\der{g}{q_b}\right|_{q+\epsilon_cp}
(\delta_{ab}-\frac{n_ap_b}{(np)}),
\\ 
\der{g}{p_a}=\left.\der{g}{q_b}\right|_{q+e_1p}(\epsilon_c\delta_{ab}+\der{\epsilon_c}{p_a}p_b)=\left.\der{g}{q_b}\right|_{q+\epsilon_cp}
(\delta_{ab}-\frac{n_ap_b}{(np)})\epsilon_c,
\eea
which gives
\begin{eqnarray} 
\der{g}{p_a}=\epsilon_c\der{g}{q_a}.
\label{eq:pq}
\end{eqnarray}

Now we can consider the four different partial derivatives required for the Jacobian:
\[
J=
\begin{pmatrix}
\der{q'}{q} & \der{q'}{p} \\
\der{p'}{q} & \der{p'}{p} \\
\end{pmatrix},
\]
whose determinant gives the change in the Euclidean measure $d^Nqd^Np$ due to the given
phase space transformation.
Introducing 
\[
Z_{ab}\equiv \der{p'_a}{q_b}.
\]
one incorporates all terms which arise from the $q$ 
dependence of the normalvector and $\Delta S$. 
In case of a straight wall with constant potential jump and $h=0$ 
this matrix vanishes. 
(Clearly, for QCD with
overlap fermions this object is very hard to calculate;
they usually require the 
diagonalization of the whole $H_W$ matrix ).
Using Eq. \ref{eq:pq} one can recognize the $Z$ matrix
in the other three components of $J$.
Denoting
\bea
X_{ab}=Q_{ab}+\left(1-\frac{2\Delta S}{(np)^2}\right)^{1/2} n_an_b
+\frac{h_an_b}{(np)} \\
Y_{ab}=Q_{ab}+\left(1-\frac{2\Delta S}{(np)^2}\right)^{-1/2} n_an_b
\eea
The useful property of $X$ and $Y$ that the determinant of their product is
\bea
\det (XY)=\det\left[\delta_{ab}+ \left(1-\frac{2\Delta
S}{(np)^2}\right)^{-1/2}\frac{h_an_b}{(np)}\right] = 1+ \left(1-\frac{2\Delta
S}{(np)^2}\right)^{-1/2} \frac{(hn)}{(np)}
\eea
which means that it is trivial for the $(hn)=0$ case.  In terms of the $X$,
$Y$ and $Z$ matrices the Jacobian is very simple.  We can split it into
2 parts: the first term contains all $X$ and $Y$ factors and has
determinant one
and all $Z$
factors are in the second term:
\bea
\label{eq:jac}
J=
\begin{pmatrix}
X & \epsilon_cX+(\epsilon/2-\epsilon_c)Y \\
0 & Y
\end{pmatrix}+
\begin{pmatrix}
(\epsilon/2-\epsilon_c)Z & (\epsilon/2-\epsilon_c)\epsilon_cZ \\
Z & \epsilon_c Z
\end{pmatrix}.
\eea
Let us introduce $J'$ as the product of $J$ and the inverse of its first
term. 
Simple algebra gives:
\bea
J'
=\begin{pmatrix}
1 & 0 \\
0 & 1
\end{pmatrix}\otimes \mathbbm{1}
+E\otimes \epsilon_cY^{-1}Z,
\eea
where $E$ is defined as
\[
E=\begin{pmatrix}
-1 & -\epsilon_c \\
1/\epsilon_c &  1
\end{pmatrix}.
\]
$E$ has an eigenvector $v_1 \propto (\epsilon_c,-1)$ with zero eigenvalue. 
The $v_2\propto (1,\epsilon_c)$ vector is orthogonal
to $v_1$ and
has the property to give zero in the product $v_2^TEv_2=0$. In the orthonormal 
basis given by $v_1$ and
$v_2$ $J'$ has the form:
\bea
J'
=
\begin{pmatrix}
1 & 0 \\
0 & 1
\end{pmatrix}
\otimes \mathbbm{1}
+ 
\begin{pmatrix}
0 & v_1^TEv_2 \\
0 & 0
\end{pmatrix}
\otimes \epsilon_cY^{-1}Z,
\eea
thus $\det J'=1$. Since $J$ and $J'$ differs only in a matrix 
with determinant one, we arrive
\[
\det J=1,
\]
thus the transformation Eq. \ref{eq:transform} preserves
the integration measure. 

The transformation for  {\bf\it reflection} is given by
\begin{eqnarray} 
\label{eq:transform1} q'=q+\epsilon_c
p+(\epsilon/2-\epsilon_c)p' \\ 
p'=p-2n(np) \nonumber + h.
\end{eqnarray}
$h$ can be chosen as in Eq. \ref{eq:lfh}, but now we have $F_-=F_+$, since at
reflection the ${\rm sgn}$ function does not change sign.  One can
obtain the Jacobian of reflection by simply making
the 
\bea
\left(1-\frac{2\Delta S}{(np)^2}\right)^{1/2} \to -1
\eea
substitution in the Jacobian of the refraction (Eq. \ref{eq:jac}). Then it is
easy to see that the $\det J=1$ holds for the reflection case, too.

Finally let us consider a {\bf\it modified reflection}, which makes only
$O(\epsilon^2)$ error in the energy conservation (see Sec. \ref{sec:rr}). The 
phase space transformation can be written as:
\bea
q'=q+\epsilon_c p + \epsilon_c p' \\
\nonumber p'=p-2n(np)+h,
\eea
The $h$ which is needed to ensure energy conservation upto
$O(\epsilon^2)$ is the following
\bea
h=-2\epsilon_c  QF_-.
\eea
This comes from Eq. \ref{eq:lfh} and using that the inward and outward updates
now take the same time ($\epsilon_c$). $h$ automatically satisfies $(hn)=0$.
The Jacobian is very similar to the Jacobian of the reflection procedure
above (ie. the one obtained from Eq. \ref{eq:jac} with the $(~~~)^{1/2} \to -1$
substitution):
\bea
J=
\begin{pmatrix}
X_1 & \epsilon_cX_1+\epsilon_cY\\
0 & Y
\end{pmatrix}
+
\begin{pmatrix}
\epsilon_c Z& \epsilon_c^2 Z\\
Z & \epsilon_c Z
\end{pmatrix}.
\eea
Instead of $\epsilon/2-\epsilon_c$ we have $\epsilon_c$ everywhere and the 
$X$ matrix is substituted by $X_1$:
\bea
[X_1]_{ab}=\delta_{ab}-\frac{(2Qp+h)_an_b}{(np)}.
\eea
$X_1$ has a trivial determinant $\det X_1=1$, since $(nh)=(n,Qp)=0$.  
From
here the
proof goes in the same way as above. One concludes to $\det J=-1$,
where the minus sign\footnote{
In the previous reflection recipe, $\det X = -1$ was also true, so 
all together one ended up with $\det J =1$.}
comes from $\det Y =-1$.

The proofs for the $SU(3)$ cases were carried out in a completely analogous
way. The only difference was the appearance of factors
associated with the group structure of $SU(3)$ which all canceled in 
the final result. Thus, we conclude that the suggested modifications of the
leap-frog conserve the integration measure. 

\section{Appendix: Classical motion on an $SU(3)$ manifold}

In this appendix we briefly discuss the Hamiltonian formulation of
a system, which coordinates are elements of a $G=SU(3)$ group. In
particular we will provide formulas to calculate the Jacobian of
some map in the phase space. Some parts of the appendix closely follow Ref. \cite{Kennedy:1989ae}. 

\subsection{Differential geometry on a Lie-group}

If the coordinates of a system are elements of a Lie-group manifold ($g_a \in
G$), then $T_gG$ is the space of tangent vectors at point $g$, this is the
vector space of velocities (with local coordinates $\dot{g}_a$). 

Let us consider a few relevant
mappings which arise due to the Lie-group structure of $G$.
There is 
a natural mapping called the right translation 
\bea
R_g:G\to G \quad \quad \quad h \to hg,
\eea
the corresponding derivative mapping 
$R_{g*}(h):T_hG\to T_{hg}G$ is a linear transformation which 
has the following matrix in local coordinates:
\bea
\nonumber
(R_{g*}(h))_{ab}=\der{(hg)_a}{h_b}.
\eea
The pullback of $R_g$ 
is in
certain sense going in backward direction as in the case of the 
derivative mapping, 
since
\bea
R_g^*(h):T^*_{hg}G\to T^*_{h}G \quad \quad \quad \alpha(hg)
\to \beta(h) : \beta(h)\langle v \rangle=\alpha(hg)\langle R_{g*}(h)v \rangle, 
\eea
for all $v$ vectors in the tangent space $T_hG$. Here $\alpha$ and $\beta$ 
are 1-forms, linear functionals acting on vectors. The group element dependence 
is indicated in the $(\dots)$ parentheses, whereas the vector, which they act on, is in the $\langle \dots \rangle$ bracket.

A {\tt vector field} is right invariant, if $v(hg)=R_{g*}(h) v(h)$ is fulfilled.  There
is a one to one correspondence between right invariant vector fields and the
elements of the Lie-algebra of the group ($v(g) \leftrightarrow v(1) \in LA(G)$), thus they
are elements of a linear space.  The Lie-bracket of two vector fields ($v$ and
$w$) measures the noncommutativity of two flows (one parameter $G\to G$ maps,
whose derivatives are the vector fields themselves).  It is again a vector
field: $[v,w]=u$, or in local coordinates it is $u_a=w_b\partial_b v_a - v_b \partial_b w_a$.
For right invariant vector fields the bracket is also right
invariant, thus if ${r_A}$ is a basis in the linear space of right invariant
vector fields, then 
\bea [r_A,r_B]=c_{AB}^C r_C.  \eea

A {\tt 1-form field} is right invariant, if $R_g^*\alpha=\alpha$, that is
$\alpha(hg)\langle R_{g*}(h) v \rangle=\alpha(h)\langle v \rangle$ for all $v$ vectors in tangent space
$T_hG$.  There is a one to one correspondence between right invariant 1-form
fields and 1-forms over the tangent space at the unit element ($\alpha(g)
\leftrightarrow \alpha(1)$, so that $\alpha(1)\langle v \rangle=\alpha(g)\langle R_{g*}v \rangle$). 
In order to prove an important identity for right invariant 1-forms, we need a
little preparation. If $\alpha = \alpha_a(g) dg_a$ is a 1-form field, then its derivative
is $d\alpha = \partial_b \alpha_a dg_b \wedge dg_a$. Its pullback corresponding to a mapping $R$
is $R^*\alpha = \alpha_b(k) \partial_a k_b dg_a$ with the $k=R(g)$ notation. Then
\bea
\nonumber
d(R^*\alpha)=(\partial_d\alpha_b(k) \partial_c k_d \partial_a k_b + \alpha_b(k) \partial_c\partial_a k_b )
dg_c \wedge dg_a = \\
\nonumber
= \partial_d \alpha_b(k) R^* ( dg_d \wedge dg_b ) = R^* d\alpha,
\eea
where we have used the antisymmetric property of the wedge product.
Using the above equation it is easy to
see that the derivative of a right invariant 1-form is also right invariant:
\bea
\nonumber
R_g^* d\alpha = d R_g^* \alpha = d \alpha.
\eea
This means that if we take $\varrho_A$ as a basis in the space of right invariant 1-forms
\footnote{It is normalized so, that $\varrho_A(r_B)=\delta_{AB}$ is satisfied
at the identity. Due to right invariance, the normalization will hold on the whole group.}, then
$d\varrho_A$ should be expressible in terms of $\varrho_B \wedge \varrho_C$. So let us calculate
the 2-form $d\varrho_A$ on two basis vectors in local coordinates:
\bea
\nonumber
d\varrho^A\langle r^B,r^C \rangle=\partial_b \varrho_{a}^A (dg_b \wedge dg_a)\langle r^B,r^C \rangle=
(\partial_b \varrho_{a}^A -\partial_a\varrho_{b}^A)r_{b}^Br_{a}^C=\\
\nonumber
r_{b}^B\partial_b(\varrho_{a}^Ar_{a}^C)-r_{a}^C\partial_a(\varrho_{b}^Ar_{b}^B)-
r_{b}^B\varrho_{a}^A\partial_br_{a}^C + r_{a}^C\varrho_{b}^A\partial_ar_{b}^B.
\eea
In parentheses we have $\delta_{AC}$ and $\delta_{AB}$ due to the normalization, 
therefore only the last two term remains. These two gives $-\varrho^A\langle [r^B,r^C] \rangle$, which
yields the following result (Maurer-Cartan structure equation):
\bea
d\varrho_A=-\frac{1}{2}c_{BC}^A\varrho_B \wedge \varrho_C.
\eea

\subsection{Hamiltonian dynamics}

The Lagrangian of the system is a real valued function on the tangent bundle ($L:
TG \to \mathcal{R}$). The derivative of the Lagrangian in the direction of the
velocities is a differential form, which maps from $T_gT_gG \sim T_gG$ to the
real numbers (ie. it is an element of the cotangent bundle $T^*G$).  Its local
coordinates are $\der{L}{\dot{g}_a}$, which are identified as the canonical
momenta ($p_a$). Since the momenta are coordinates of linear forms on $TG$, the
Hamiltonian phase space is the manifold $T^*G$ with local coordinates
$\{g_a,p_a\}$.

The $T^*G$ manifold is symplectic, ie. we have a 2-form $\omega$ on $T^*G$
which has vanishing derivative:
\bea
\omega\equiv d\left( \sum_a \varrho_a p_a \right) \quad \quad \quad \Longrightarrow d\omega=0.
\eea
According to the Maurer-Cartan equation, the 
\bea
\omega = \sum_a \varrho_a \wedge dp_a + \frac{1}{2} p_a c_{bc}^a \varrho_b \wedge \varrho_c 
\eea
relation holds.
From the symplectic structure follows, that 
there is a one to one correspondence between vector fields ($v$) and 1-form fields ($\alpha$): 
\bea
\nonumber
v \leftrightarrow \alpha \quad \quad \quad \alpha\langle w \rangle=\omega\langle v,w \rangle.
\eea
for all $w$ vectors.

The equations of motion arise through a Hamiltonian function ($H$) {\it and} the symplectic structure.
The change in the Hamiltonian is described by the derivative 1-form $dH$. Along the vector field $h$, which
corresponds to the 1-form $dH$ through the symplectic structure, the Hamiltonian is conserved:
\bea
\nonumber
dH\langle h \rangle=\omega\langle h,h \rangle=0.
\eea  
In order to determine $h$ we use the right invariant vector and 1-form basis on the group. In this basis 
the Hamiltonian vector field $h$ and an arbitrary vector field $v$ has the following form:
\bea
\label{eq:basis}
h= h_a r_a + \bar{h}_a \der{}{p_a}, \quad \quad \quad
v= v_a r_a + \bar{v}_a \der{}{p_a}.
\eea
The derivative 1-form of the Hamiltonian $dH$ 
can be written as
\bea
\nonumber
dH= dH\langle r_a \rangle \varrho_a + \der{H}{p_a} dp_a,
\eea
where $dH\langle r_a \rangle$ is just the $r_a$ directional derivative of $H$.
Now it is easy to see that 
\bea
\nonumber
\omega\langle h,v \rangle=h_a\bar{v}_a-\bar{h}_av_a+c^{a}_{bc}p_ah_bv_c 
\quad \quad {\rm and} \quad
dH\langle v\rangle=v_adH\langle r_a \rangle+\bar{v}_a\der{H}{p_a}
\eea
holds. Equating coefficients of $v_a$ and $\bar{v}_a$ we get the result for $h$:
\bea
h=\der{H}{p_a}r_a+(c^{c}_{ba}p_c\der{H}{p_b}-dH\langle r_a \rangle)\der{}{p_a}.
\eea 
The integral curve corresponding to the vector field $h$ describes the motion of the system
in the phase space as the time ($t$) goes on.
The equations of motion are the differential equations for $\{g_a,p_a\}$ coordinates 
which is solved by the integral curve:
\bea
\label{eq:eom}
\dot{p}_a(t)= dp_a\langle h\rangle=-dH\langle r_a \rangle+c^{c}_{ba}p_c\der{H}{p_b}\quad \quad {\rm and} \quad
\dot{g}_a(t)= dg_a\langle h\rangle=\der{H}{p_b}dg_a\langle r_b \rangle.
\eea

\subsection{Volume element and phase space maps}

The integral curve\footnote{For simplicity the notation of the integral curve is $g(t)$ instead of $\{g(t),p(t)\}$.}
corresponding to the Hamiltonian vector field preserves the symplectic
structure $g(t)^*\omega=\omega$, which means $\omega(g(t))\langle g(t)_*v,g(t)_*w \rangle=\omega(g(0))\langle v,w \rangle$. Moreover
higher "wedge" powers are also preserved, notably the largest one 
\bea
\Omega=\omega^d= \omega \wedge \omega \wedge \dots \wedge \omega = 
\varrho_1 \wedge \varrho_2 \dots \wedge \varrho_d \wedge dp_1 \wedge dp_2 \dots \wedge dp_d,
\eea
with $d$ being the dimension of the group. $\Omega$ is the volume element of the phase space, it is 
the wedge product of the Haar-measure of the group and a Euclidean volume element ($d^dp$). The $g(t)^*\Omega=\Omega$
property is usually called area conservation.

Let us consider a phase space map $f:T^*G \to T^*G$ with $\{g,p\}$ coordinates mapped to $\{G,P\}$. 
Since $\Omega$ is the only $2d$-form on $T^*G$, $f^*\Omega$ is proportional to $\Omega$. The 
proportionality constant describes the change in an infinitesimal phase space volume under the map $f$.
By definition
\bea
\label{eq:fomega}
(f^*\Omega)(gp)\langle v_{(1)},\dots,v_{(2d)} \rangle = \Omega(GP)\langle f_*v_{(1)},\dots,f_*v_{(2d)} \rangle,
\eea
with $f_*$ being the derivative mapping of $f$, $v_{(a)}$'s are arbitrary vectors in $T_{gp}(T^*G)$ tangent space.
In the usual basis (see Eq. \ref{eq:basis}) an $f_*v$ vector can be written as 
\bea
\nonumber
f_*v=v_a f_*r_a + \bar{v}_a f_*\der{}{p_a}.
\eea
Eq. \ref{eq:fomega} is actually a $2d$ dimensional determinant, in which we have to deal with the following types of objects:
\bea
\nonumber
\varrho_a\langle f_*v\rangle  = v_b \varrho_a \langle f_*r_b\rangle  + \bar{v}_b \varrho_a\langle f_*\der{}{p_b}\rangle  \quad {\rm and} \quad
dP_a\langle f_*v\rangle  = v_b dP_a \langle f_*r_b\rangle  + \bar{v}_b dP_a\langle f_*\der{}{p_b}\rangle .
\eea
Based on these relations the determinant of Eq. \ref{eq:fomega} is
\bea
\nonumber
\Omega(GP)\langle f_*v_{(1)},\dots,f_*v_{(2d)}\rangle=\det (Jv_{(1)}, Jv_{(2)}, \dots, Jv_{(2d)} )=
(\det J) \Omega(gp)\langle v_{(1)},\dots,v_{(2d)} \rangle,
\eea
where $J$ hypermatrix was introduced as:
\bea
\label{eq:jacobian}
J=
\begin{pmatrix}
\varrho_a\langle f_*r_b\rangle  & \varrho_a\langle f_*\der{}{p_b}\rangle  \\
dP_a\langle f_*r_b\rangle  & dP_a\langle f_*\der{}{p_b}\rangle  \\
\end{pmatrix}.
\eea
$\det J$ is the proportionality constant that we were looking for.

\subsection{Formulas in matrix representation}
In practice the dynamics is treated in terms of matrices instead of independent real parameters.
The group variables are represented by unitary matrices. A possible parametrization is $U(g)=\exp(g_aT_a)$ with
$T_a$ traceless, antihermitian matrix basis. The momentum becomes a traceless, antihermitian matrix $\Pi(p)=p_aT_a$.
Let us consider the $r^A$ directional derivative of $U(g)$:
\bea
\label{eq:dur}
\nonumber
dU\langle r^A \rangle=\der{U}{g_b} r^A_b(g) = \der{U}{g_b}dg_b\langle R_{g*}(1)r^A(1) \rangle =\\
=\der{U}{g_b}\left. \der{(hg)_b}{h_c}\right|_{h=0}r^A_c(1)=\left. \der{U(hg)}{h_c}\right|_{h=0} r^A_c(1)=T_cU r^A_c(1)=T_AU.
\eea
We have used the right invariance, the form of $R_{g*}$ in local coordinates, the explicit form of $U(g)$ and
finally we have fixed the local coordinates of the $r^A$ basis at the identity ($r^A_c(1)=\delta_{Ac}$). 
We will also need an equation similar to the above
\bea
\label{eq:durho}
\nonumber
\der{g^b}{U^T}\varrho^A_b(g)=\left.\der{(hg)^b}{U^T(hg)} \der{h^A}{(hg)_b}\right|_{h=0}=
\left.\der{h^A}{U^T(hg)}\right|_{h=0}=\\
=\left.\der{U(h)}{U^T(hg)}\der{h^A}{U(h)}\right|_{h=0}=-U^{\dagger}T^A.
\eea
We have used the local coordinate version of the right invariant 1-forms and the 
orthogonality property of the $T_A$ matrix basis (${\rm tr}(T_AT_B)=-\delta_{AB}$).

For simplicity we will assume the following Hamiltonian, when deriving the equations of motion:
$H=\frac{1}{2}\sum_a p_a^2 + S(g)$. Using this Hamiltonian and Eq. \ref{eq:dur} the equations of motion of Eq. \ref{eq:eom}
can be transformed into the simple, well-known form:
\bea
\nonumber
\dot{U}=\der{U}{g_a}\dot{g}_a=\der{H}{p_a}dU\langle r_a \rangle=\Pi U,\\
\dot{\Pi}=-T^a{\rm tr}(\der{S}{U^T}dU\langle r_a \rangle) =-T^a {\rm tr} (T^a U \der{S}{U^T}) = -\mathcal{A}(U\der{S}{U^T})
\eea
with $\mathcal{A}$ traceless, antihermitian matrix projector.

Finally let us calculate in the matrix representation the Jacobian of $f$ phase space function, which
maps the $\{U(g),\Pi(p)\}$ variables
to $\{\bar{U}(G),\bar{\Pi}(P)\}$. Lets take the first element of the $J$
hypermatrix in Eq. \ref{eq:jacobian} and use Eq. \ref{eq:dur} and Eq. \ref{eq:durho} to
eliminate the components of the right invariant fields:
\bea
\nonumber
\varrho^A\langle f_*r^B\rangle=\varrho^A_c(G) \der{G_c}{g_d} r^B_d(g) = \varrho^A_c(G) \der{G_c}{\bar{U}_{\alpha\beta}}
\der{\bar{U}_{\alpha\beta}}{U_{\gamma\delta}}r^B_d(g)\der{U_{\gamma\delta}}{g_d}=\\
=-(\bar{U}^\dagger T_A)_{\beta\alpha}\der{\bar{U}_{\alpha\beta}}{U_{\gamma\delta}}(T_BU)_{\gamma\delta}
\eea
Similar calculation yields the other matrix elements of $J$:
\bea
\nonumber
\varrho^A\langle f_*\der{}{p^B}\rangle=-(\bar{U}^\dagger T_A)_{\beta\alpha}\der{\bar{U}_{\alpha\beta}}{\Pi_{\gamma\delta}}(T_B)_{\gamma\delta},\\
\nonumber
dP^A\langle f_*r^B\rangle=-(T_A)_{\beta\alpha}\der{\bar{\Pi}_{\alpha\beta}}{U_{\gamma\delta}}(T_BU)_{\gamma\delta},\\
dP^A\langle f_*\der{}{p^B}\rangle=-(T_A)_{\beta\alpha}\der{\bar{\Pi}_{\alpha\beta}}{\Pi_{\gamma\delta}}(T_B)_{\gamma\delta}.
\eea

\chapter{Dynamical staggered fermions}

It is known for a long time, that for high enough temperatures and/or densities
the quarks and gluons are liberated from confinement, the chiral symmetry is
restored: the so called quark-gluon plasma phase of the matter is created. 

There is a huge literature of this transition: theoretical works based on the
symmetries of QCD, analytical and numerical calculations in QCD like models and
lattice QCD.  It is worth emphasizing, that the only known way to obtain the
properties of the quark-gluon plasma from first principles of the theory is
lattice QCD. However until recently lattice result were usually burdened by
large systematical errors: extrapolation to the physical quark mass, finite
volume effects, missing continuum extrapolations. 

Lattice QCD recently has entered a new era, where we are facing a huge
reduction of these systematics. In this chapter we describe the details and
results of a large scale simulation, where we attempted to eliminate (almost)
all systematics of previous lattice calculations. Thus these results can be
considered as the final ones, where the only remaining is to crosscheck against
the work of other groups or against different lattice discretizations.  The
amount of computer work is tremendous, we used $O(10^{19})$ floating point
operations on the fastest supercomputers of the world. The algorithmic and
theoretical improvements still continue, so as the increase in the speed of the
computers.  We hope that one day these results will be just as easy to obtain
as getting the value of eg. $\sin(1.0)$ using a pocket calculator today.   

We emphasize that extensive experimental work is currently being done with
heavy ion collisions to study the QCD transition (most recently at the
Relativistic Heavy Ion Collider, RHIC). Moreover there is 
rich perspective for the future: the heavy ion program is expected to start
in 2009 at the Large Hadron Collider (LHC) in Geneva and in 2011 at the Facility for Ion and 
Antiproton Research (FAIR) in Darmstadt.
Both for the cosmological transition
and for RHIC, the net baryon densities are quite small, and so the baryonic
chemical potentials ($\mu$) are much less than the typical hadron masses
($\approx$45~MeV at RHIC and negligible in the early Universe).  A calculation
at $\mu$=0 is directly applicable for the cosmological transition and most
probably also relevant for the transition at RHIC. 

Let us remark here, that even if the finite temperature equilibrium state of
QCD is soon going to be solved, there are still important areas with only
moderate or no progress. Most notably there is the equilibrium state at finite
$\mu$, the well-known sign problem prohibited calculations for many years. The
breakthrough of \cite{Fodor:2001au,Fodor:2001pe} has opened new possibilities
(for a recent review of this subfield see \cite{Schmidt:2006us}), still many
questions remain unanswered.

In Sec. 1 we present the definition of our lattice action, the numerical
details of the algorithm used for the simulations and finally the concept of
line of constant physics (LCP). In Sec. 2 we give a detailed list of our
simulation points, whereas Sec. 3 is for the physics results.

\section{Setting up the simulations}
\subsection{The lattice action}
First we give our definition for the Symanzik improved gauge and for the
stout-link improved fermionic action.  We demonstrate that our choice of
stout-link improved staggered fermionic action has  small taste violation, when
compared to other staggered actions used in the literature to determine the equation of state (EoS)
of QCD. 

Isotropic lattice couplings are used, thus the lattice spacings are identical
in all directions.  The lattice action we used has the following form:
\begin{eqnarray}
\label{action} S & = & S_g + S_f,\\ S_g & = & \sum_x
\frac{\beta}{3} (  c_0  \sum_{\mu>\nu} W_{\mu,\nu}^{1\times 1}(x) + c_1
\sum_{\mu\ne\nu} W_{\mu,\nu}^{1\times 2}(x) ),\\ S_f & = & \sum_{x,y} \{
\overline{\eta}_{ud}(x) [\Dsl(U^{stout})_{xy}+m_{ud}\delta_{x,y}]^{-1/2}
\eta_{ud}(y) \nonumber\\ &  & \mbox{\hspace{2pt}} +\mbox{\hspace{2pt}}
\overline{\eta}_{s}(x) [\Dsl(U^{stout})_{xy}+m_{s}\delta_{x,y}]^{-1/4}
\eta_{s}(y)\}, 
\end{eqnarray} 
where $W_{\mu,\nu}^{1\times 1}$, $W_{\mu,\nu}^{1\times 2}$ are real parts of
the traces of the ordered products of link matrices along the $1\times 1$,
$1\times 2$ rectangles in the $\mu$, $\nu$ plane.  The coefficients satisfy
$c_0+8c_1=1$ and $c_1=-1/12$ for the tree-level Symanzik improved action.
$\eta_{ud}$ and $\eta_s$ are the pseudofermion fields for $u$, $d$ and $s$
quarks.  $\Dsl(U^{stout})$ is the four-flavor staggered Dirac matrix with
stout-link improvement \cite{Morningstar:2003gk}.  Let us also note here, that
we use the 4th root trick in Eq. (\ref{action}), which might lead to problems
of locality.

Our staggered action at a given $N_t$ yields the same limit for the pressure at
infinite temperatures as the standard unimproved action. There are various
techniques improving the high temperature scaling. However one also has to take
into account, that with highly improved actions (which contain far neighbor
interactions) smaller ($N_t \ge 8$) lattice spacings will not be available. In
this case one risks to have large lattice artefacts coming from the scale
setting procedure.

Staggered fermions have an inconvenient property: they violate taste symmetry
at finite lattice spacing. Among other things this violation results in a
splitting in the pion spectrum, which should vanish in the continuum limit.
The stout-link improvement makes the staggered fermion taste symmetry violation
small already at moderate lattice spacings.  We found that a stout-smearing
level of $N_{smr}$=2 and smearing parameter of $\rho$=0.15 are the optimal
values of the smearing procedure.

\subsection{The algorithm}
\label{ssec:ralg}
\begin{figure}
\begin{center}
\epsfig{file=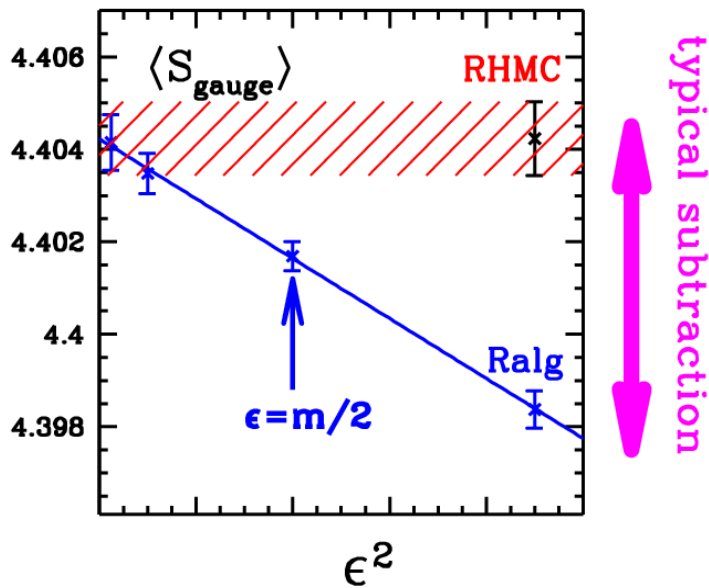,width=10cm}
\caption{\label{fi:sub} Average gauge action densities in various simulations. The blue points
are from inexact R-algorithm simulations, one can clearly see the stepsize ($\epsilon$) dependence.
Previous thermodynamical calculations used the R-algorithm, with the stepsize set to the half of 
the light quark mass. The black point is obtained using the RHMC algorithm with large stepsize.
Note, that it is unsafe to use the R-algorithm for equation of state calculations, since the
typical size of the subtraction (ie. gauge action difference on zero and non-zero temperature
lattices) is in the order of the systematical error of the R-algorithm.}
\end{center}
\end{figure}
The equation of state calculation is an extreme high precision measurement.
There are many things which can spoil it, one of them is the systematic error
coming from the algorithm. Before our work the R-algorithm was used exclusively
for staggered thermodynamical calculations, where in principle one has to make
an extrapolation in the intrinsic parameter of the algorithm (stepsize). These
extrapolations were never carried out, in the best cases there were attempts to
estimate the systematic errors. The $N_t=6$ equation of state has forced us to
change.  Here the measured quantity (action density difference at zero and
finite temperature lattices) has the same magnitude as the systematic error
(see Fig. \ref{fi:sub}.), which is clearly an unsafe situation.  Fortunately the
development of exact staggered algorithms were in a prospective phase at that
time (rational hybrid Monte-Carlo \cite{Clark:2003na} and
polynomial hybrid Monte-Carlo \cite{Frezzotti:1997ym}). We decided to use the RHMC, the algorithm
which is nowadays obligatory in lattice thermodynamics. It was worth changing,
the exact RHMC algorithm is significantly faster than the R-algorithm, and one
can get rid of one systematical error.

The RHMC technique approximates the fractional powers of the Dirac operator by
rational functions. Since the condition number of the Dirac operator changes as
we change the mass, one should determine the optimal rational approximation for
each quark mass. Note however, that this should be done only once, and the
obtained parameters of these functions can be used in the entire configuration
production. Our choices for the rational approximation were as good as few
times the machine precision for the whole range of the eigenvalues of the Dirac
operator.  We have also introduced multiple time scales in the algorithm
\cite{Clark:2006fx}.

The time consuming parts of the computations were carried out in single
precision. This might effect the algorithm in a negative way at two places:
firstly the reversibility of the trajectory is lost, secondly the precision
problems in the accept/reject step might result a bad distribution.  

The reversibility violation is considerably larger than single precision
accuracy, even if every step was carried out upto single precision through the
trajectory. This is due to the chaotic nature of the QCD equations of motion.
The usual way out is to use double precision arithmetics everywhere. However it
turns out that at several places single precision accuracy is tolerable, if at
some critical places high enough precision is chosen.  We make the force
calculation (which is the most time consuming part) in single precision however
the link and the momentum are updated in a higher precision scheme (in turned
out that we need at least 80bit precision on larger volumes). In this case the
reversibility will be exact in single precision. The link and the momentum
might differ after going forth and back along a trajectory, but only in high
precision. The forces will be bit by bit the same in the forward and backward
directions, which ensures that the two cannot deviate from each other.

For the other problem (leakage of precision in the accept/reject step) one can
use mixed precision inverters, which work in single precision for most of the
time. Here one adds intermediate double precision steps, with which one can achieve even
double precision accuracy. To be on the safe side on one of our largest
lattices we have cross-checked the results with a fully double precision
calculation, the results were the same within the errorbars.  We were also
constantly monitoring the $\langle \exp(-\Delta H) \rangle$ expectation value,
and find no statistically significant deviation from $1$.

We based our code on the publicly available MILC lattice gauge theory code,
however several parts were (re)written by ourselves. Most of the code was
written in two independent copies, the two versions agree upto machine precision.
These (among other things) include the staggered matrix
multiplication, solvers, smearing and measurement routines. We have developed
the code for four different architectures (Intel P4, AMD Opteron, Nvidia
Graphics Card, IBM Blue Gene L), each version required careful optimization.
For some details of the implementations see eg. \cite{Fodor:2002zi,Egri:2006zm}.

\subsection{Line of constant physics (LCP)}
\label{ssec:lcp}
\begin{figure}
\epsfig{file=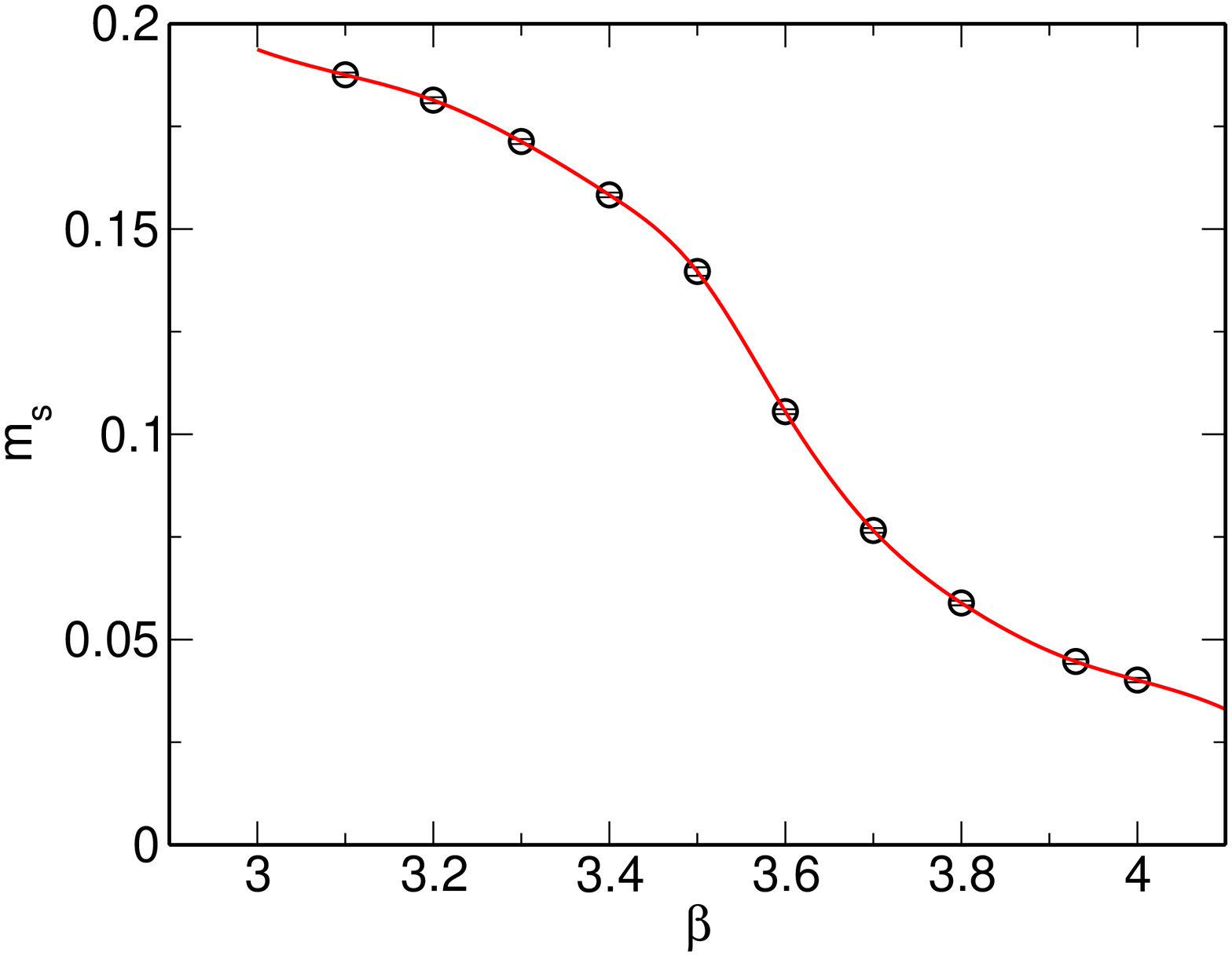,width=7.2cm}
\hspace{1cm}
\epsfig{file=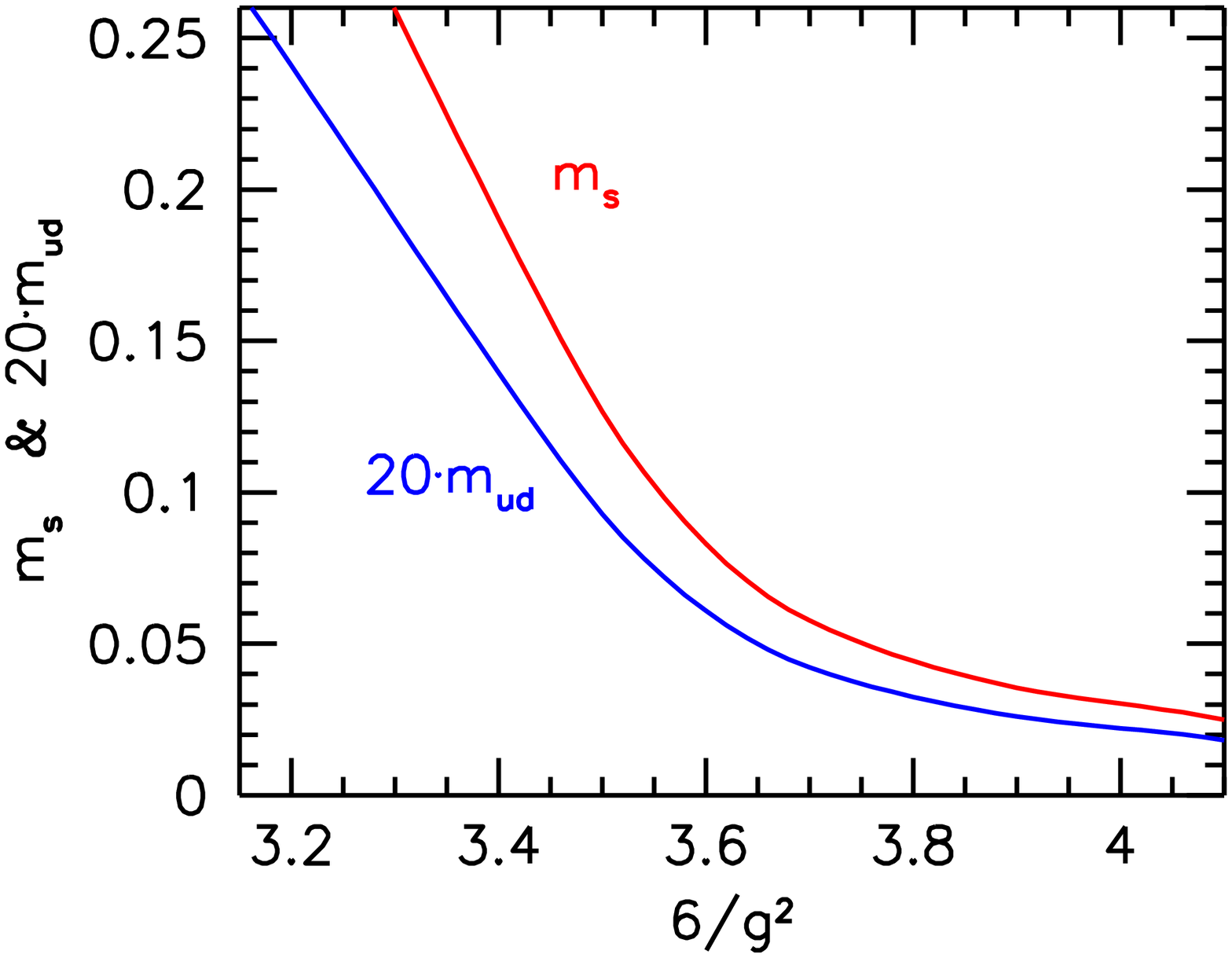,width=8cm}
\caption{\label{fig:LCP}
The line of constant physics. The left panel shows the strange mass as a
function of $\beta$ along LCP1. LCP1 is an approximate LCP, it was obtained by
using the $\phi$ and K masses (see text). LCP2 is a refinement of LCP1.
The right panel shows the strange
mass (red) and 20 times the light quark mass (blue) in lattice units as functions of
$\beta=6/g^2$ along LCP2. 
}
\end{figure}
Let us discuss the determination of the LCP.  The LCP is defined as
relationships between the bare lattice parameters ($\beta$ and lattice bare
quark masses $m_{ud}$ and $m_s$). These relationships express that the physics
(e.g. mass ratios) remains constant, while changing any of the parameters.
It is important to emphasize that the LCP is unambiguous (independent of the
physical quantities, which are used to define the above relationships) only in
the continuum limit ($\beta\rightarrow\infty$).  For our lattice spacings
fixing some relationships to their physical values means that some other
relationships will slightly deviate from the physical one.  In thermodynamics
the relevance of LCP comes into play when the temperature is changed by
$\beta$ parameter. Then adjusting the mass parameters ($m_{ud}$ and $m_s$) is
an important issue, neglecting this in simulations can lead to several \%
error in the EoS \cite{Csikor:2004ik}.

A particularly efficient (however only approximate, see later) way to obtain an LCP is by using simulations with 
three degenerate flavors with lattice quark mass $m_q$.
The leading order chiral perturbation theory implies 
the mass relation for $s\bar{s}$ mesons.
The strange quark mass is tuned accordingly, as
\begin{equation}
 m_{PS}^2/m_{V}^2|_{m_q=m_s} = (2m_K^2-m_\pi^2)/m_\phi^2,
  \label{eq:tl chpt}
\end{equation}
where $m_{PS}$ and $m_{V}$ are the pseudoscalar and vector meson 
masses in the simulations with three degenerate quarks.
The light quark mass is calculated using the ratio $m_{ud}=m_s/25$
obtained by experimental mass input in the chiral perturbation theory.
We obtain $m_s(\beta)$ as shown in the left panel of Fig. \ref{fig:LCP}. 
This (approximate) line of constant physics is 
called LCP1 later, the equation of state calculations
were carried out along this line.

Our approach using Eq. (\ref{eq:tl chpt}) is appropriate if in the $n_f$=2+1
theory the vector meson mass depends only weakly on the light quark masses and
the chiral perturbation theory for meson masses works upto the strange quark
mass.  After applying the LCP1 we cross-checked the obtained spectrum of the
$n_f$=2+1 simulations.  These simulations showed, however, that the hadron mass
ratios slightly differ from their physical values on the 5--10\% level.  In
order to eliminate all uncertainties related to an unphysical spectrum, we
determined a new line of constant physics.  The new LCP (which is called LCP2
afterwards) was defined by fixing $m_K/f_K$ and $m_K/m_\pi$ to their
experimental values (right panel of Fig. \ref{fig:LCP}). The more precise LCP2 was used for simulations
to determine the order of the QCD transition, and to measure the transition
temperature in physical units.

We have also carried out $n_f=2+1$ flavor $T=0$ simulations on LCP2.  Chiral
extrapolation to the physical pion mass led to $m_K/f_K$ and $m_K/m_\pi$
values, which agree with the experimental numbers on the 2\% level.
(Differences resulting from various fitting forms and finite volume corrections
were included in the systematics.) This is the accuracy of LCP2. 

\begin{figure}[h!]
\centerline{\includegraphics*[width=12cm,height=15cm]{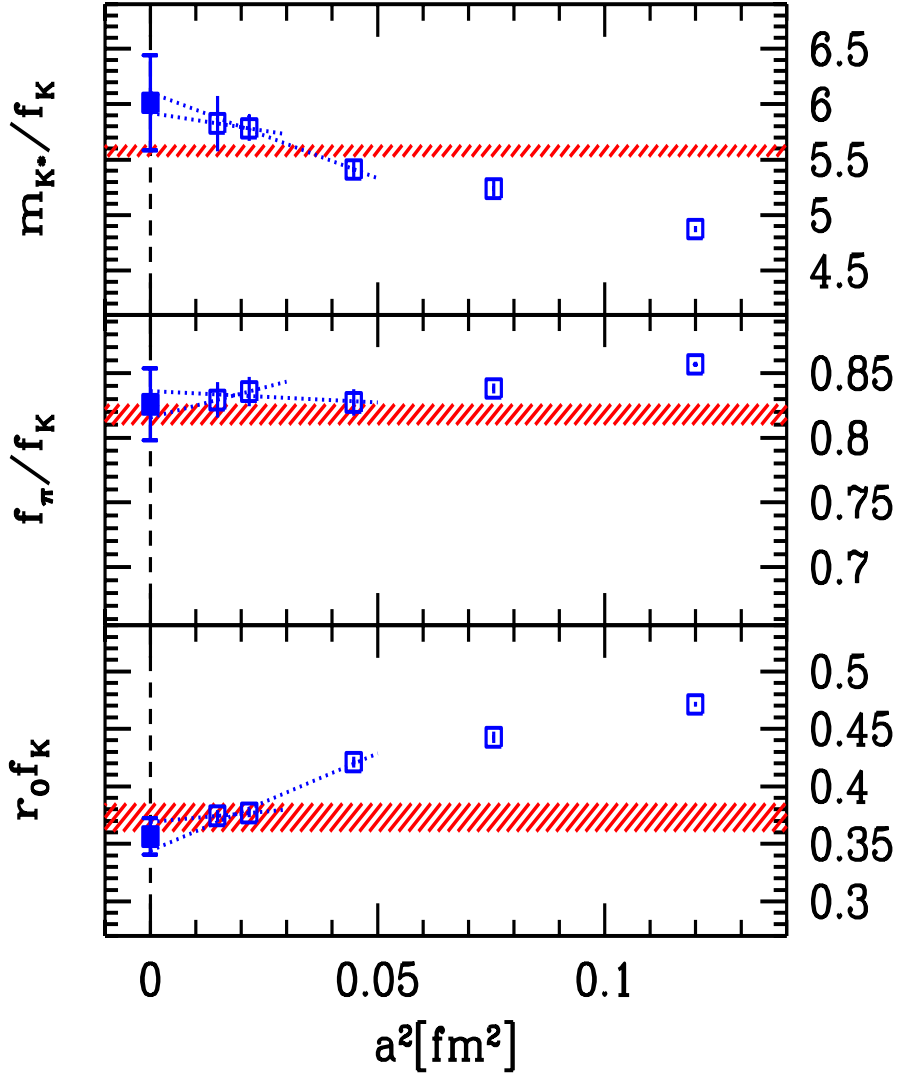}}
\caption{\label{fig:spect}
Scaling of the mass of the $K^*(892)$ meson, the pion decay constant and $r_0$ 
towards the continuum limit.
As a continuum value (filled boxes) we took the average of the continuum extrapolations obtained using our 2 and our 3
finest lattice spacings. The difference was taken as a systematic uncertainty, which is included in the shown errors.
The quantities are plotted in units of the kaon decay constant. 
In case of the upper two panels the bands indicate the physical values of the ratios and their
experimental uncertainties. For $r_0$ (lowest panel) in the absence of direct
experimental results
we compare
our value with the $r_0f_K$
obtained by the MILC, HPQCD and UKQCD collaborations 
\cite{Aubin:2004fs,Gray:2005ur}.
}
\end{figure}
In order to be sure that our results are safe from ambiguous determination
of the overall scale, and to prove that we are really in the $a^2$
scaling region,  we carried out a continuum extrapolation for three
additional quantities which could be similarly good to set the scale (we normalized
them by $f_K$, for $f_K$ determination in staggered QCD see \cite{Aubin:2004fs}).  Fig. \ref{fig:spect} shows the measured values of
$m_{K^*}/f_K$, $f_\pi/f_K$ and $r_0f_K$, at different lattice spacings and
their continuum extrapolation. Our three continuum predictions are in complete
agreement with the experimental results (note, that $r_0$ can not be measured
directly in experiments; in this case the original experimental input is the bottonium spectrum
which was used by the MILC, HPQCD and UKQCD collaborations to calculate
$r_0$ on the lattice \cite{Aubin:2004fs,Gray:2005ur}).

It is important to emphasize
that at lattice spacings given by $N_t$=4 and 6 the overall 
scales determined by $f_K$ and
$r_0$ are differing by $\sim$20-30\%, which is most probably true for any
other staggered formulation used for thermodynamical calculations. Since
the determination of the overall scale has a $\sim$20-30\% ambiguity, the
value of $T_c$ can not be determined with the required accuracy.

\section{Simulation points}

For our thermodynamical calculations we have used two LCPs: LCP1 and LCP2. The
LCP1 can be considered as an approximate LCP, this was used for the equation of
state calculation. The LCP2 was determined using the $n_f=2+1$ simulations
carried out along LCP1, it can be considered as a refinement of LCP1.  We used
it to determine the order of phase transition and the transition temperature.  In
this section we list the simulation points along the two LCPs. 

\subsection{Along LCP1}
The determination of the EoS needs quite a few simulation points. Results
are needed on finite temperature lattices ($N_t$=4 or 6) and on zero
temperature lattices ($N_t \gg$ 4 or 6) at several $\beta$ values (we
used 16 different $\beta$ values for $N_t$=4 and 14 values for $N_t$=6). Since
our goal is to determine the EoS for physical quark masses we have to
determine quantities in this small physical quark mass limit
(we call these $\beta$ dependent bare light quark masses $m_{ud}(phys)$). 
\begin{table}
\begin{center}
\begin{tabular}{|c|c|c|c|c|c||c|c|c|c|c|c|}
\hline
$\beta$    & $m_s$ & T=0  & \# & T$\neq$0& \# &
$\beta$    & $m_s$ & T=0  & \# & T$\neq$0& \#  \\ 
\hline 
3.000 & 0.1938 & 16$^3$$\cdot$16& 4  &12$^3$$\cdot$4& 9 &3.450 & 0.1507 & 16$^3$$\cdot$32&	29  & 18$^3$$\cdot$6 & 120	\\
3.150 & 0.1848 & 16$^3$$\cdot$16& 4  &12$^3$$\cdot$4& 9 &3.500 & 0.1396 & 16$^3$$\cdot$32&	33  & 18$^3$$\cdot$6 & 156	\\
3.250 & 0.1768 & 16$^3$$\cdot$16& 4  &12$^3$$\cdot$4& 9 &3.550 & 0.1235 & 16$^3$$\cdot$32&	30  & 18$^3$$\cdot$6 & 133	\\
3.275 & 0.1742 & 16$^3$$\cdot$16& 4  &12$^3$$\cdot$4& 9 &3.575 & 0.1144 & 16$^3$$\cdot$32&	28  & 18$^3$$\cdot$6 & 151	\\
3.300 & 0.1713 & 16$^3$$\cdot$16& 4  &12$^3$$\cdot$4& 9 &3.600 & 0.1055 & 16$^3$$\cdot$32&	31  & 18$^3$$\cdot$6 & 158	\\
3.325 & 0.1683 & 16$^3$$\cdot$16& 4  &12$^3$$\cdot$4& 9 &3.625 & 0.0972 & 16$^3$$\cdot$32&	33  & 18$^3$$\cdot$6 & 144	\\
3.350 & 0.1651 & 16$^3$$\cdot$16& 4  &12$^3$$\cdot$4& 9 &3.650 & 0.0895 & 16$^3$$\cdot$32&	30  & 18$^3$$\cdot$6 & 160	\\
3.400 & 0.1583 & 16$^3$$\cdot$16& 3  &12$^3$$\cdot$4& 9 &3.675 & 0.0827 & 16$^3$$\cdot$32&	32  & 18$^3$$\cdot$6 & 178 	\\
3.450 & 0.1507 & 16$^3$$\cdot$32& 29  &12$^3$$\cdot$4& 9 &3.700 & 0.0766 & 16$^3$$\cdot$32&	33  & 18$^3$$\cdot$6 & 174	\\
3.500 & 0.1396 & 16$^3$$\cdot$32& 33  &12$^3$$\cdot$4& 9 &3.750 & 0.0666 & 16$^3$$\cdot$32&	35  & 18$^3$$\cdot$6 & 140	\\
3.550 & 0.1235 & 16$^3$$\cdot$32& 30  &12$^3$$\cdot$4& 9 &3.800 & 0.0589 & 20$^3$$\cdot$40&	26  & 18$^3$$\cdot$6 & 158	\\
3.600 & 0.1055 & 16$^3$$\cdot$32& 31  &12$^3$$\cdot$4& 9 &3.850 & 0.0525 & 20$^3$$\cdot$40&	23  & 18$^3$$\cdot$6 & 157	\\
3.650 & 0.0895 & 16$^3$$\cdot$32& 30  &12$^3$$\cdot$4& 9 &3.930 & 0.0446 & 24$^3$$\cdot$48&	6   & 18$^3$$\cdot$6 & 171	\\
3.700 & 0.0766 & 16$^3$$\cdot$32& 33  &12$^3$$\cdot$4& 9 &4.000 & 0.0401 & 28$^3$$\cdot$56&	4   & 18$^3$$\cdot$6 & 166	\\
3.850 & 0.0525 & 20$^3$$\cdot$40& 23  &12$^3$$\cdot$4& 9 & & & & & & \\
4.000 & 0.0401 & 28$^3$$\cdot$56& 4   &12$^3$$\cdot$4& 9 & & & & & & \\
\hline                                     
\end{tabular}
\end{center}
\caption{                                  
\label{ta:points}
Summary of our simulation points along LCP1. For the physical light quark masses 
(we call them $m_{ud}(phys)$) 
25 times smaller values were taken than for the strange mass. T$\neq$0
simulations were performed with the above $m_s$ and $\beta$ pairs,
and at 5 different $m_{ud}$ values: \{1,3,5,7,9\}$\cdot m_{ud}(phys)$.
T=0 simulations were performed with the above $m_s$ and $\beta$ pairs,
but at 4 different $m_{ud}$ values: \{3,5,7,9\}$\cdot m_{ud}(phys)$.
The total number of trajectories divided by 100 are collected
in the \# columns. 
The left column shows the $N_t$=4, whereas the right column shows
the $N_t$=6 data. (For an explanation of our labeling see the text.)
}
\end{table}

For our finite temperature simulations ($N_t$=4,6) we used physical quark
masses.  The spatial sizes were always at least 3 times the temporal sizes.
For the whole $\beta$ range on $N_t=4$ we checked that by increasing the
$N_s/N_t$ ratio from 3 to 4 the results remained the same within our
statistical uncertainties. 

In the chirally broken phase (our zero temperature simulations, thus lattices
for which $N_t \gg$ 4 or 6, belong always to this class) chiral perturbation
theory can be used to extrapolate by a controlled manner to the physical light
quark masses.  Therefore 
for most of our simulation points\footnote{In the $\beta=3.0..3.4$ range the
$T=0$ simulations were carried out at $m_{ud}(phys)$.}
we used four pion masses ($m_\pi\approx$250, 320, 380
and 430 MeV), which were somewhat larger than the physical one. (To simplify
our notation in the rest of this section we label these points as 3,5,7 and 9
times $m_{ud}(phys)$.) It turns out that the chiral condensates at all the
four points can be fitted by linear function of pion mass squared with good
$\chi^2$. (Later we will show, that only the chiral condensate is to be
extrapolated to get the EoS at the physical quark mass.) The volumes were
chosen in a way, that for three out of these four quark masses the spatial
extentions of the lattices were approximately equal  or larger than four times
the correlation lengths of the pion channel.  We checked for a few $\beta$
values that increasing the spatial and/or temporal extensions of the lattices
results in the same expectation values within our statistical uncertainties.
(For 3$\cdot m_{ud}(phys)$ values the spatial lengths of the lattices were
only three times the correlation length of the pion channel.  However,
excluding this point from the extrapolations, the results do not change.) 

A detailed list of our
simulation points at zero and at non-zero temperature lattices
are summarized in Table \ref{ta:points}. 

\subsection{Along LCP2}
In order to perform the necessary renormalizations of the measured quantities
and to fix the scale in physical units we carried out $T=0$ simulations on our
new LCP (c.f. Table \ref{tab:T0}).  Six different $\beta$ values were used.
Simulations at T=0 with physical pion masses are quite expensive and in our
case unnecessary (chiral perturbation theory provides a controlled
approximation at vanishing temperature).  Thus, for each $\beta$ value we used four
different light quark masses, which resulted in pion masses somewhat larger
than the physical one (the $m_\pi$ values were approximately 250~MeV, 320~MeV,
380~MeV and 430~MeV), whereas the strange quark mass was fixed by the LCP
at each $\beta$.
The lattice sizes were chosen to satisfy the $m_\pi N_s\ge4$ condition.
However, when calculating the systematic uncertainties of meson masses and decay constants, we
have taken finite size corrections into account using continuum finite volume
chiral perturbation theory \cite{Colangelo:2005gd} (these corrections were around or less than 1\%).
We have simulated between 700 and 3000 RHMC trajectories for each point in Table \ref{tab:T0}.

\begin{table}
\begin{center}
\begin{tabular}{|c|c|c|c|}
\hline
$\beta$ & $m_{s}$ & $m_{ud}$ & lattice size\\
\hline
\hline
$3.330$ & 0.23847 & 0.02621 	&  $12^3\cdot24$ \\
  	&     	  & 0.04368   	&  $12^3\cdot24$\\
  	&     	  & 0.06115   	&  $12^3\cdot24$\\
  	&     	  & 0.07862   	&  $12^3\cdot24$\\
\hline                                               
$3.450$ & 0.15730 &  0.01729 	&  $16^3\cdot32$\\
  	&     	  &  0.02881 	&  $12^3\cdot28$\\
  	&    	  &  0.04033 	&  $12^3\cdot28$\\
  	&     	  &  0.05186 	&  $12^3\cdot28$\\
\hline                                               
$3.550$ & 0.10234 &   0.01312	&  $16^3\cdot32$ \\
  	&     	  &   0.01874	&  $16^3\cdot32$	  \\
  	&     	  &   0.02624	&  $12^3\cdot28$	  \\
  	&     	  &   0.03374	&  $12^3\cdot28$	  \\
\hline
 $3.670$ &  0.06331 	&  0.00928& $24^3\cdot32$  \\
 	  &  	   	&  0.01391&  $16^3\cdot32$  \\
 	  &  		&  0.01739&  $16^3\cdot32$ \\
 	  &  		&  0.02203&  $14^3\cdot32$ \\
\hline
 $3.750$ &  0.05025 	&  0.00736& $24^3\cdot32$   \\
 	  &  		&  0.01104&  $24^3\cdot32$   \\
 	  &  		&  0.01473&  $16^3\cdot32$   \\          
 	  &  		&  0.01841&  $16^3\cdot32$  \\
\hline

\end{tabular}          
\end{center}
\caption{\label{tab:T0}
Lattice parameters and sizes of our zero temperature simulations. The strange quark mass
is varied along the LCP as $\beta$ is changed. The light quark masses,
listed at each ($\beta$,$m_s$) values, correspond 
approximately to $m_\pi$ values of 250~MeV, 320~MeV,  380~MeV and 430~MeV.}
\end{table}

The T$\neq$0 simulations (c.f. Table \ref{tab:T}) were carried out along our LCP
(that is at physical strange and light quark masses,
which correspond to $m_K$=498~MeV and $m_\pi$=135~MeV) at four different
sets of lattice spacings ($N_t=4,6,8$ and $10$) and on three different volumes
($N_s/N_t$ was ranging between 3 and 6). We have observed moderate finite
volume effects on the smallest volumes for quantities which are supposed to depend 
strongly on light quark masses (e.g. chiral susceptibility). To determine
the transition point we used $N_s/N_t\ge 4$, for which we did not
observe any finite volume effect. The number of RHMC trajectories were between 1500 and 8000
for each parameter set (the integrated autocorrelation time was smaller 
or around 10 for all our runs).

\begin{table}
\begin{center}
\begin{tabular}{|c|c|c|}
\hline
temporal size ($N_t$) & $\beta$ range & spatial sizes ($N_s$)\\
\hline
\hline
$4$   &	$3.20-3.50$	&	$12,16,24$	\\
$6$   &	$3.45-3.75$	&	$18,24,32$	\\
$8$   &	$3.57-3.76$	&	$24,32,40$	\\
$10$  &	$3.63-3.86$	&	$28,40,48$	\\
\hline	
\end{tabular}
\end{center}
\caption{\label{tab:T}Summary of the T$\neq$0 simulation points.}
\end{table}

\newpage
\section{Improvement over previous results}

As we have already mentioned in the introduction, there are many lattice results 
on QCD thermodynamics. In this section we highlight the points,
where we have made improvements on previous calculations.

\subsubsection*{Physical quark masses}

We decided to use physical values for the quark masses.  Owing to the
computational costs this is a great challenge in lattice QCD. Previous analyses
used computationally less demanding non-physically large quark masses.

On the one hand, results with Wilson fermions \cite{AliKhan:2000iz,AliKhan:2001ek} were
obtained with pion masses $m_\pi \gtrsim 540$ MeV when approaching the
thermodynamical limit (since lattice QCD can give only dimensionless
combinations, it is more precise to say that $m_\pi$/$m_\rho\ge$ 0.6,
where $m_\rho$ is the mass of the rho meson). 

On the other hand, in staggered simulations one can afford considerably smaller
quark masses. The MILC collaboration \cite{Bernard:2004je,Bernard:2006nj} is currently using
two light quark masses (0.1 and 0.2 times $m_s$), the Bielefeld-Brookhaven-Columbia-RIKEN
collaboration is studying thermodynamics down to a pion mass of $\approx$ 320
MeV on $N_t$=4 and 6 lattices \cite{Cheng:2006qk}.  However these numbers should
be taken with a grain of salt.  Staggered fermions suffer from taste violation.
Therefore there is a large (usually several hundred MeV), unphysical mass splitting
between this lightest pion state and the higher lying other pion states. This
mass splitting results in an unphysical spectrum. The artificial pion mass
splitting disappears only in the continuum limit. For some choices of the
actions the restoration of the proper spectrum happens only at very small
lattice spacings, whereas for other actions somewhat larger lattice spacings
are already satisfactory.

The finite temperature transition is related to the spontaneous breaking of the
chiral symmetry (which is driven by the pion sector) and the three physical
pions have masses smaller than the transition temperature, thus the numerical
value of $T_c$ could be sensitive to the unphysical spectrum.  Furthermore, the
order of the transition depends on the quark mass. In three-flavor QCD for
vanishing quark masses the transition is of first-order. For intermediate
masses it is most probably a crossover. For infinitely heavy quark masses the
transition is again first-order.  Therefore the physical quark masses should be
used directly.

It is also important to mention that though at $T$=0 chiral perturbation theory
provides a technique to extrapolate to physical $m_\pi$, unfortunately no such
controllable method exists around $T_c$. 

\subsubsection*{Continuum limit}
The second ingredient is to remove the uncertainty associated with the lattice
discretization.  Discretization errors disappear in the continuum limit;
however, they strongly influence the results at non-vanishing lattice spacing. 

For lattice spacings which are smaller than some approximate limiting value the
dimensionless ratio of different physical quantities have a specific dependence
on the lattice spacing (for staggered QCD the continuum value is approached in
this region by corrections proportional to the square of the lattice spacing).
For these lattice spacings we use the expression: $a^2$ scaling region.
Clearly, results for at least three different lattice spacings are needed to
decide, whether one is already in this scaling region or not (two points can
always be fitted by $c_0$+$c_2$$a^2$, independently of possible large higher
order terms).  Only using the $a^2$ dependencies in the scaling region, is it
possible to unambiguously define the absolute scale of the system. Outside the
scaling region\footnote{Note, that outside the scaling region even a seemingly
small lattice spacing dependence can lead to an incorrect result.  An infamous
example is the Naik action \cite{Naik:1986bn} in the Stefan-Boltzmann limit:
$N_t$=4 and 6 are consistent with each other with a few \% accuracy, but since
they are not in the scaling region they are 20\% off the continuum value.}
different quantities lead to different overall scales, which lead to ambiguous
values for e.g. $T_c$.  

In three-flavour unimproved staggered QCD, using a lattice spacing of about
0.28~fm, the first-order and the crossover regions are separated by a
pseudoscalar mass of $m_{\pi,c}\approx300$~MeV. Studying the same
three-flavour theory with the same lattice spacing, but with an improved p4
action  (which has different discretization errors)  we obtain
$m_{\pi,c}\approx70$~MeV. In the first approximation, a pseudoscalar mass of
140~MeV (which corresponds to the numerical value of the physical pion mass)
would be in the first-order transition region, whereas using the second
approximation, it would be in the  crossover region.  The different
discretisation uncertainties are solely responsible for these qualitatively
different results \cite{Karsch:2003va,Endrodi:lat07,Phil:lat07}. 

In summary the proper approach is to extrapolate to vanishing lattice spacings
using lattices which are alrady in the scaling regime.  We approach the scaling
region by using four different sets of lattice spacing, which are defined as
the transition region on $N_t$=4,6,8 and 10 lattices. The results show (not
surprisingly) that the coarsest lattice with $N_t$=4 is not in the $a^2$
scaling region, whereas for the other three a reliable continuum limit
extrapolation can be carried out.  In case of the equation of state we only
have two lattice spacings ($N_t=4$ and $6$), for the continuum limit we have to
wait for results on finer lattices.

\subsubsection*{Lattice artefacts for $T=0$ and for $T\to \infty$}

\begin{figure}
\begin{center}
\hspace{2cm}
\epsfig{file=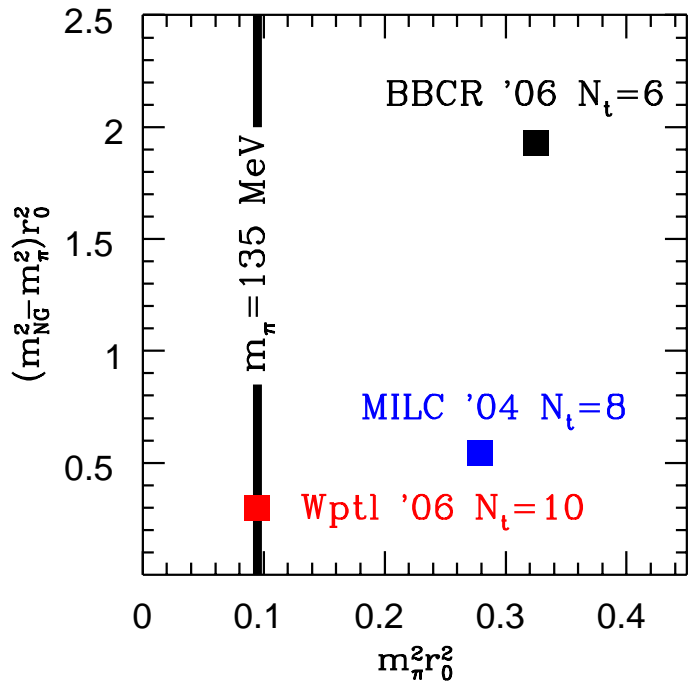}
\hspace{2cm}
\caption{
 Pion mass splitting $\Delta m_\pi^2=m_{\rm NG}^2-m_\pi^2$ as a 
 function of $m_\pi^2$ in units of $r_0$ for different works on lattice
 thermodynamics with staggered quarks.
 The lattice spacings are the same as those at the
 finite temperature transition point. 
The mass of the Goldstone pion is denoted by $m_\pi$, that of the first
 non-Goldstone mode is by $m_{\rm NG}$. The vertical black line corresponds to the physical value of
$m_\pi^2$.
The taste violation of our stout-link improved action
is smaller than that of any other actions used in the literature.
}
\label{fig:taste}
\end{center}
\end{figure}

For the staggered formulation of quarks the physically almost degenerate pion
triplet has an unphysical non-degeneracy (so-called taste violation). This mass
splitting $\Delta m_{\pi}^2$ vanishes in the continuum limit as
$a\rightarrow$0. Due to our smaller lattice spacing and particularly due to our
stout-link improved action the splitting $\Delta m_{\pi}^2$ is much smaller
than that of the previously or currently used staggered actions in
thermodynamics.  In order to illustrate the advantage of the stout-link action
Fig. \ref{fig:taste} compares the taste violation in different approaches of
the literature, which have been used for staggered thermodynamics.  Results on
the pion mass splitting for p4 improved (used by Bielefeld-Brookhaven-Columbia-RIKEN collaboration
\cite{Cheng:2006qk}), asqtad
improved (used by MILC collaboration \cite{Bernard:2004je,Bernard:2006nj}) and stout-link improved (this
work) staggered fermions are shown. The parameters were chosen to be the ones
used by the different collaborations at the finite temperature transition
point.

At infinitely large temperatures improved actions (p4 \cite{Heller:1999xz} or
asqtad \cite{Lepage:1998vj,Orginos:1999cr} action) show considerably smaller
discretizations errors, than the standard staggered action (used by this work).
However our choice of action is about an order of magnitude faster than e.g.
p4, we decided to use this less improved action, with which our CPU resources
made it possible to study several lattice spacings ($N_t$=4 and 6 for the
equation of state and $N_t=4,6,8$ and $10$ for determining the order of the
phase transition and the transition temperature). This turned out to be extremely
beneficial, when converting the transition temperature into physical units.  In
particular the $T=0$ simulations -which are used to do this conversion- have
very large lattice artefacts at $N_t=4$ and $6$ lattice spacings and can not be
used for controlled continuum extrapolations. The high-temperature improvement
is not designed to reduce these artefacts.

\subsubsection*{Setting the physical scale}

An additional problem appears if we want to give dimensionful predictions with
a few percent accuracy.  As we already emphasized lattice QCD predicts
dimensionless combinations of physical observables. For dimensionful
predictions one calculates an experimentally known dimensionful quantity, which
is used then to set the overall scale.  In many analyses the overall scale is
related to some quantities which strictly speaking do not even exist in full
QCD (e.g. the mass of the rho eigenstate and the string tension are not well
defined due to decay or string breaking). A better, though still not
satisfactory possibility is to use quantities, which are well defined, but can
not be measured directly in experiments. Such a quantity is the heavy
quark-antiquark potential (V), or its characteristic distances: the $r_0$ or
$r_1$ parameters of V \cite{Sommer:1993ce} ($r^2 d^2 V/dr^2$=1.65 or 1, for
$r_0$ or $r_1$, respectively).  For these quantities intermediate lattice
calculations and/or approximations are needed to connect them to measurements.
These calculations are based on bottonium spectroscopy.  This procedure leads
to further, unnecessary systematic uncertainties. 

The ultimate solution is to use quantities, which can be measured directly in
experiments and on the lattice. We use the decay constant of the kaon
$f_K$=159.8~MeV, which has about 1\% measurement error.  Detailed additional
analyses  were done by using the mass of the $K^*(892)$ meson $m_{K^*}$, the
pion decay constant $f_\pi$ and the value of $r_0$, which all show that we are
in the $a^2$ scaling regime and our choice of overall scale is unambiguous 
(see subsection \ref{ssec:lcp}).

\subsubsection*{Algorithm}

In previous staggered thermodynamics simulations the inexact R-algorithm was
used exclusively to simulate three quark flavours.  This algorithm has an
intrinsic parameter, the stepsize which, similarly to the lattice spacing, has
to be extrapolated to zero.  None of the previous staggered lattice
thermodynamic studies carried out this extrapolation. Using the R-algorithm
without stepsize extrapolation leads to uncontrolled systematic errors. Instead
of using the approximate R-algorithm this work uses the exact RHMC-algorithm
(rational hybrid Monte-Carlo) \cite{Clark:2003na}.

\section{Order of the QCD transition}
The nature of the QCD transition affects our understanding of the
Universe's evolution~(see Ref.~\cite{Schwarz:2003du} for example). 
In a strong 
first-order phase transition the quark--gluon plasma supercools before 
bubbles of hadron gas are formed. 
These bubbles grow, collide and merge, during which gravitational 
waves could be produced~\cite{Witten:1984rs}. Baryon-enriched 
nuggets could remain 
between the bubbles, contributing to dark matter.
The hadronic phase is the initial condition for nucleosynthesis, so 
inhomogeneities in this phase could have a strong 
effect on nucleosynthesis~\cite{Applegate:1985qt}. 
As the first-order phase transition weakens, these effects become less 
pronounced. Our calculations provide strong evidence that 
the QCD transition is a crossover and thus the 
above scenarios ---and many others---  are ruled out.
 
There are some QCD results and model calculations to determine the order of the
transition at $\mu$=0 and $\mu$$\neq$0 for different fermionic contents
(compare
refs~\cite{Pisarski:1983ms,Celik:1983wz,Kogut:1982rt,Gottlieb:1985ug,Brown:1988qe,Fukugita:1989yb,Halasz:1998qr,Berges:1998rc,Schaefer:2004en,Herpay:2005yr}).
Unfortunately, none of these approaches can give an unambiguous answer for the
order of the transition for physical values of the quark masses. The only known
systematic technique which could give a final answer is lattice QCD.  

There are several lattice results for the order of the QCD transition (for the
two most popular lattice fermion formulations see refs~\cite{Brown:1990ev}
and~\cite{AliKhan:2000iz}), although they have unknown systematics.  As we have
already emphasized in the previous section from the lattice point of view there
are two important 'ingredients' to eliminate these systematic uncertainties:
one has to use physical quark masses and carry out a continuum extrapolation.  

Our goal is to identify the nature of the transition for physical quark masses
as we approach the continuum limit.  We will study the finite size scaling of
the lattice chiral susceptibilities
$\chi(N_s,N_t)$=$\partial^2$/$(\partial$$m_{ud}^2)$($T/V$)$\cdot\log Z$, where
$m_{ud}$ is the mass of the light u,d quarks and $N_s$ is the spatial
extension.  This susceptibility shows a pronounced peak around the transition
temperature ($T_c$). For a real phase transition the height of the
susceptibility peak increases and the width of the peak decreases when we
increase the volume.  For a first-order phase transition the finite size
scaling is determined by the geometric dimension, the height is proportional to
$V$, and the width is proportional to $1/V$. For a second-order transition the
singular behaviour is given by some power of $V$, defined by the critical
exponents. The picture would be completely different for an analytic crossover.
There would be no singular behaviour and the susceptibility peak does not get
sharper when we increase the volume; instead, its height and width will be $V$
independent for large volumes.

\begin{figure}[p]
\centerline{\includegraphics*[bb=20 430 580 710,width=13cm]{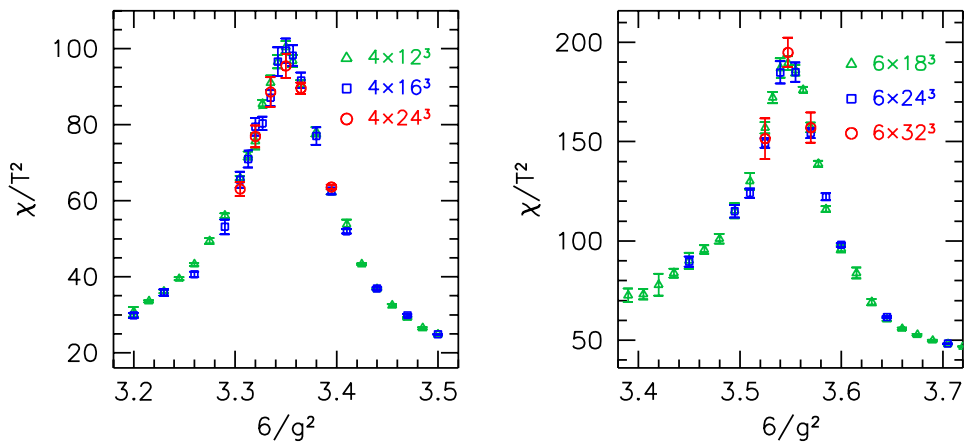}}
\caption{\label{fig:susc_46}
Susceptibilities  for the light quarks for $N_t$=4 (left panel) and for $N_t$=6
(right panel) as a function of $6/g^2$, where $g$ is the gauge coupling ($T$
grows with $6/g^2$).  The largest volume is eight times bigger than the
smallest one, so a first-order phase transition would predict a susceptibility
peak that is eight times higher (for a second-order phase transition the
increase would be somewhat less, but still dramatic). Instead of such a
significant change we do not observe any volume dependence. Error bars are
s.e.m.
}
\end{figure}

Fig. \ref{fig:susc_46} shows the susceptibilities for the light quarks for
$N_t=$4 and 6, for which we used aspect ratios $r=N_s/N_t$ ranging from 3 to 6
and 3 to 5, respectively.  A clear signal for an analytic crossover for both
lattice spacings can be seen. However, these curves do not say much about the
continuum behaviour of the theory. In principle a phenomenon as unfortunate as
that in the three-flavour theory could occur~\cite{Karsch:2003va}, in which the
reduction of the discretization effects changed the nature of the transition
for a pseudoscalar mass of $\approx$140~MeV.

\begin{figure}[p]
\centerline{\includegraphics*[bb=20 500 592 690,width=17cm]{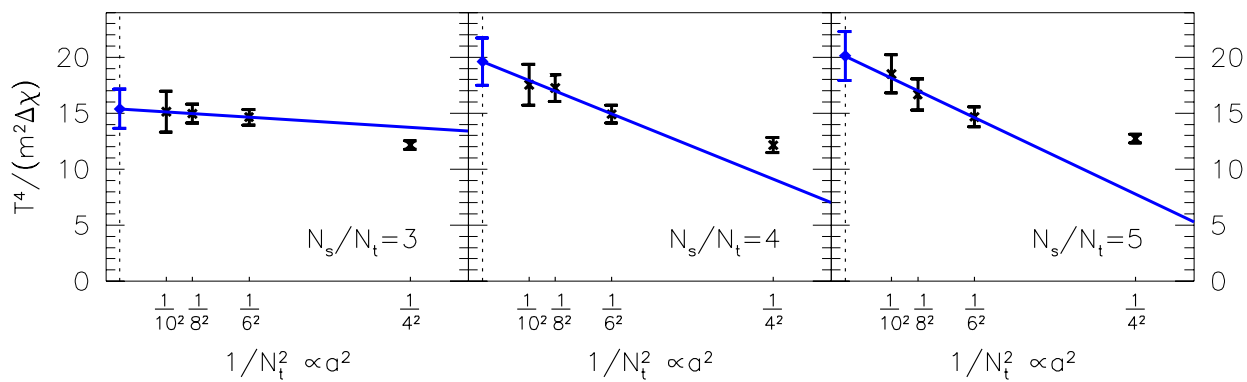}}
\caption{\label{fig:cont_ex}
Normalized susceptibilities $T^4/(m^2\Delta\chi)$ 
for the light quarks for aspect
ratios r=3 (left panel) r=4 (middle panel) and r=5 (right panel)
as functions of the lattice spacing. Continuum 
extrapolations are carried out for all three physical volumes and the
results are given by the leftmost blue diamonds. Error bars are s.e.m with 
systematic estimates.
}
\end{figure}
Because we are interested in genuine temperature effects we subtract the $T$=0
susceptibility and study only the difference between $T$$\neq$0 and $T$=0 at
different lattice spacings.  To do it properly, when we approach the continuum
limit the renormalization of $\chi$ has to be performed. This leads to
$m^2$$\Delta$$\chi$, which we study (for the details see subsection \ref{ssec:chi}). 

\begin{figure}[t]
\centerline{\includegraphics*[width=10cm,bb=0 170 592 614]{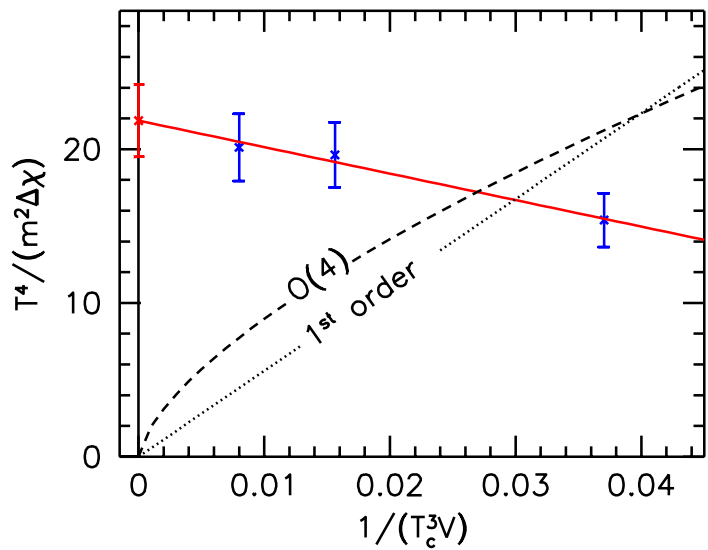}}
\caption{\label{fig:cont_scal}
Continuum extrapolated susceptibilities $T^4/(m^2\Delta\chi)$ 
as a function of 1/$(T_c^3V)$.
For true phase transitions the infinite volume extrapolation should
be consistent with zero, whereas
for an analytic crossover the infinite volume extrapolation gives
a non-vanishing value. The continuum-extrapolated
susceptibilities show no phase-transition-like volume dependence, though
the volume changes by a factor of five.
The V$\rightarrow$$\infty$ extrapolated value is 22(2)
which is
11$\sigma$ away from zero. For illustration, we fit the expected 
asymptotic
behaviour for first-order and O(4) (second order) 
phase transitions shown by dotted and dashed lines,
which results in chance probabilities of 
$10^{-19}$ ($7\times10^{-13}$), respectively. Error bars are s.e.m with 
systematic estimates.
}
\end{figure}
To give a continuum result for the order of the transition we carry out a
finite size scaling analysis of the dimensionless quantity
$T^4/(m^2\Delta\chi)$ directly in the continuum limit. For this study we need
the height of the susceptibility peaks in the continuum limit for fixed
physical volumes.  The continuum extrapolations are done using four different
lattice spacings ($N_t$=4,6,8 and 10). The volumes at different lattice
spacings are fixed in units of $T_c$, and thus $VT_c^3$=$3^3$,$4^3$ and $5^3$
were chosen.  (In three cases the computer architecture did not allow us to
take the above ratios directly. In these cases, we used the next possible
volume and interpolated or extrapolated. The height of the peak depends weakly
on the volume, so these procedures were always safe.) Altogether we used twelve
different lattice volumes ranging from $4\cdot12^3$ to $10\cdot48^3$ at $T>0$.
For the $T=0$ runs lattice volumes from $24\cdot12^3$ up to $56\cdot28^3$ were
used.  The number of trajectories were between 1500 and 8000 for $T>0$ and
between 1500 and 3000 for $T=0$, respectively.  Fig. \ref{fig:cont_ex} shows the
continuum extrapolation for the three different physical volumes. The $N_t$=4
results are slightly off but the $N_t$=6,8 and 10 results show a good
$a^2$$\propto$$1/N_t^2$ scaling.

Having obtained the continuum values for $T^4/(m^2\Delta\chi)$ at fixed
physical volumes, we study the finite size scaling of the results.
Fig. \ref{fig:cont_scal} shows our final results.  The volume dependence strongly
suggests that there is no true phase transition but only an analytic crossover
in QCD.

\section{Transition temperature}

There are several results in the literature for $T_c$ using both staggered and
Wilson fermions
\cite{Karsch:2000kv,Bernard:1997an,AliKhan:2000iz,Bornyakov:2004ii,Bernard:2004je,Cheng:2006qk}.
There is an additional limitation of these, which has not been mentioned
before.  This problem is related to an implicit assumption about a a real
singularity, thus ignoring the analytic cross-over feature of the finite
temperature QCD transition.

\begin{figure}[p]
\centerline{\includegraphics*[width=12cm]{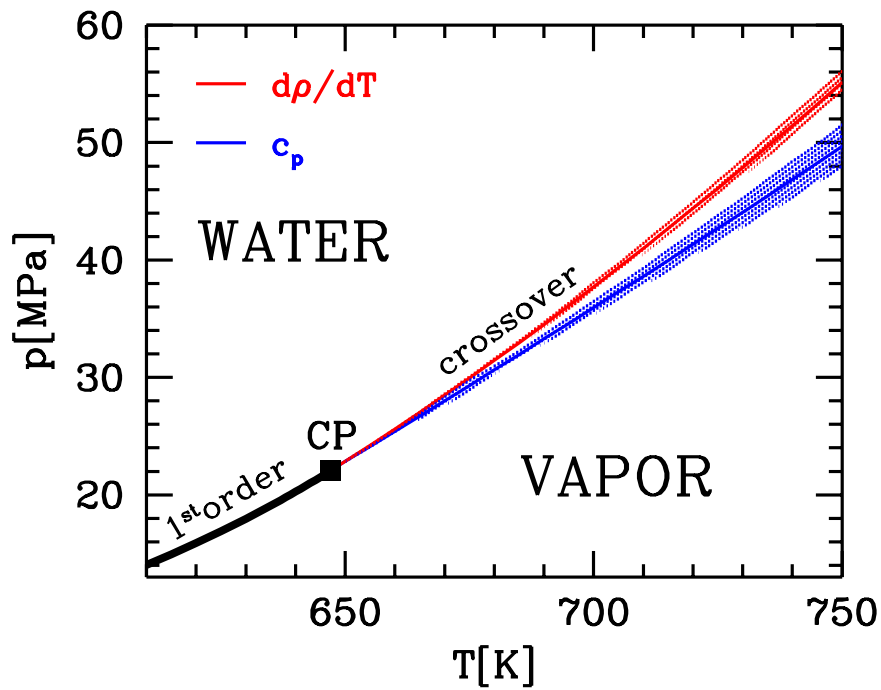}}
\caption{\label{fig:steam}
The phase diagram of water around its critical point (CP). For pressures below the critical
value ($p_c$) the transition is first order, for $p>p_c$ values there is a rapid crossover.
In the crossover region the transition temperatures defined from different quantities are not 
necessarily equal. This can be seen for the temperature derivative of the density ($d\rho/dT$) and
the specific heat ($c_p$). The bands show the experimental uncertainties (see \cite{Spang}).
}
\end{figure}

As we have seen before the QCD transition at non-vanishing temperatures is an
analytic cross-over. Since there is no singular temperature dependence
different definitions of the transition point lead to different values. The
most famous example for this phenomenon is the water-vapor transition, for
which the transition temperature can be defined by the peaks of $d\rho/dT$
(temperature derivative of the density) and $c_p$ (heat capacity at fixed
pressure). For pressures ($p$) somewhat less than $p_c=22.064$~MPa the
transition is of first order, whereas at $p=p_c$ the transition is second
order. In both cases the singularity guarantees that both definitions of the
transition temperature lead to the same result.  For $p>p_c$ the transition is
a rapid cross-over, for which e.g. both $d\rho/dT$ and $c_p$ show pronounced
peaks as a function of the temperature, however these peaks are at different
temperature values. Fig. \ref{fig:steam} shows the phase diagram based on
\cite{Spang}. Analogously, there is no unique transition temperature in QCD. 

Our goal is to eliminate all the above limitations and give the full answer.
We determine $T_c$ using the sharp changes of the temperature (T) dependence of
renormalized dimensionless quantities obtained from the chiral condensate
($\langle \bar \psi \psi \rangle$), quark number susceptibility ($n_q$) and
Polyakov loop ($P$).  We expect that all three quantities result in different
transition points (similarly to the case of the water, c.f. Fig.
\ref{fig:steam}). 

\subsection{Chiral susceptibility}
\label{ssec:chi}

The chiral susceptibility of the light quarks ($\chi$) is defined as 
\begin{equation} 
\chi_{\bar{\psi}\psi}=\frac{T}{V}\frac{\partial^2}{\partial m_{ud}^2} \log Z= 
-\frac{\partial^2}{\partial m_{ud}^2}f, 
\end{equation}
where $f$ is the free energy density. Since both the bare quark mass and
the free energy density contain divergences, $\chi_{\bar{\psi}\psi}$ has to be renormalized.

The renormalized quark mass can be written as $m_{R,ud}=Z_m\cdot m_{ud}$.
If we apply a mass independent renormalization then we have
\begin{equation} 
m_{ud}^2\frac{\partial^2}{\partial
m_{ud}^2}=m_{R,ud}^2\frac{\partial^2}{\partial m_{R,ud}^2}. 
\end{equation}
The free energy has additive, quadratic divergencies. They can be removed by
subtracting the free energy at $T=0$ (this is the usual renormalization
procedure for the free energy or pressure), which leads to $f_R$. Therefore, we have the following
identity: 
\begin{equation}
m_{ud}^2\frac{\partial^2}{\partial
m_{ud}^2}\left(f(T)-f(T=0)\right)= m_{R,ud}^2\frac{\partial^2}{\partial
m_{R,ud}^2}f_R(T). 
\end{equation} 
the right hand side contains only renormalized quantities, which
can be determined by measuring the susceptibilities of the left hand side 
(for the above expression we use the shorthand notation
$m_{ud}^2 \cdot \Delta\chi_{\bar{\psi}\psi}$).
In order to obtain a dimensionless quantity
it is natural to  normalize the above quantity by $T^4$
(which minimizes the final errors). 
Alternatively, one can use 
combinations of $T$ and/or $m_\pi$ to construct dimensionless
quantities (though these conventions lead to larger errors). 
Since the transition is a cross-over (c.f. discussion d of our Introduction)
the maxima of $m_{ud}^2/m_\pi^2 \cdot \Delta\chi_{\bar{\psi}\psi}/T^2$ or
$m_{ud}^2/m_\pi^4 \cdot \Delta\chi_{\bar{\psi}\psi}$ give somewhat different values for
$T_c$. 

\begin{figure}[p]
\centerline{\includegraphics*[height=15cm,width=8cm]{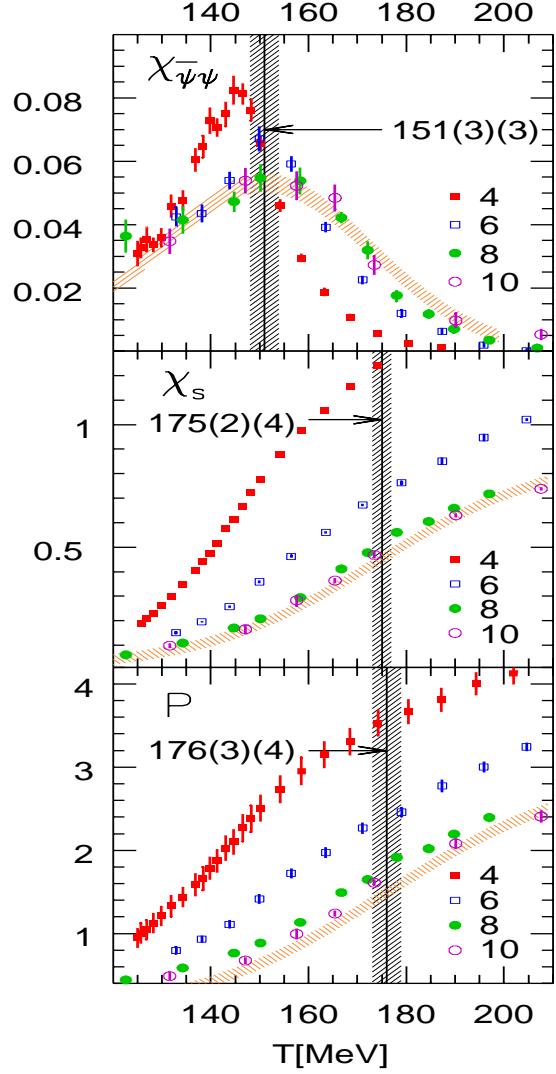}}
\caption{\label{fig:susc}
{Temperature dependence of the renormalized
chiral susceptibility ($m^2\Delta \chi_{\bar{\psi}\psi}/T^4$), the strange
quark number susceptibility ($\chi_s/T^2$) 
and the renormalized Polyakov-loop ($P_R$) in the transition region. The different symbols show the results for $N_t=4,6,8$ and
$10$ lattice spacings (filled and empty boxes for $N_t=4$ and $6$, filled and open circles for $N_t=8$
and $10$).
The vertical bands indicate the corresponding transition temperatures and 
its uncertainties coming from the T$\neq$0 analyses. This error is 
given by the number in the first parenthesis, whereas the error of the 
overall scale determination is indicated by the number in the second 
parenthesis. The orange bands show our continuum limit estimates for the 
three renormalized quantities as 
a function of the temperature with their uncertainties.}
}
\end{figure}

The upper panel of 
Fig. \ref{fig:susc} shows the temperature dependence of the 
renormalized chiral susceptibility
for different temporal extensions ($N_t$=4,6,8 and 10). 
For small enough lattice spacings, thus close to the continuum limit,
these curves should coincide.  As it can be seen, the $N_t=4$
result has considerable lattice artefacts,
however the two smallest lattice
spacings ($N_t=8$ and $10$) are already consistent with each other,
suggesting that they are also consistent with the continuum limit 
extrapolation (indicated by the orange band). The curves exhibit pronounced peaks. We define the 
transition temperatures by the position of these peaks. We 
fitted a second order expression to the peak to obtain its position. 
The slight
change due to the variation of the fitting range is taken as a systematic
error. The left panel of Fig. \ref{fig:tc} shows the transition 
temperatures in
physical units for different lattice spacings obtained from the
chiral susceptibility. As it can be seen
$N_t$=6,8 and 10 are already in the scaling region, thus a safe
continuum extrapolation can be carried out.
The extrapolations based on $N_t=6,8,10$ fit and
$N_t=8,10$ fit are consistent with each other. For our final result we
use the average of these two fit results (the difference between them
are added to our systematic uncertainty). 
Our T=0 simulations resulted in a $2\%$ error on the overall scale.
Our final result for the transition temperature based on the chiral
susceptibility reads:  
\begin{equation} 
T_c(\chi_{\bar{\psi}\psi})=151(3)(3) {\rm ~MeV},
\end{equation} 
where the first error comes from the T$\neq$0, the second from
the T=0 analyses.  

We use the second derivative of the chiral susceptibility ($\chi''$) at the peak position
to estimate the width of the peak ($(\Delta T_c)^2 = - \chi(T_c)/\chi''(T_c)$).
For the continuum
extrapolated width we obtained:
\begin{equation}
\Delta T_c(\chi_{\bar{\psi}\psi})=28(5)(1) {\rm ~MeV.}
\end{equation}
  
Note, that for a real phase transition (first or second order), the
peak would have a vanishing width
(in the thermodynamic limit), 
yielding a unique value for the transition
temperature (which then would be called critical temperature). Due to the crossover nature of the transition there is no such
value, there is a range ($151 \pm 28$ MeV) where the transition phenomena
takes place. Other quantities than the chiral susceptibility could
result in transition temperatures within this range.

The MILC collaboration also reported a continuum result on the transition
temperature based on the chiral susceptibility \cite{Bernard:2004je}. Their result is 
169(12)(4)~MeV. Note, that
their lattice spacings were not as small as ours
(they used $N_t$=4,6 and 8), their aspect ratio was quite small
($N_s$/$N_t$=2), they used non-physical quark masses (their smallest
pion mass at T$\neq$0 was $\approx$220~MeV),   
the non-exact R-algorithm was applied for the simulations
and they did not use the renormalized susceptibility, but
they looked for the peak in the bare $\chi_{\bar{\psi}\psi}/T^2$.
Using $T^4$ as a normalization prescription (as we did) 
the transition temperature would decrease their $T_c$ values
by approximately $9$~MeV.
Note, that their continuum extrapolation resulted in a quite
large error. Taking into account their uncertainties
our result and their result agree on the 1-sigma level.

\subsection{Quark number susceptibility}
\begin{figure}[t]
\centerline{\includegraphics*[width=16cm,height=6cm,bb=18 520 589 704]{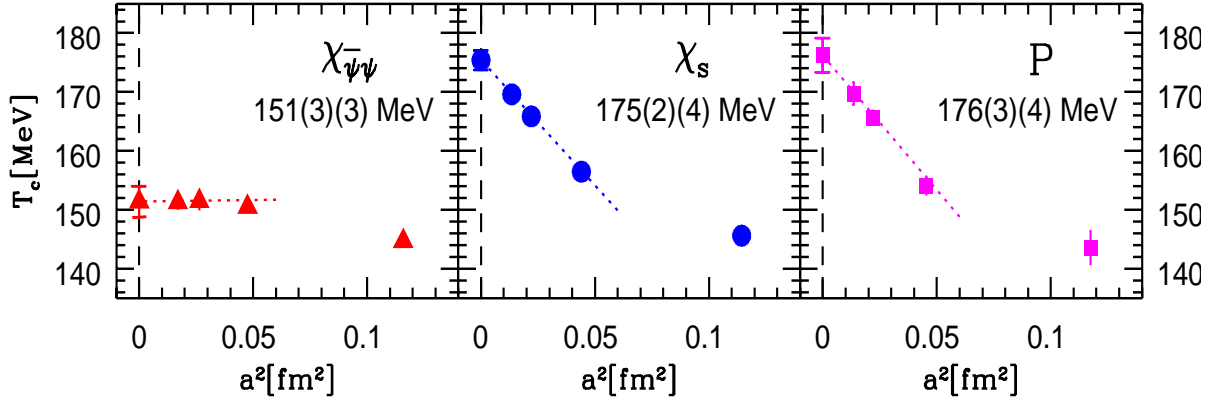}}
\caption{\label{fig:tc}
Continuum limit of the transition temperatures obtained from the renormalized chiral
susceptibility ($m^2\Delta \chi_{\bar{\psi}\psi}/T^4$), 
strange quark number susceptibility ($\chi_s/T^2$) and renormalized
Polyakov-loop ($P_R$). 
}
\end{figure}

For heavy-ion experiments the quark number susceptibilities are 
quite useful, since
they could be related to event-by-event fluctuations.
Our second transition temperature is obtained from the strange quark number
susceptibility,  which is defined via 
\cite{Bernard:2004je}
\begin{equation}
\frac{\chi_{s}}{T^2}=\frac{1}{TV}\left.\frac{\partial^2 \log Z}{\partial \mu_{s} ^2
}\right|_{\mu_{s}=0}, 
\end{equation}
where $\mu_s$ is the strange quark chemical potential (in
lattice units). Quark number susceptibilities have the convenient property,
that they automatically have a proper continuum limit, there is no need for
renormalization.

The middle panel of
Fig. \ref{fig:susc} shows the temperature dependence of the
strange quark number susceptibility
for different temporal extensions ($N_t$=4,6,8 and 10).
For small enough lattice spacings, thus close to the continuum limit,
these curves should coincide again (our continuum limit estimate is 
indicated by the orange band).

As it can be seen, the $N_t=4$
results are quite off, however the two smallest lattice
spacings ($N_t=8$ and $10$) are already consistent with each other,
suggesting that they are also consistent with the continuum limit
extrapolation. This feature indicates, that they are closer to 
the continuum result than our statistical uncertainty.

We defined the transition temperature as the peak in the temperature
derivative of the strange quark number susceptibility, 
that is the inflection point of the
susceptibility curve. The position was determined by two independent ways,
which yielded the same result. In the first case we fitted a cubic
polynomial on the susceptibility curve, while in the second case we
determined the temperature derivative numerically from neighboring points
and fitted a quadratic
expression to the peak. The slight
change due to the variation of the fitting range is taken as a systematic
error. The middle panel of Fig. \ref{fig:tc} shows the transition
temperatures in
physical units for different lattice spacings obtained from the
strange quark number susceptibility. As it can be seen
$N_t$=6,8 and 10 are already in the $a^2$ scaling region, thus a safe
continuum extrapolation can be carried out.
The extrapolations based on $N_t=6,8,10$ fit and
$N_t=8,10$ fit are consistent with each other. For our final result we
use the average of these two fit results (the difference between them
is added to our systematic uncertainty).
The continuum extrapolated value for the transition temperature 
based on the strange quark number susceptibility is
significantly higher than the one from the chiral susceptibility. The 
difference is 24(4)~MeV.  For the transition temperature in the continuum 
limit one gets:
\begin{equation}
T_c(\chi_s)=175(2)(4) {\rm ~MeV},
\end{equation}
where the first (second) error is from the T$\neq$0 (T=0) temperature
analysis (note, that due to the uncertainty of the
overall scale, the difference is more precisely determined than the
uncertainties of 
$T_c(\chi_{\bar{\psi}\psi})$
and $T_c(\chi_s)$ would suggest).
\footnote{ 
A continuum extrapolation using only the two 
coarsest lattices ($N_t=4$ and $6$) 
yielded
$T_c \sim 190$~MeV \cite{Katz:2005br}, 
where an approximate LCP (LCP1) was used,
if the lattice spacing is set by $r_0$.
}
Similarly to the 
chiral susceptibility analysis, the curvature at the peak can be used
to define a width for the transition. 
\begin{equation}
\Delta T_c(\chi_s)= 42(4)(1) {\rm ~MeV}.
\end{equation}

\subsection{Polyakov loop}

In pure gauge theory the order parameter of the confinement transition is 
the Polyakov-loop: 
\begin{equation} 
P=\frac{1}{N_s^3}\sum_{\bf x} {\rm tr} 
[U_4({\bf x},0) U_4({\bf x},1) \dots U_4({\bf x},N_t-1)]. 
\end{equation} 
P acquires a non-vanishing expectation value in the deconfined phase, 
signaling the spontaneous breakdown of the Z(3) symmetry. When fermions 
are present in the system, the physical interpretation of the 
Polyakov-loop expectation value is more complicated (see e.g..
\cite{Kratochvila:2006jx}). 
However, its absolute value can be related to the quark-antiquark 
free energy at infinite separation: 
\begin{equation} 
|\langle P \rangle |^2 = 
\exp(-\Delta F_{q\bar{q}}(r\to \infty)/T). 
\end{equation} 
$\Delta F_{q\bar{q}}$ is the difference of the free energies 
of the quark-gluon plasma with and without the quark-antiquark pair.

The absolute value of the Polyakov-loop vanishes in the continuum 
limit. It needs renormalization. This can be 
done by renormalizing the free energy of the quark-antiquark pair
\cite{Kaczmarek:2002mc}. 
Note, that QCD at T$\neq$0 has only the  
ultraviolet divergencies which are already present at T=0. 
In order to remove these divergencies at a given lattice spacing
we used a simple renormalization condition \cite{Fodor:2004ft}: 
\begin{equation}
V_R(r_0)=0,
\end{equation} 
where the potential is measured at T=0 from Wilson-loops. The above
condition fixes the additive term in the potential at a given lattice
spacing. This additive term can be used at the same lattice spacings
for the potential obtained from Polyakov loops, or equivalently
it can be built in into the definition of the renormalized 
Polyakov-loop.
\begin{equation} 
|\langle P_R \rangle | = |\langle 
P \rangle | \exp(V(r_0)/(2T)), 
\end{equation}
where $V(r_0)$ is the unrenormalized potential  
obtained from Wilson-loops.

The lower panel of
Fig. \ref{fig:susc} shows the temperature dependence of the
renormalized Polyakov-loops
for different temporal extensions ($N_t$=4,6,8 and 10).
The two smallest lattice
spacings ($N_t=8$ and $10$) are approximately in 1-sigma agreement
 (our continuum limit estimate is indicated by the orange band).

Similarly to the strange quark susceptibility case
we defined the transition temperature as the peak in the temperature
derivative of the Polyakov-loop, that is the inflection point of the
Polyakov-loop curve. To locate this point and determine its uncertainties
we used the same two methods,
which were used to determine $T_c(\chi_s)$.  
The right panel of Fig. \ref{fig:tc} shows the transition
temperatures in
physical units for different lattice spacings obtained from the
Polyakov-loop. As it can be seen
$N_t$=6,8 and 10 are already in the scaling region, thus a safe
continuum extrapolation can be carried out. The extrapolation
and the determination of the systematic error were done as for
$T_c(\chi_s)$.
The continuum extrapolated value for the transition temperature 
based on the renormalized Polyakov-loop is
significantly higher than the one from the chiral susceptibility. The 
difference is 25(4)~MeV.  For the transition temperature in the continuum 
limit one gets:
\begin{equation}
T_c(P)=176(3)(4) {\rm ~MeV},
\end{equation}
where the first (second) error is from the T$\neq$0 (T=0) temperature
analysis (again, due to the uncertainties of the
overall scale, the difference is more precisely determined than the
uncertainties of $T_c(\chi)$ and $T_c(P)$ suggest). Similarly to the
chiral susceptibility analysis, the curvature at the peak can be used
to define a width for the transition.
\begin{equation}
\Delta T_c(P)=38(5)(1) {\rm ~MeV}.
\end{equation}

\subsection{Comparison with the recent result of the BBCR collaboration}
\begin{figure}[h!]
\centerline{\includegraphics*[width=18cm,bb=90 620 492 816]{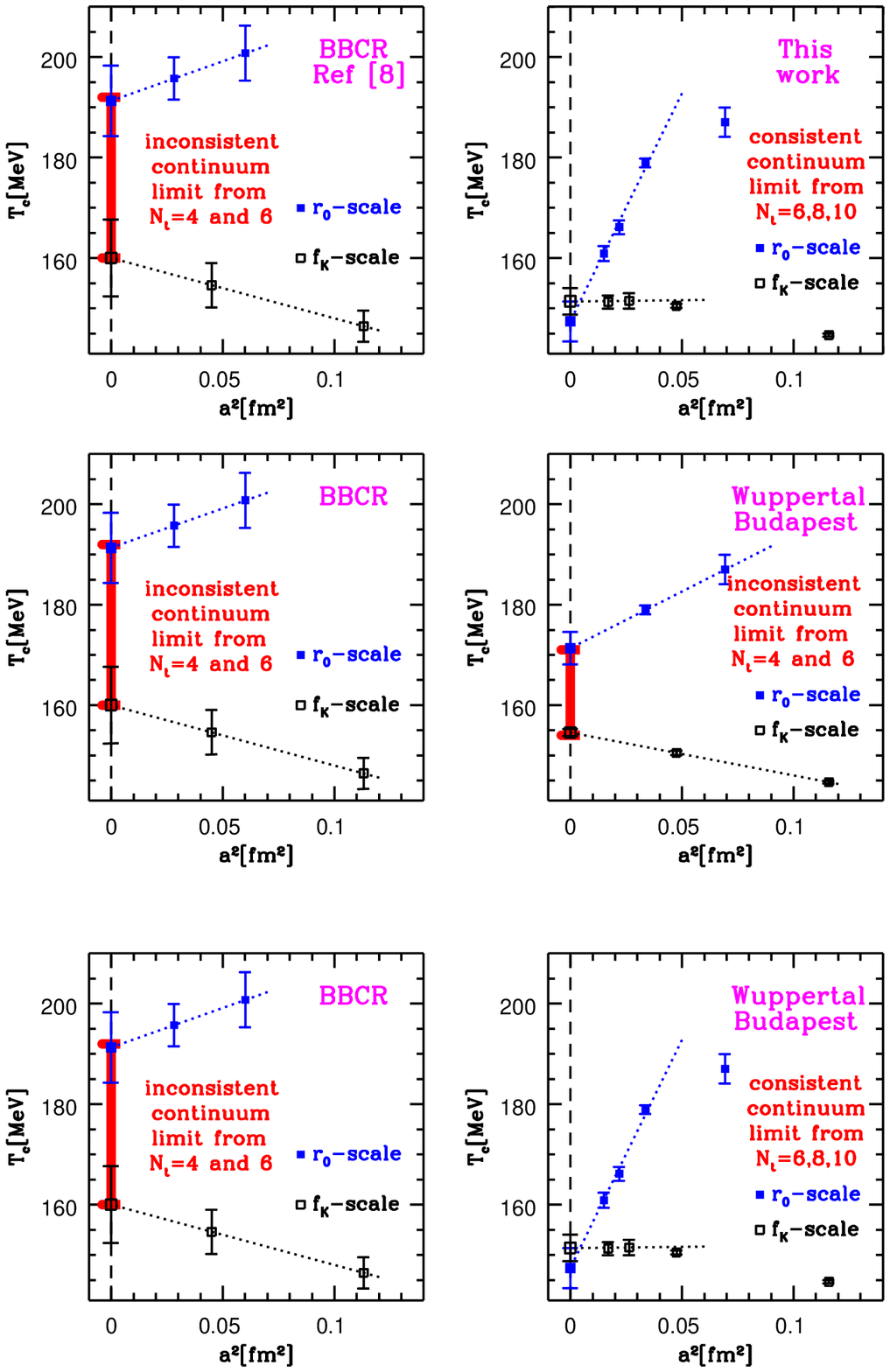}}
\caption{\label{fig:note}
Resolving the discrepancy between the transition temperature of Ref.
\cite{Cheng:2006qk} and that of the present work (see text). 
The major part of the difference can be traced
back to the unreliable continuum extrapolation of \cite{Cheng:2006qk}.  Left
panel: In Ref. \cite{Cheng:2006qk} $r_0$ was used for scale setting
(filled boxes), however using the kaon decay constant 
(empty boxes) leads to different transition temperatures even after performing 
the
continuum extrapolation.  Right panel: in our work the extrapolations based on
the finer lattices are safe, using the two different scale setting methods one
obtains consistent results.  
}
\end{figure}

Let us comment here on an independent study on $T_c$ based on large scale
simulations of the  Bielefeld-Brookhaven-Columbia-RIKEN group
\cite{Cheng:2006qk}.  The p4fat3 action was used, which is designed to give
very good results in the (T$\to$$\infty$) Stefan-Boltzmann limit
(their action is not optimized at T=0, which is needed e.g. to set the scale).
The overall scale was set by $r_0$.  The $T_c$ analysis based on the chiral
susceptibility peak gave in the continuum limit $T_c(\chi)$=192(7)(4)~MeV.
(The second error, 4~MeV,  estimates the uncertainty of the continuum limit
extrapolation, which we do not use in the following, since we attempt to give a
more reliable estimate on that.) This result is in obvious contradiction with
our continuum result from the same observable, which is
$T_c(\chi)$=151(3)(3)~MeV. For the same quantity (position of chiral
susceptibility peak with physical quark masses in the continuum limit) one
should obtain the same numerical result independently of the lattice action.
Since the chance probability that we are faced with a statistical fluctuation
and both of the results are correct is small, we attempted to understand the
origin of the discrepancy.  We repeated some of their simulations and analyses.
In these cases a complete agreement was found. In addition to their T=0
analyses we carried out an $f_K$ determination, too. This $f_K$ was used to
extend their work, to use an LCP based on $f_K$  and to determine $T_c$ in
physical units. 

We summarize the origin of the contradiction between our findings and theirs.
The major part of the difference can be explained by the fact, that the lattice
spacings of  \cite{Cheng:2006qk} are too large ($\gtrsim$0.20~fm), thus they are
not in the $a^2$ scaling regime, in which a justified continuum extrapolation
could have been done.  Setting the scale by different dimensionful quantities
should lead to the same result. However at their lattice spacings the overall
scales obtained by $r_0$ or by $f_K$ can differ by $\gtrsim$20\%, and even the
continuum extrapolated $r_0f_K$ value of these scales is about 4--5$\sigma$
away from the value given by the literature \cite{Aubin:2004fs,Gray:2005ur}.
This scale ambiguity appears in $T_c$, too (though other uncertainties of
\cite{Cheng:2006qk},  e.g. coming from the determination of the peak-position,
somewhat hide its high statistical significance).  We used their $T_c$ values
fixed by their $r_0$ scale, and in addition we converted their peak position of
the chiral susceptibility to $T_c$ setting the scale by $f_K$. (In order to
ensure the possibility of a consistent continuum limit --independently of the
actual physical value of $r_0$-- we used for both $r_0$ and $f_K$ the results
of \cite{Aubin:2004fs,Gray:2005ur} as Ref. \cite{Cheng:2006qk} did it for
$r_0$.) Setting the overall scale by $f_K$ predicts a much smaller $T_c$ at
their lattice spacings than doing it by $r_0$ (see left panel of Fig.
\ref{fig:note}). Even after carrying out the continuum extrapolation the
difference does not vanish ($\sim 30$~MeV), which means that the lattice
spacings $\gtrsim$0.20~fm used by \cite{Cheng:2006qk} are not in the scaling
regime.  Thus, results obtained with their lattice spacings can not give a
consistent continuum limit for $T_c$.

In our case not only $N_t=4$ and $6$ temporal extensions were used, but more
realistic $N_t$=8 and $10$ simulations were carried out, which led to smaller
lattice spacings. These calculations are already in the $a^2$ scaling regime
and a safe continuum extrapolation can be done. For our lattice spacing
different scale setting methods give  consistent results. This is shown on the
right panel of Fig. \ref{fig:note} (independently of the scale setting one
obtains the same $T_c$) and also justified with high accuracy by Fig.
\ref{fig:spect}, where $r_0f_K$ converges to the physical value on our finer
lattices.  As it can be seen on the plot, using only our $N_t=4$ and 6 results
would also give an inconsistent continuum limit. This emphasizes our conclusion
that lattice spacings $\gtrsim$ 0.20~fm can not be used for consistent continuum
extrapolations. 

The second, minor part of the difference comes from the different definitions
of the transition temperatures related to the chiral susceptibility.  We use the
renormalized chiral susceptibility with $T^4$ normalization to obtain the peak
position, which yields $\sim 9$~MeV smaller transition temperature than the bare
susceptibility normalized by $T^2$ of Ref. \cite{Cheng:2006qk}. 

\section{Equation of state}
The equilibrium description as a function
of the temperature is given by the equation of state (EoS).
The complete determination of the EoS needs non-perturbative inputs,
out of which the lattice simulation is the most 
systematic approach. 
  
The EoS has been determined in the continuum limit for the pure gauge theory
\cite{Boyd:1996bx,Okamoto:1999hi,Namekawa:2001ih}.  In this case --quenched
simulations-- the simulations are particularly easy as there is no fermionic
degree of freedom.  The simulated systems is equivalent to that where all the
fermions are infinitely heavy, thus, is far from the physical situation. (Note,
however, that even in this relatively simple case there is still a few \%
difference between the different approaches.) 

The situation in the unquenched case (QCD with dynamical quarks) is a bit
barouqe. There are many results with different flavor content, fermion
formulation, quark masses. None of them have used the proper physical quark
content and none of them has attempted to carry out a continuum extrapolation.
Moreover staggered fermion studies were always using inexact R-algorithm (which
can result in uncontrolled systematics, see subsection \ref{ssec:ralg}).
Let us give here a brief review of the literature. 
\begin{itemize}
\item 
There are published results for two-flavor QCD using unimproved
staggered \cite{Blum:1994zf,Bernard:1996cs}, and improved Wilson fermions
\cite{AliKhan:2001ek}. 
\item 
There are results available for the 2+1 flavour case,
among which the study done by Karsch, Laermann and Peikert in the year 2000
\cite{Karsch:2000ps}, using p4-improved staggered fermions, has often been used
as the best result of EoS from lattice QCD. An additional drawback of this result, 
that the concept
of the LCP was ignored from the calculation.
If 
Karsch,Laermann and Peikert had cooled down e.g. two of their systems one at 
$T$=3$T_c$ and one at $T$=0.7$T_c$ down to T=0, the first system 
would have had approximately 4 times larger quark masses -two times larger pion masses- 
than the second one; this unphysical
choice is known to lead systematics, which are comparable to the 
difference between the interacting and non-interacting plasma
\item 
Recently the MILC collaboration studied the equation of state along LCP's
with two light quark masses (0.1 and 0.2 times $m_s$) at $N_t$=4 and 6 lattices
using asqtad improved staggered fermionic action \cite{Bernard:2006nj}.
\end{itemize}

There are ongoing thermodynamics projects improving on previous results.  The
joint Bielefeld-Brookhaven-Columbia-MILC-RIKEN collaboration (hotQCD
collaboration) have presented new results at the lattice conference \cite{Heide:lat07}. Still the
results are only available for two different lattice spacings ($N_t=4$ and $6$)
and for unphyiscal quark masses.

There are two important further issues with the equation of state, which have
been overlooked in previous calculations. Firstly the heavier quarks can have
significant contribution to the pressure even at few times the transition
temperature (exploratory investigations \cite{Cheng:lat07} show a significant jump in the pressure
around $\sim 2\cdot T_c$ due to the charm quark). Secondly there is no serious
practical obstacle to extend the EoS calculations well beyond the usual $4-5
\cdot T_c$ \cite{Szabo:lat07}. This opens the possibility to find the missing connection between
conventional perturbation theory and nonperturbative methods.

In this section we present our first step towards the final solution of the EoS
with proper physical quark content.  Though we have the EoS on two different
sets of lattice spacings ($N_t=4$ and $6$) and one might attempt to do a
continuum extrapolation, it is fair to say that another set of
lattice spacings is needed ($N_t$=8). One of the reasons is, that in the
hadronic phase, where the integration for the pressure starts, the lattice
spacing is larger than 0.3 fm.  In this region the lattice artefacts can not be
really controlled (and in this deeply hadronic case it does not really help
that an action is very good at asymptotically high temperatures in the free
non-interacting gas limit).

\subsection{Integral technique}
We shortly review the integral technique to obtain the pressure
\cite{Engels:1990vr}. For large homogeneous systems the pressure
is proportional to the logarithm of the partition
function: 
\begin{align}
\label{eq:pa}
pa^4=\frac{Ta}{V/a^3}\log Z(T,V)=\frac{1}{N_tN_s^3}\log Z(N_s,N_t;\beta,m_q).
\end{align}
(Index `q' refers to the ${ud}$ and $s$ flavors.)
The volume and temperature are connected to the spatial and temporal
extensions of the lattice:
\begin{align}
V=(N_sa)^3, && T=\frac{1}{N_ta}.
\end{align}  
The divergent zero-point energy has to be 
removed by subtracting the zero 
temperature ($N_t\to \infty$) part of 
Eq. (\ref{eq:pa}). In practice the zero 
temperature subtraction 
is performed by using lattices with finite, but large $N_{t}$ (called $N_{t0}$, see Table \ref{ta:points}). So the normalized 
pressure
becomes:
\begin{align}
\frac{p}{T^4}=N_t^4\left[ \frac{1}{N_tN_s^3}\log
 Z(N_s,N_t;\beta,m_q) - \frac{1}{N_{t0}N_{s0}^3}\log 
Z(N_{s0},N_{t0};\beta,m_q) \right].
\end{align}
With usual Monte-Carlo techniques one cannot measure $\log Z$ directly, but only its
derivatives with respect to the bare parameters of the lattice action.
Having determined the partial derivatives one
integrates in the
multi-dimensional parameter space:
\begin{equation}\label{integral}
\frac{p}{T^4}=
N_t^4\int^{(\beta,m_q)}_{(\beta_0,m_{q0})}
d (\beta,m_q)\left[
\frac{1}{N_tN_s^3}
\left(\begin{array}{c}
{\partial \log Z}/{\partial \beta} \\
{\partial \log Z}/{\partial m_{q}}
\end{array} \right )-
\frac{1}{N_{t0}N_{s0}^3}
\left(\begin{array}{c}
{\partial \log Z_0}/{\partial \beta} \\
{\partial \log Z_0}/{\partial m_q}
\end{array} \right )
\right],
\end{equation}
where $Z/Z_0$ are shorthand notations for $Z(N_s,N_t)/Z(N_{s0},N_{t0})$.
Since the integrand is a gradient, the result is by
definition independent of
the integration path. We need the pressure
along the LCP, thus it is convenient to measure the derivatives of $\log Z$
along the LCP and perform the integration over
this line in the $\beta$, $m_{ud}$ and $m_s$ parameter space. 
The lower limits of the integrations (indicated by $\beta_0$ and $m_{q0}$)
were set sufficiently below the transition point. By this choice
the pressure gets
independent of the starting point (in other words it vanishes
at small temperatures).  In the case of $2+1$ flavor staggered QCD
the derivatives of $\log Z$ with
respect to $\beta$ and $m_q$ are proportional to 
the expectation value of the gauge action ($\langle S_g \rangle$ c.f. Eq. (\ref{action})) and to the 
chiral condensates ($\langle \bar{\psi}\psi_q \rangle $), respectively. 
Eq. (\ref{integral}) can be rewritten appropriately and the pressure
is given by (in this formula we write out explicitely the flavors):
\begin{equation}\label{pmu0}
\frac{p}{T^4}=
N_t^4\int^{(\beta,m_{ud},m_s)}_{(\beta_0,m_{ud0},m_{s0})}
d (\beta,m_{ud},m_s)\left[
\frac{1}{N_tN_s^3}
\left(\begin{array}{c}
\langle{\rm -S_g/\beta}\rangle \\
\langle\bar{\psi}\psi_{ud}\rangle \\
\langle\bar{\psi}\psi_{s}\rangle
\end{array} \right) 
-
\frac{1}{N_{t0}N_{s0}^3}
\left(\begin{array}{c}
\langle{\rm -S_g/\beta}\rangle_0 \\
\langle\bar{\psi}\psi_{ud}\rangle_0 \\
\langle\bar{\psi}\psi_{s}\rangle_0
\end{array} \right)\right], 
\end{equation}
where $\langle \dots \rangle _0$ means averaging on a $N_{s0}^3\cdot N_{t0}$ lattice.

The integral method was originally introduced for the pure gauge case, for
which the integral is one dimensional, it is performed along the $\beta$ axis.
Many previous studies for staggered dynamical QCD (e.g.
\cite{Bernard:1996cs,Engels:1996ag,Karsch:2000ps}) used a one-dimensional
parameter space instead of performing it along the LCP. Note, that for full
QCD the integration should be performed along a LCP path in a
multi-dimensional parameter space. 
\begin{table} \label{ta:c_values}
\hspace{2cm}
\begin{center}
\begin{tabular}{|c||c|c|c|}
\hline
$N_t$   & p/$T^4$   & $c_s^2$   & $\chi/T^2$     \\
\hline
4       & 9.12    & 1/3   &  2.24 \\
6       & 7.86   &  1/3   &  1.86 \\
$\infty$   & 5.21   & 1/3  & 1 \\
\hline
\end{tabular}
\end{center}
\hspace{2cm}
\caption{
Summary of the results for the 2+1 flavor pressure, 
speed of sound and 1 flavor quark number susceptibility in the non-interacting
Stefan-Boltzmann limit. $\epsilon/T^4$ is 3 times, whereas $s/T^3$ is
4 times the normalized value of the pressure ($p/T^4$) in the Stefan-Boltzmann limit.
The first two lines gives the results
for $N_t$=4,6 and the third line contains the results 
in the continuum
(in the thermodynamic limit). 
}
\end{table}

Using appropriate thermodynamical relations one can obtain any thermal properties
of the system. For example the energy density ($\epsilon$), entropy density ($s$)
and
speed of sound ($c_s^2$)
can be derived as
\begin{align}
\label{eq:escs}
\epsilon = T(\partial p/\partial T)-p, && s = (\epsilon + p) T, && c_s^2
=\frac{dp}{d\epsilon}.
\end{align}
To be able to do theses derivatives one has to know the temperature along the LCP.
Since the temperature is connected to the lattice spacing as
\(
T=(N_t a)^{-1
},
\)
we need a reliable estimate on $a$.
The lattice spacings at different points of the LCP are
determined by first matching the static potentials for different $\beta$ values at an intermediate distance
for $m_{ud}=\{3,5\}m_{ud}(phys)$ quark masses, then extrapolating
the results to the physical quark mass.
Relating these distances to physical observables
(determining the overall scale in physical units) will be the topic of
a subsequent publication. 
We show the results as a function of $T/T_c$. The transition
temperature ($T_c$) is defined by the inflection point of the
isospin number 
susceptibility ($\chi_{I}$, see later).

To get the energy density the literature usually uses another quantity, namely
$\epsilon$-3$p$, which can be also directly measured on the lattice.  In our
analysis it turned out to be more appropriate to calculate first the pressure
directly from the raw lattice data (Eq. (\ref{pmu0})) and then determine the
energy density and other quantities from the pressure (Eq. (\ref{eq:escs})).
The reasons for that can be summarized as follows. As we discussed we perform
T$\neq$0 simulations with physical quark masses, whereas the subtraction terms
from T=0 simulations are extrapolated from larger quark masses. This sort of
extrapolation is adequate for the chiral condensates, for which chiral
perturbation techniques work well. Thus, one can choose an integration path
for the T=0 part of the pressure, which moves along a LCP at some larger
$m_{ud}$ (e.g. 9 times $m_{ud}(phys)$) and then at fixed $\beta$ goes down to
the physical quark mass. No comparable analogous technique is available for
the combination $\epsilon$-3$p$. 

We have also calculated the pressure for the larger quark masses. Plotting it
as a function of the temperature the differences between them are significant.
As a function of $T/T_c$ these differences are smaller, but still remain
statistically significant in the $1.2...2.0 T_c$ region.  
Note that statements on the mass dependence are only qualitative
since such an analysis requires the careful matching of the scales at
different quark masses.

\subsection{Physics results}
\begin{table}
\begin{tabular}{|c|c|c|c||c|c|c|c|}
\hline
$\beta$    & $T/T_c$ & $p/T^4$ (raw)& $p/T^4$ (scaled) &
$\beta$    & $T/T_c$ & $p/T^4$ (raw)& $p/T^4$ (scaled) \\ \hline 
3.000 &$0.90$&$0.12(0.02)$&$0.07(0.01)$& 3.450 &  $0.80$&$0.07(0.11)$&$0.05(0.08)$ \\
3.150 &$0.95$&$0.32(0.07)$&$0.19(0.04)$& 3.500 &  $0.87$&$0.23(0.11)$&$0.15(0.08)$ \\
3.250 &$0.98$&$0.59(0.10)$&$0.34(0.06)$& 3.550 &  $0.96$&$0.59(0.12)$&$0.39(0.08)$ \\
3.275 &$0.99$&$0.73(0.10)$&$0.42(0.06)$& 3.575 &  $1.02$&$0.91(0.12)$&$0.60(0.08)$ \\
3.300 &$1.01$&$0.91(0.10)$&$0.52(0.06)$& 3.600 &  $1.07$&$1.29(0.13)$&$0.86(0.08)$ \\
3.325 &$1.04$&$1.13(0.10)$&$0.65(0.06)$& 3.625 &  $1.14$&$1.69(0.13)$&$1.12(0.09)$ \\
3.350 &$1.06$&$1.39(0.09)$&$0.79(0.05)$& 3.650 &  $1.20$&$2.10(0.14)$&$1.40(0.09)$ \\
3.400 &$1.14$&$2.04(0.10)$&$1.16(0.06)$& 3.675 &  $1.28$&$2.51(0.14)$&$1.66(0.10)$ \\
3.450 &$1.23$&$2.79(0.10)$&$1.59(0.06)$& 3.700 &  $1.35$&$2.88(0.15)$&$1.91(0.10)$ \\
3.500 &$1.34$&$3.56(0.11)$&$2.04(0.07)$& 3.750 &  $1.52$&$3.50(0.15)$&$2.32(0.10)$ \\
3.550 &$1.49$&$4.32(0.12)$&$2.47(0.07)$& 3.800 &  $1.70$&$3.99(0.16)$&$2.65(0.11)$ \\
3.600 &$1.66$&$4.96(0.12)$&$2.83(0.07)$& 3.850 &  $1.90$&$4.36(0.16)$&$2.89(0.11)$ \\
3.650 &$1.86$&$5.46(0.12)$&$3.12(0.07)$& 3.930 &  $2.24$&$4.82(0.17)$&$3.19(0.11)$ \\
3.700 &$2.09$&$5.84(0.12)$&$3.34(0.07)$& 4.000 &  $2.55$&$5.14(0.17)$&$3.41(0.11)$ \\
3.850 &$2.93$&$6.57(0.15)$&$3.75(0.09)$&       & 	&            &  \\
4.000 &$3.93$&$6.97(0.16)$&$3.98(0.09)$&       & 	&            &  \\
\hline                                     
\end{tabular}                              
\caption{                                  
\label{ta:p_values}
Numerical values of the pressure for all of our simulation points.
The left column shows the $N_t$=4, whereas the right column shows
the $N_t$=6 data. Both the raw values and the ones scaled by 
$c_{cont}/c_{N_t}$ are given.
}
\end{table}

Let us present the results. In order to show how the different quantities
scale with the lattice spacing we show always $N_t$=4,6 results on the same
plot. In addition, in order to make the relationship with the continuum limit
more transparent we multiply the raw lattice results at finite temporal
extensions ($N_t$=4,6) with $c_{cont}/c_{N_t}$, where the c values are the
results in the free non-interacting plasma (Stefan-Boltzmann limit).  These c
values are summarized in Table \ref{ta:c_values} for the pressure, speed of
sound, and for the quark number susceptibility at $N_t$=4,6 and in the
continuum limit. By this multiplication  the lattice thermodynamic quantities
should approach the continuum Stefan-Boltzmann values for extreme large
temperatures.

Table \ref{ta:p_values} contains our most important numerical results. We
tabulated the raw and normalized pressure values for both lattice
spacings and for all of our simulation 
points. This data set and Eq. (\ref{eq:escs}) were used to obtain the following
figures.  Fig. \ref{fig:eos_pe} shows the equation of state on $N_t$=4,6 lattices.
The pressure (left panel) and $\epsilon$ (right panel) are presented as a
function of the temperature.  The Stefan-Boltzmann limit is also shown.  
Fig. \ref{fig:eos_sc} shows the entropy density (left panel) and the speed of
sound (right panel), which can be obtained by using the pressure and energy
density data (c.f.  $sT$=$\epsilon$+$p$ and $c_s^2$=$dp$/$d\epsilon$) of the
previous Fig. \ref{fig:eos_pe}. Clearly, the uncertainties of the pressure and
those of the energy density cumulate in the speed of sound, therefore it is less
precisely determined.

\begin{figure}
\begin{center}
\epsfig{file=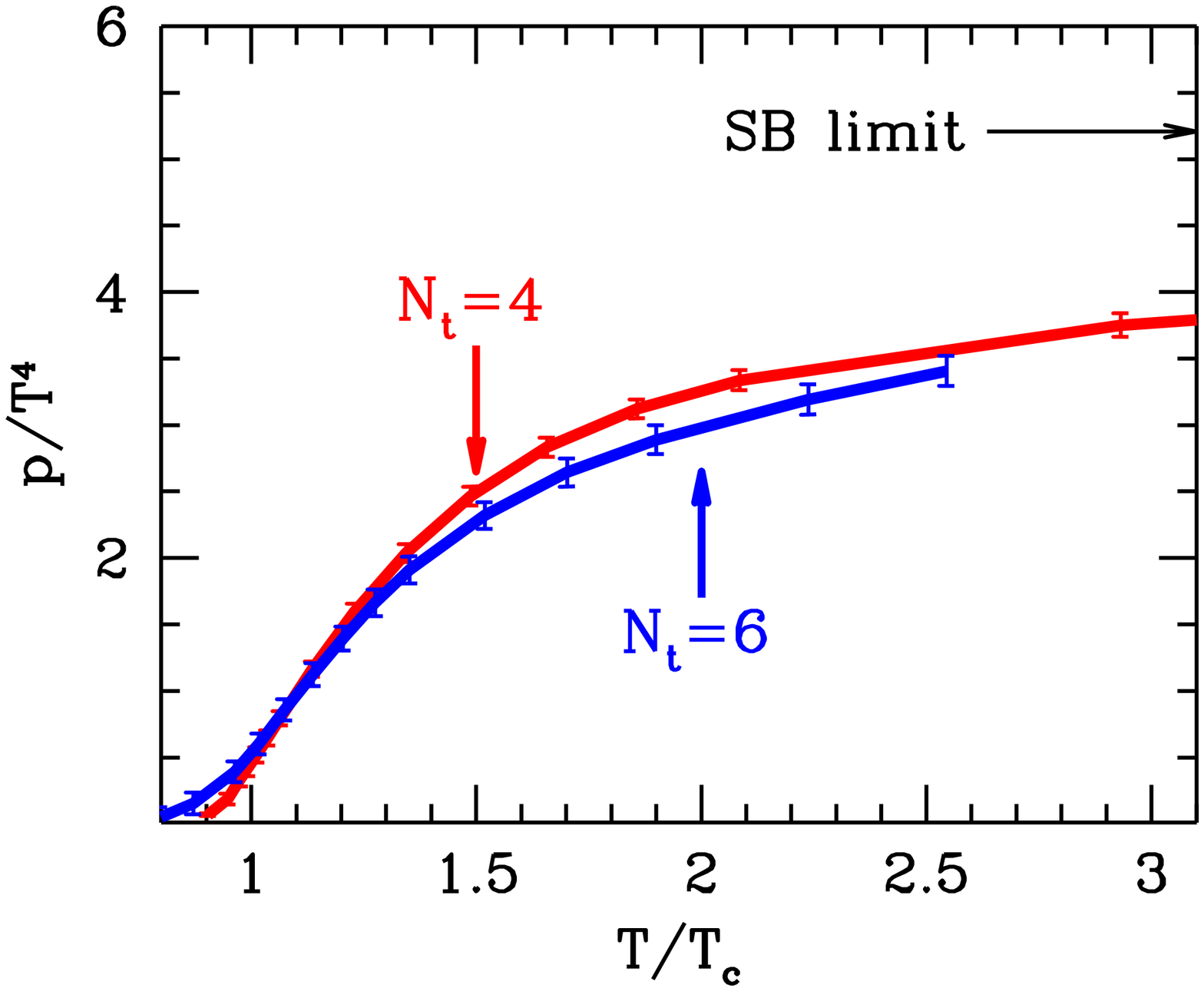,width=7.2cm}
\epsfig{file=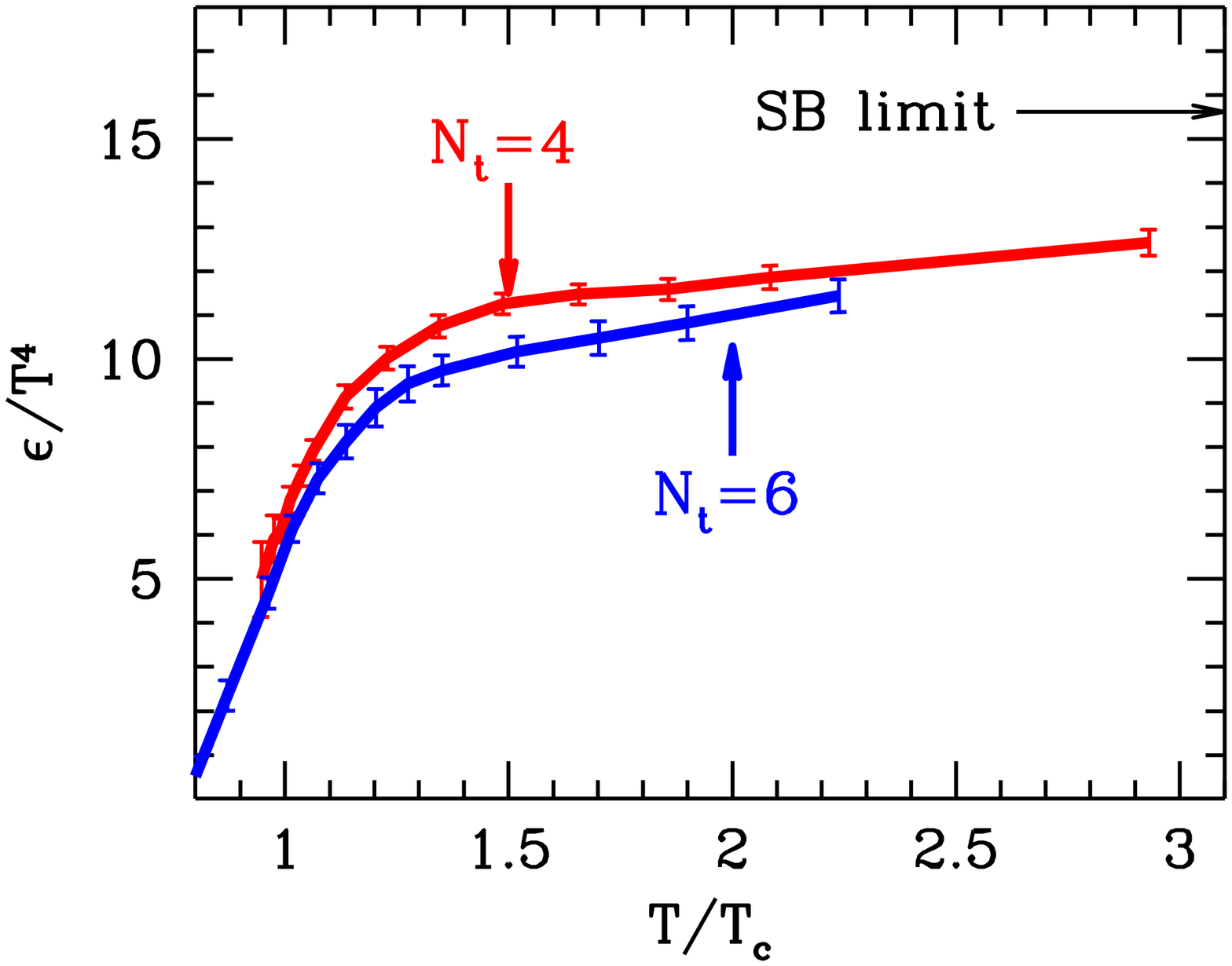,width=7.2cm}
\end{center}
\caption{ \label{fig:eos_pe}
a.) The left panel shows the pressure $p$,
as a function of the temperature. Both $N_t$=4 (red, upper curve)
and $N_t$=6 (blue, lower curve) data are obtained along the LCP. They 
are normalized by $T^4$ and scaled by $c_{cont}/c_{N_t}$ (see text
and Table \ref{ta:c_values}). 
In order to lead the eye lines connect the data points.
b.) The right panel is the energy density ($\epsilon$), red (upper) and blue (lower) for $N_t$=4 and
6 respectively. This result was obtained directly from the pressure. 
}
\end{figure}
\begin{figure}[t!]
\begin{center}
\epsfig{file=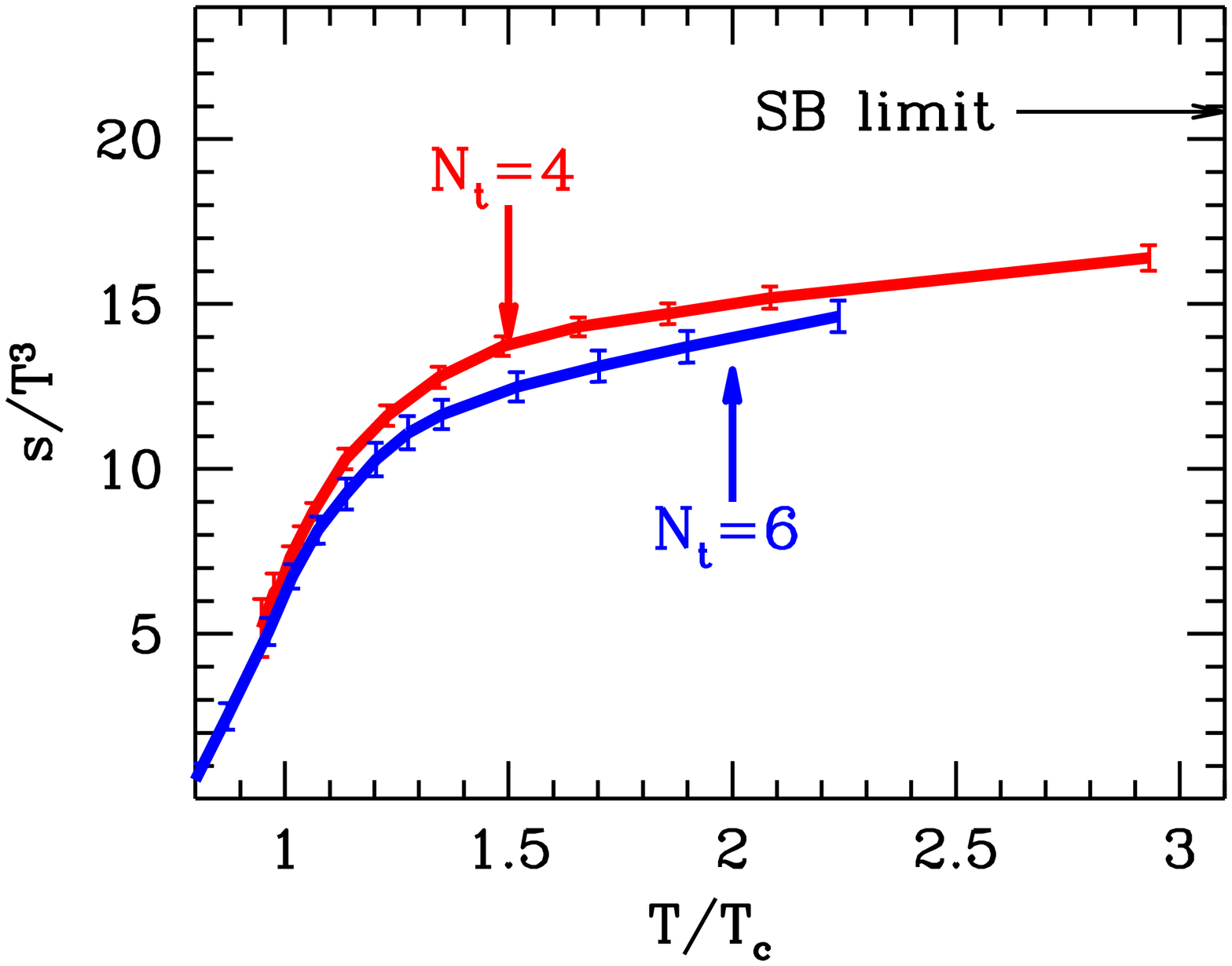,width=7.2cm}
\epsfig{file=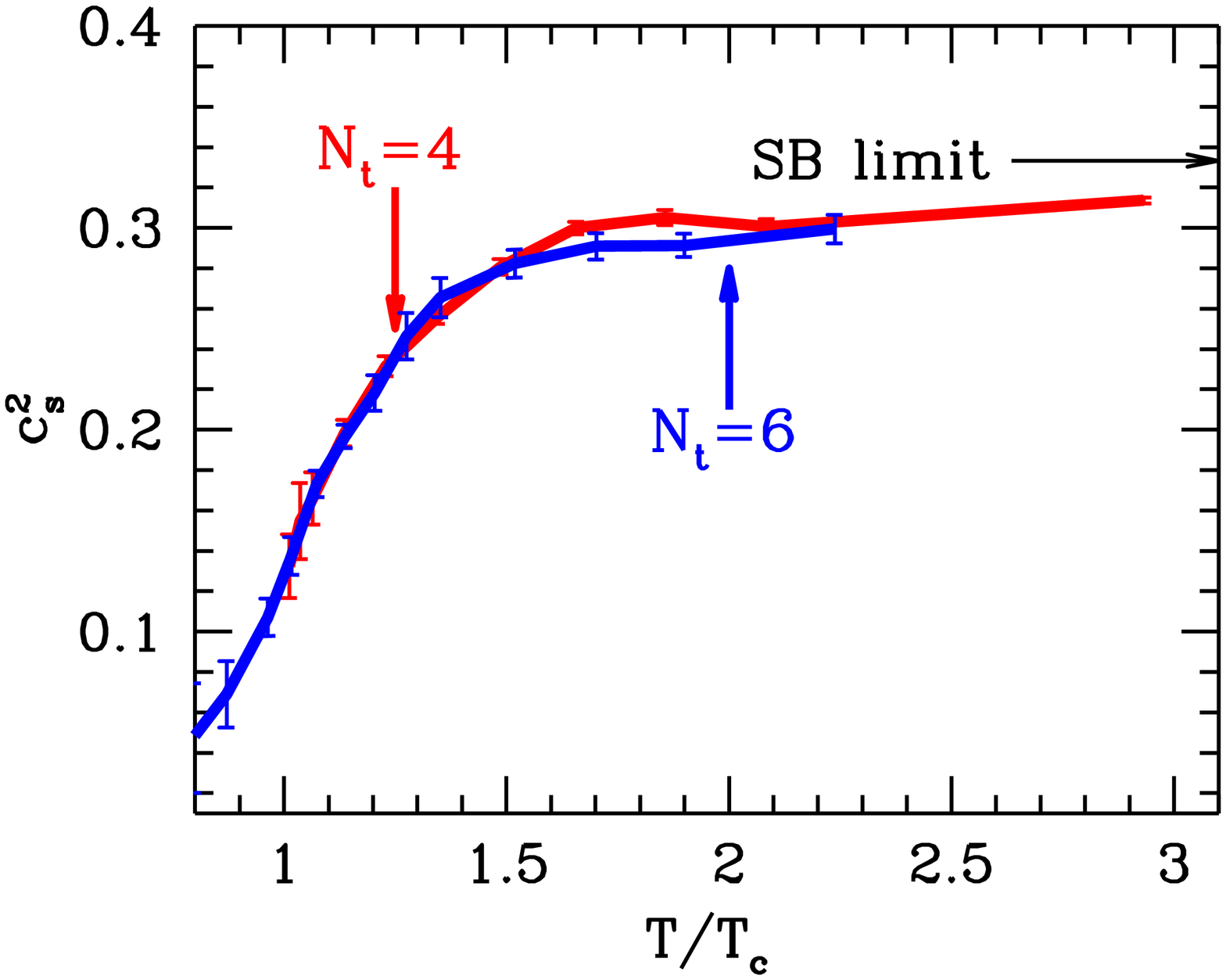,width=7.2cm}\\
\epsfig{file=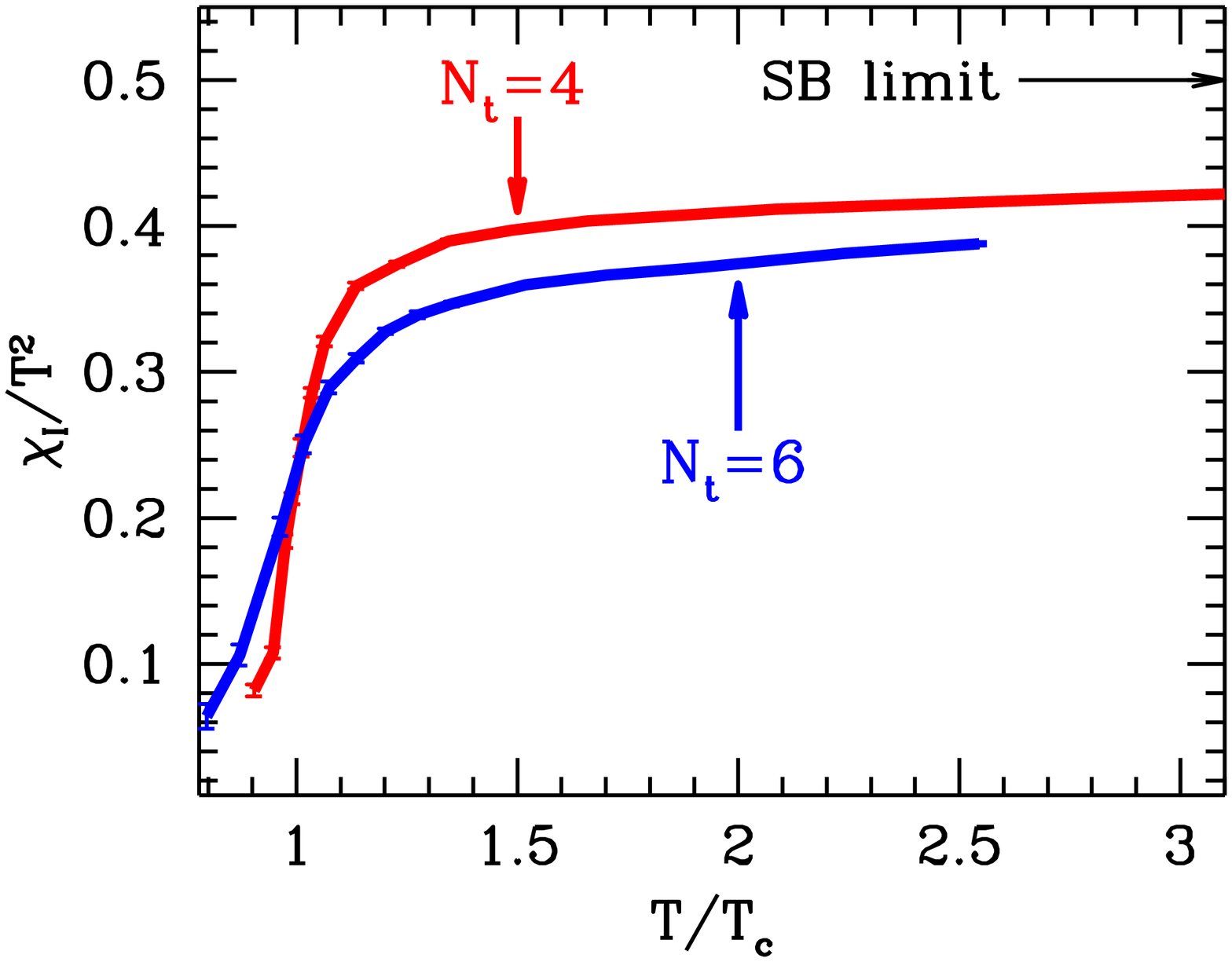,width=7.2cm}
\epsfig{file=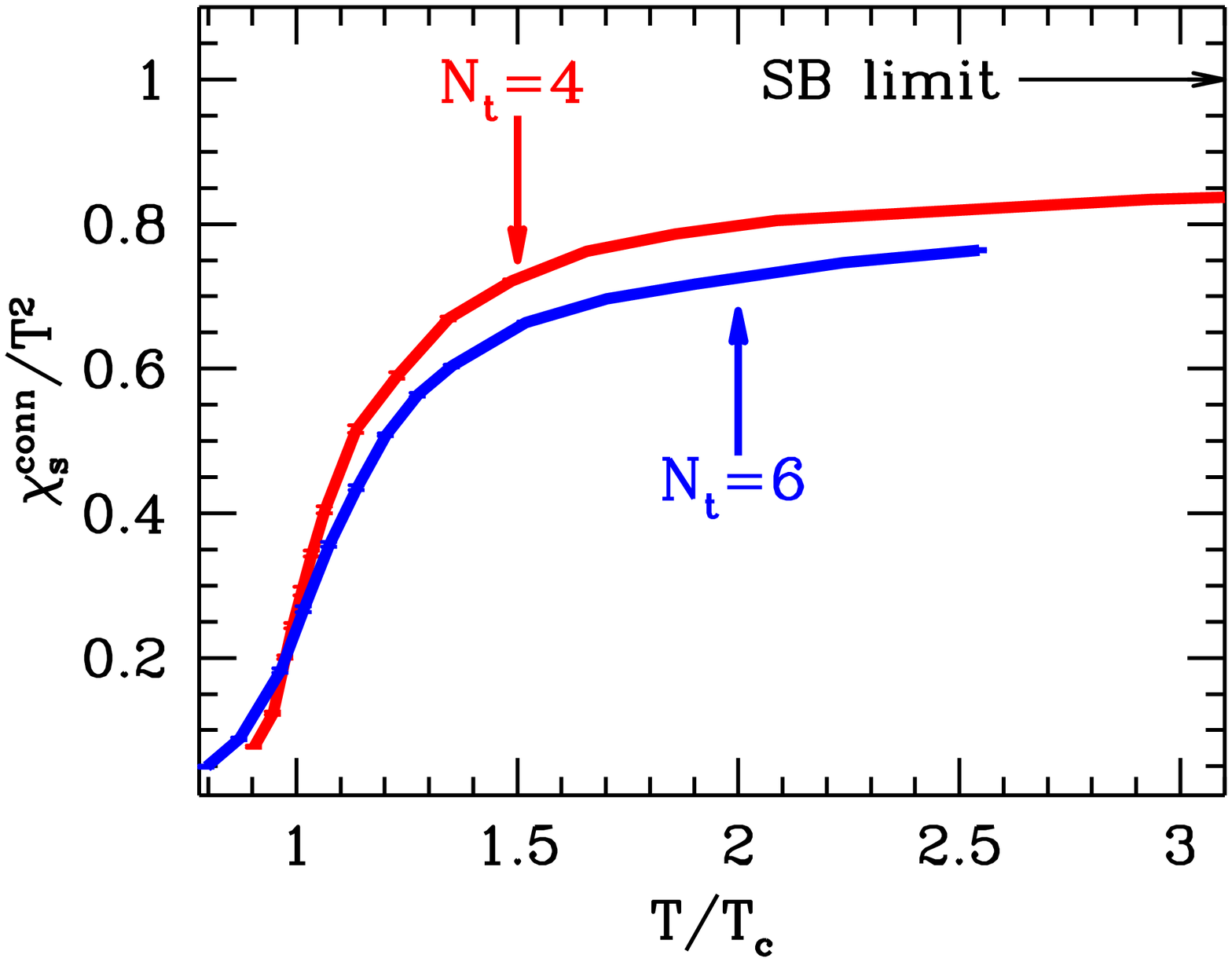,width=7.2cm}
\end{center}
\caption{\label{fig:eos_sc}
The 
entropy density (upper left panel, normalized by $T^3$) and the
speed of sound (upper right panel). 
The isospin susceptibility 
(lower left panel, normalized by $T^2$) and the connected part of the strangeness susceptibility
(lower right panel, normalized by $T^2$). The labeling is the same as for Fig. \ref{fig:eos_pe}.
}
\end{figure}

Light and strange quark number susceptibilities ($\chi_{ud}$ and $\chi_s$) are defined via
\cite{Bernard:2004je}
\begin{align}
\frac{\chi_{q}}{T^2}=\frac{N_t}{N_s^3}\left.\frac{\partial^2 \log Z}{\partial \mu_{q} ^2
}\right|_{\mu_{q}=0}, 
\end{align}
where $\mu_{ud}$ and $\mu_s$ are the light and strange quark chemical
potentials (in lattice units). With the help of the quark number operators 
\[Q_{q}=\frac{1}{4}\frac{\partial }{\partial
\mu_q}
\log \det (\Dsl+m_{q}),\] 
the susceptibilities can be written as
\[
\frac{\chi_{q}}{T^2}=\frac{N_t}{N_s^3}\left(\langle Q_q^2
\rangle_{\mu_q=0} + \left\langle \frac{\partial
Q_q}{\partial \mu_q}\right\rangle_{\mu_q=0}\right).
\]
The first term is usually referred as disconnected, the second as connected part. 
The connected part of the light quark number susceptibility is 2 times
the susceptibility of the isospin number ($\chi_I$). 
It is presented on the left panel of Fig. \ref{fig:eos_sc}. 
For our statistics and evaluation method the disconnected 
parts are all consistent with zero and their value is far smaller than those
of the connected parts. 
The right panel of Fig. \ref{fig:eos_sc} contains the connected part of the strange number
susceptibility.

\chapter{Summary}

The focus of this work was implementing and applying dynamical fermions in
lattice QCD.  

The first part was exploring the unknown territory of dynamical algorithms for
the overlap fermion. The conventional dynamical algorithm (Hybrid Monte Carlo)
fails to work for this type of fermion. The failure was identified, it is due
to the change in the topological charge. We have proposed a possible workaround
for this problem, and shown that with this modification the algorithm works
reasonably. There were further modifications necessary to increase the
performance to an acceptable level. At the end we have determined the
topological susceptibility as the function of quark mass. The result was the
first in lattice QCD which has shown the suppression of the susceptibility for
small quark masses at finite lattice spacing.  The simulations however
considerably more expensive than the ones with other fermion formulations.
Therefore it still seems to be more beneficial to investigate the current
algorithms or to come up with new ones than starting larger scale physics
projects with dynamical overlap fermions.

The second part of the work was the determination of bulk properties of the
finite temperature QCD matter using dynamical improved staggered fermions. This
was a large scale project and attempted to give final answers based on a first
principles approach. In order to achieve our goal we have done the simulations
for physical values of the quark masses and carried out a continuum
extrapolation wherever possible.  Firstly we have determined the nature of the
transition based on a finite size scaling analysis of a susceptibility type
quantity. The transition turned out to be a smooth crossover, that is no sign
of singularity has been found in the thermodynamical limit. Secondly we have
determined some typical temperatures in physical units where this crossover
takes place. Since there is no singularity, there is no unique temperature
which can be identified as a critical temperature.  We have calculated transition
temperatures from various quantities: peak of the chiral susceptibility,
inflection point of the quark number susceptibility and inflection point of the
Polyakov loop. The second two gave significantly higher temperature values,
than the first one.  Thirdly we have presented result on the equation of state
towards the continuum limit.  From the currently available lattice spacings a
reliable continuum extrapolation cannot be carried out, this is left for the future.

\section*{Acknowledgments}

First of all I would like to thank the help of my advisor Zolt\'an Fodor.  His
very good sense in choosing the right topics was inevitable to accomplish this
thesis. I am also very grateful to S\'andor Katz. They both have a great part
in my results with lots of ideas, hints.

In the overlap project we had a nice collaboration with Gy\H oz\H o Egri, I
thank him. It was a pleasure to work together at various stages of the
staggered project with Yasumichi Aoki and Gerg\H o Endr\H odi. I thank the
discussions with Christian Hoelbling, D\'aniel N\'ogr\'adi, Stefan Krieg,
Tam\'as Kov\'acs, Anna T\'oth and B\'alint T\'oth.

The numerical computations for this thesis were carried out on the following
supercomputers: on BlueGene/L at FZ J\"ulich, on PC clusters at Bergische
Universit\"at, Wuppertal and on PC clusters at E\"otv\"os University, Budapest.

\bibliographystyle{unsrt}
\bibliography{thesis}

\end{document}